\documentclass[journal,final,twoside]{IEEEtran}
\usepackage{cite}

\usepackage{amssymb}
\usepackage{blindtext}

\usepackage{graphicx}

\usepackage[font=small]{caption}
\usepackage[labelformat=simple,font=footnotesize]{subcaption}

\usepackage{amsfonts,bm}
\usepackage[cmex10]{amsmath}
\usepackage[colorlinks=true, allcolors=blue]{hyperref} 
\usepackage{bbm}
\usepackage{paralist}

\DeclareMathOperator*{\argmax}{arg\,max}
\newcommand{\iP}[1]{\mathrm{P}({#1})} 
\newcommand{\E}[1]{\mathrm{E}\!\left[\,{#1}\,\right]}   
\newcommand{\iE}[1]{\mathrm{E}[\,{#1}\,]}   
\newcommand{\var}[1]{\mathrm{var}\!\left({#1}\right)}   
\newcommand{\ivar}[1]{\mathrm{var}({#1})}   
\newcommand{\bias}{\operatorname{bias}}
\newcommand{\Frac}[2]{{{#1}/{#2}}}  
\newcommand{\eqlabel}[1]{ \stackrel{(#1)}{=} }
\newcommand{\geqlabel}[1]{ \stackrel{(#1)}{\geq} }
\def\smid{\,|\,}  
\def\sMid{\,;\,}  
\newcommand{\openCaserl}{\left\{ \begin{array}{@{\,}rl}}
\newcommand{\closeCase} {\end{array} \right.}
\DeclareMathOperator{\Poisson}{Poisson}

\def\Ltilde{\widetilde{L}}
\def\mtilde{\widetilde{m}}
\def\Mtilde{\widetilde{M}}
\def\Ttilde{\widetilde{T}}
\def\Xtilde{\widetilde{X}}
\def\Ytilde{\widetilde{Y}}

\def\CR{Cram{\'e}r--Rao}

\def\etaBaseline{\widehat{\eta}_{\rm baseline}}
\def\etaOracle{\widehat{\eta}_{\rm oracle}}
\def\etaCTQM{\widehat{\eta}_{\rm CTQM}}
\def\etaCTLQM{\widehat{\eta}_{\rm CTLQM}}
\def\etaCTML{\widehat{\eta}_{\rm CTML}}
\def\etaDTQM{\widehat{\eta}_{\rm DTQM}}
\def\etaDTLQM{\widehat{\eta}_{\rm DTLQM}}
\def\etaDTML{\widehat{\eta}_{\rm DTML}}

\def\PposMfull{{1 - e^{-\eta}}}
\def\PposM{\rho}

\newlength{\threewidewidth}
\begin{document}
\title{Time-Resolved Focused Ion Beam Microscopy:
Modeling, Estimation Methods, and Analyses}

\author{Minxu~Peng,
        John~Murray-Bruce,
        and~Vivek~K~Goyal
\thanks{M. Peng and V. K. Goyal are with the Department of Electrical and Computer Engineering,
Boston University, Boston, MA 02215 USA (e-mail: mxpeng@bu.edu; v.goyal@ieee.org).}%
\thanks{J. Murray-Bruce is with the Department of Computer Science and Engineering, University of South Florida, Tampa, FL 33620 USA (e-mail: murraybruce@usf.edu).}%
\thanks{This work was supported in part by the US National Science Foundation under Grant No.\ 1422034 and Grant No.\ 1815896.}}

\maketitle

\begin{abstract}
In a focused ion beam (FIB) microscope,
source particles interact with a small volume of a sample to generate
secondary electrons that are detected,
pixel by pixel,
to produce a micrograph.
Randomness of the number of incident particles causes excess variation in the micrograph,
beyond the variation in the underlying particle--sample interaction.
We recently demonstrated that joint processing of multiple time-resolved measurements from a single pixel
can mitigate this effect of source shot noise
in helium ion microscopy.
This paper is focused on establishing a rigorous framework for understanding the potential for this approach.
It introduces idealized continuous- and discrete-time abstractions of FIB microscopy
with direct electron detection
and estimation-theoretic limits of imaging performance under these measurement models.
Novel estimators for use with continuous-time measurements are introduced and analyzed,
and estimators for use with discrete-time measurements are analyzed and shown to approach
their continuous-time counterparts as time resolution is increased.
Simulated FIB microscopy results are consistent with theoretical analyses and demonstrate that
substantial improvements over conventional FIB microscopy image formation are made possible by time-resolved measurement.
\end{abstract}


\section{Introduction}
\IEEEPARstart{T}{he} ability to image the structure of a sample at nanoscale resolution
using microscopes that scan samples with a focused beam of particles
is critical in material science and the life sciences.
In a scanning electron microscope (SEM)~\cite{McMullan1995}, 
a focused electron beam is raster scanned over the sample,
causing the sample to emit secondary electrons (SEs).
An SEM is capable of providing information regarding 
composition and distribution of sample components. 
Resembling an SEM, a focused ion beam (FIB) microscope~\cite{erwin1969field}
instead uses a focused beam of ions, 
such as gallium, helium, or xenon.
As one member of the FIB microscope family,
a helium ion microscope (HIM)~\cite{WardNE:06} offers many advantages 
compared to an SEM,
leading to widespread use in
semiconductor and biological imaging~\cite{emmrich2016nanopore, joens2013helium, bazargan2012electronic, wirtz2019imaging}. 
The interaction volume with the sample is much smaller 
for ions compared to that for electrons, 
resulting in
higher contrast~\cite{scipioni2009understanding}.
Higher particle mass also leads to
higher overall SE yield~\cite{ramachandra2009model, morgan2006introduction}.
Furthermore, to image insulating samples with an SEM, 
prior deposition of a conductive coating is required 
to prevent the accumulation of electron charges on the sample; 
an HIM uses an electron flood gun to prevent charge accumulation,
thus avoiding the masking of subtle features by a coating~\cite{hlawacek2014helium}.

The number of incident ions
determines the number of sample interactions that can be measured.
Hence, increasing this number---through increased beam current or increased dwell time---will ideally create
measurements
that are more informative about the mean number of SEs per incident ion, which is the sample property of interest.
However, the vastly greater mass of ions compared to electrons
(by a factor of $7.3 \times 10^3$ for helium or $1.3 \times 10^5$ for gallium)
makes sputtering much more significant for FIB microscopy than for SEM\@.
Various studies have shown how the sputtering damage induced by helium ions 
evolves with increasing numbers of incident ions~\cite{livengood2009subsurface, castaldo2009influence, orloff1996fundamental},
and this damage is often determinative of the best possible image quality. 
Dose is conventionally defined as the number of incident ions per unit area,
and dose limits to prevent significant damage have been measured for certain
materials and imaging configurations.
For example, a safe imaging dose for suspended graphene is as low as $10^{13}$ to $10^{14}$ per square centimeter~\cite{fox2013helium}.
It follows, for example, that a $(10\,\mathrm{nm})^2$
pixel should be subjected to only 10 to 100 ions.
In our abstractions, we dispense with spatial dimensions and hereafter express doses per pixel rather than per unit area;
expressed per pixel, finer resolution necessitates lower dose limits.
While randomized subsampling combined with regularized 
reconstruction can sometimes yield high-quality images from reduced doses~\cite{anderson2013sparse, donati2017compressed, boughorbel2017high},
this requires piecewise smooth image structure.
Here we restrict our attention to pixelwise acquisition and estimation methods that do not rely on such assumptions.

We recently introduced the concept of \emph{time-resolved} (TR) measurements in FIB microscopy
to mean dividing any given pixel dwell time $t$ into $n$ dwell times $t/n$
for some integer $n > 1$~\cite{peng2020source}.
Without proofs,
we gave theoretical evidence that a set of TR measurements is fundamentally more informative than a single measurement with the same total dose.
Experiments with HIM data
demonstrated mean-squared error (MSE) improvement by about a factor of 4 
at doses of 1.0 and 2.5 incident ions per pixel.
Importantly, this use of time resolution is entirely for the purpose of making the measurements more informative without increasing the total dose.
It is not for imaging of dynamic samples and hence not comparable to any previous use of time resolution in microscopy.

The Zeiss ORION NanoFab HIM used in~\cite{peng2020source},
like most commercial FIB microscopes, uses
indirect electron detection
with a scintillator and photomultiplier tube.
Although not yet prevalent, direct electron detection
offers higher
signal-to-noise ratio (SNR)
as it avoids
the statistical noises brought by
electron--photon conversion and additional readout noise~\cite{mcmullan2009detective}.
For example, Yamada et al.~\cite{yamada1991electron}
demonstrated that the SNR of direct electron detection is 2.5 times higher
than that of indirect mode
at low dose.
Furthermore, direct electron detection technology has also been applied 
to imaging in transmission electron microscopy
to improve resolution~\cite{jin2008applications}.

In this paper,
we develop comprehensive theoretical results for FIB microscopy with time-resolved direct SE detection.
Indirect detection introduces many sources of noise, including spatial nonuniformity in the scintillator response,
nonideal light transport from the scintillator to the photomultiplier tube, and variations in pulses generated by the photomultiplier tube.
Though these effects were empirically modeled in~\cite{peng2020source},
including them here would make already lengthy
expressions considerably more complicated
and more difficult to interpret.
Restricting attention to direct detection
allows us to concentrate on the implications of TR sensing
for mitigation of source shot noise,
separated from the effects of detection noise.
While some results presented here add rigor to statements in~\cite{peng2020source},
we more importantly introduce a new continuous-time abstraction for FIB microscopy
that yields to more elegant analyses while also representing the ultimate limit of this technology.
In addition to maximum likelihood (ML) estimation,
our study includes plug-in estimators
for which we can complete analytical performance analyses and
that are easily generalized to settings in which SE detection is indirect.

\subsection{Main Contributions}
\begin{itemize}
\item \emph{A new continuous-time probabilistic model for FIB microscopy 
wherein the data at any one pixel are related to a marked Poisson process.}
Conventional and continuous- and discrete-time time-resolved observation models
are different functions of the marked Poisson process.
\item \emph{Fisher information analyses.}
We show that,
at any ion dose level,
continuous-time time-resolved measurements
have Fisher information matching an upper bound
that conventional measurements meet only in a low-dose limit.
For conventional measurements,
asymptotic expressions presented without proof in~\cite{peng2020source}
are proven here.
\item \emph{Estimator analyses.}
Biases and variances of quotient-mode estimators in
both the continuous- and discrete-time settings are derived.
Convergence of the discrete-time estimator's performance to the performance of the continuous-time estimator is proven.
\end{itemize}

\subsection{Outline}
We introduce our mathematical abstraction for the operation of a FIB microscope in Section~\ref{sec:Models}.
This leads to four measurement models:
an unimplementable oracle model,
conventional measurement, and
continuous- and discrete-time time-resolved measurement.
The oracle and conventional cases are analyzed within Section~\ref{sec:Models}.
Section~\ref{sec:CTTR_measurement} develops the novel continuous-time case in detail.
Measurement distributions are derived, three estimators are introduced and simulated,
and the performance of a quotient-mode estimator is rigorously analyzed.
Section~\ref{sec:DTTR_measurement} develops the discrete-time case first introduced in~\cite{peng2020source} in detail.
Three estimators are simulated,
and the performance of the discrete-time quotient-mode estimator is rigorously analyzed.
Convergences of Fisher information and quotient-mode estimator performance to their
continuous-time counterparts are shown.
Section~\ref{sec:simulated-microscopy} compares all the estimators in a simulated HIM experiment,
demonstrating substantial improvement of the time-resolved methods over the conventional interpretation of the collected data.
Section~\ref{sec:conclusion} provides concluding comments on how mean SE yield influences
the advantages of TR methods, the time resolution necessary to capitalize on these advantages,
the roles of the various estimators, and generalizations to indirect SE detection.

Table~\ref{tab:symbol_acronyms} summarizes the variables, symbols, and acronyms used in the manuscript.

\begin{table}
    \caption{List of symbols and acronyms}
    \centering
    \footnotesize
    \begin{tabular}{rl}
    \hline
    \hline
    $\eta$        & mean secondary electron yield\\
    $\etaOracle$  & oracle estimator \eqref{eq:oracle-estimator} \\
    $\etaBaseline$ & baseline estimator \eqref{eq:eta-baseline} \\
    $\etaCTQM$    & continuous-time quotient mode estimator \eqref{eq:eta_QM_conti} \\
    $\etaCTLQM$   & continuous-time Lambert quotient mode estimator \eqref{eq:eta_LQM_conti} \\
    $\etaCTML$    & continuous-time maximum likelihood estimator \eqref{eq:eta_trml_conti} \\
    $\etaDTQM$    & discrete-time quotient mode estimator \eqref{eq:eta_DTQM} \\
    $\etaDTLQM$   & discrete-time Lambert quotient mode estimator \eqref{eq:DTLQM_expression} \\
    $\etaDTML$    & discrete-time maximum likelihood estimator \eqref{eq:eta_DTML} \\
    $\lambda$     & ion dose per pixel \\
    $\Lambda$     & dose per unit time\\
    $\rho$        & $\iP{X_i > 0} = 1 - e^{-\eta}$ \\
    $i$           & ion index \\
    $k$           & discrete time index \\
    $n$           & number of subacquisitions \\
    $p$           & $\iP{Y_k > 0} = 1 - \exp(-(\Frac{\lambda}{n})(1-e^{-\eta}))$ \\
    $t$           & dwell time\\
    $\mathcal{I}_Z( \eta \sMid \lambda )$ & Fisher information about $\eta$ in $Z$ with $\lambda$ available \\
    $L$           & number of subacquisitions with positive SEs\\
    $\Ltilde$     & zero-truncated version of $L$ \\
    $M$           & number of incident ions\\
    $\Mtilde$     & number of incident ions yielding positive SEs \\
    $T_i$         & $i$th ion incidence time \\
    $\Ttilde_i$   & incidence time of $i$th ion to yield positive SEs \\
    $X_i$         & SEs detected due to $i$th incident ion \\
    $\Xtilde_i$   & SEs detected due to $i$th ion to yield positive SEs \\
    $Y$           & total detected SEs \\
    $Y_k$         & detected SEs in $k$th subacquisition \\
    $\Ytilde_k$ & positive SE counts in a subacquisition \\
    CRB    & {\CR} bound \\
    CTTR   & continuous-time time-resolved (see observation \eqref{eq:CTTR-observation}) \\
    DTTR   & discrete-time time-resolved (see observation \eqref{eq:DTTR-observation}) \\
    FI     & Fisher information\\
    LQM    & Lambert quotient mode\\
    ML     & maximum likelihood\\
    QM     & quotient mode\\
    SE     & secondary electron\\
    TR     & time-resolved\\
    \hline
    \hline
    \end{tabular}
    \label{tab:symbol_acronyms}
\end{table}

\section{Measurement Models and Basic Analyses}
\label{sec:Models}

\subsection{Physical Abstractions}
\label{sec:physical-abstraction}
Throughout this paper, we model the incident ions
at the sample to be imaged as a Poisson process with known rate $\Lambda$ per unit time.
Imaging proceeds by raster scanning with known dwell time $t$ at each pixel.
Hence, the number of ions $M$ incident on a pixel is a Poisson random variable with known parameter $\lambda = \Lambda t$.
Since pixel area is not relevant in our abstraction, we refer to $\lambda$ as the dose.
The interaction of the $i$th incident ion with the sample causes a number $X_i$ of SEs to be detected.
All of these SE counts are mutually independent Poisson random variables with parameter $\eta$,
independent of the incident-ion Poisson process,
and estimation of the mean SE yield $\eta$ is the objective of the imaging experiment.
Our analysis is for each pixel separately, so no pixel indexing is necessary.

\begin{figure}
  \begin{center}
    \begin{subfigure}{0.95\linewidth}
      \includegraphics[width=\linewidth]{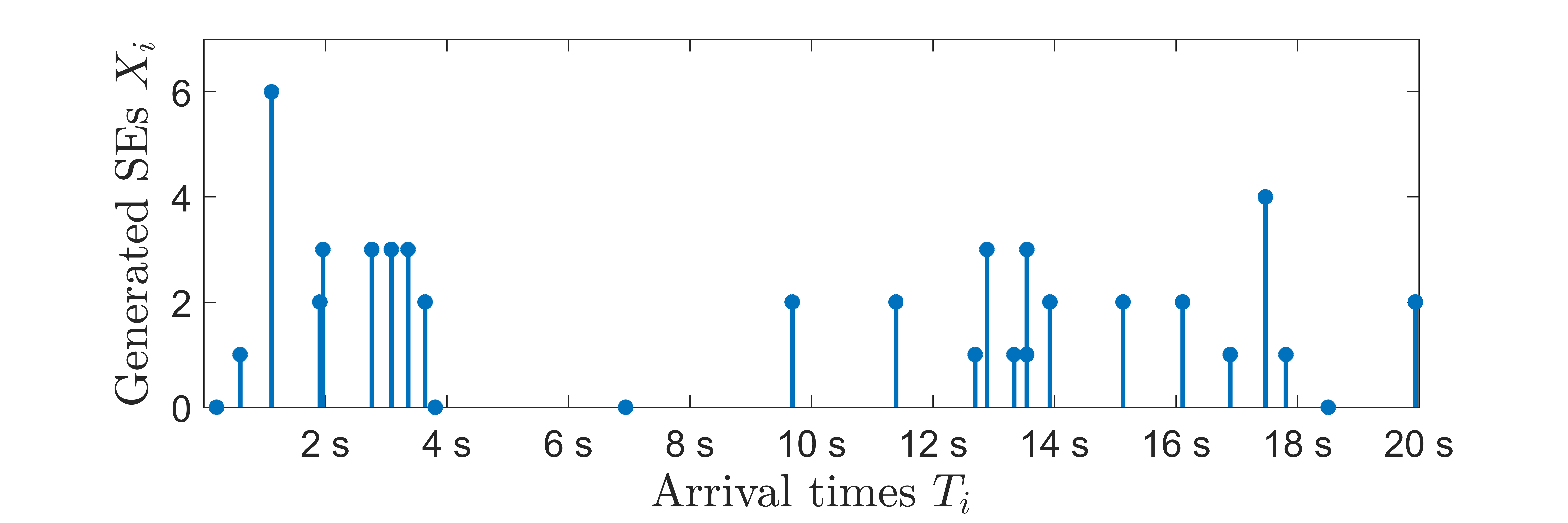}
      \subcaption{Underlying marked Poisson process.}
      \label{subfig:marked-CT}
    \end{subfigure}
    \begin{subfigure}{0.95\linewidth}
      \includegraphics[width=\linewidth]{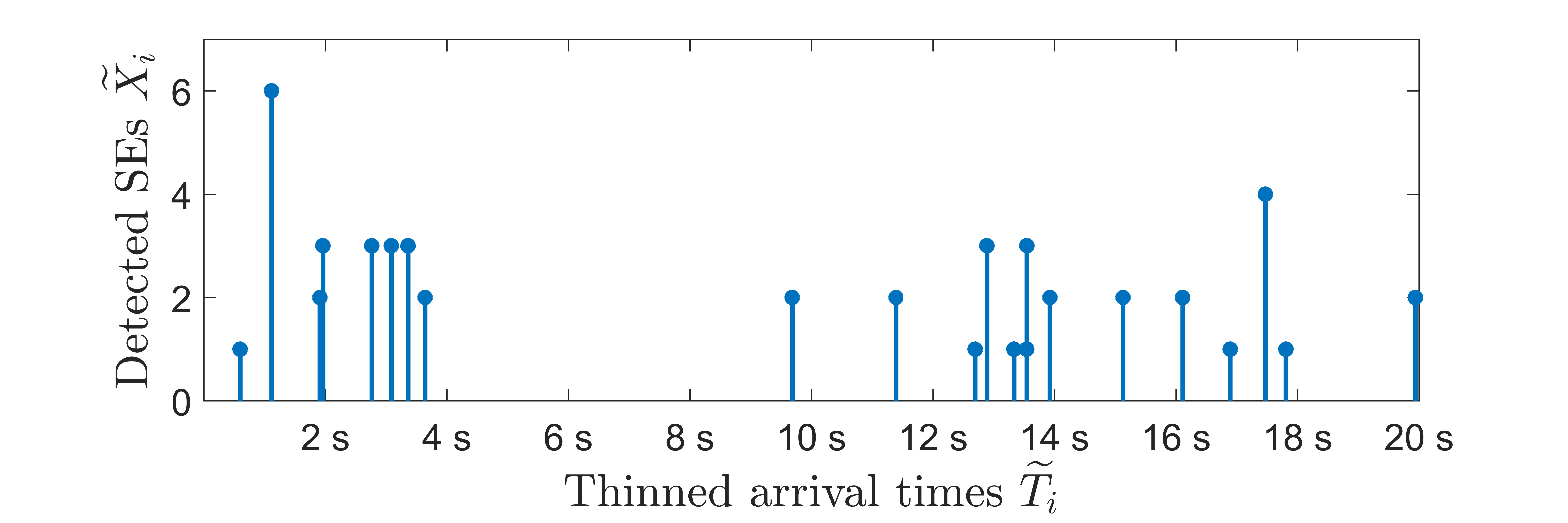}
      \subcaption{Process observed where mark is positive.}
      \label{subfig:thinned-process}
    \end{subfigure}
    \begin{subfigure}{0.95\linewidth}
      \includegraphics[width=\linewidth]{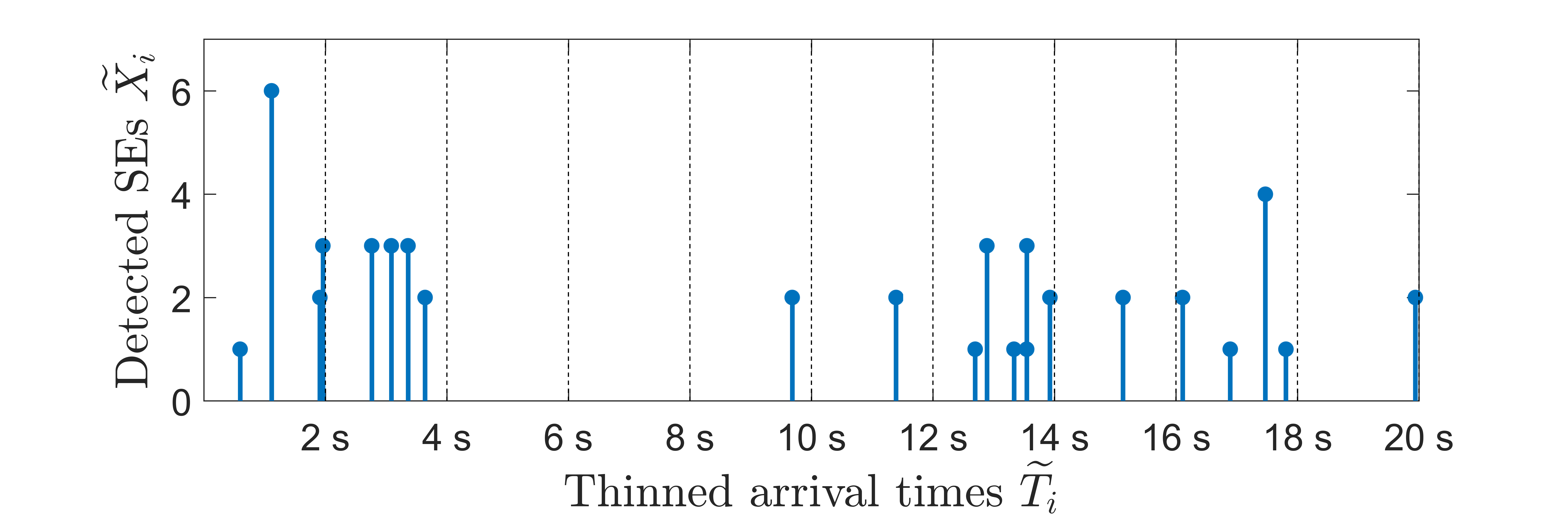}
      \subcaption{Dwell time divided into $n=10$ subintervals.}
      \label{subfig:divide-time}
    \end{subfigure}
    \begin{subfigure}{0.95\linewidth}
      \includegraphics[width=\linewidth]{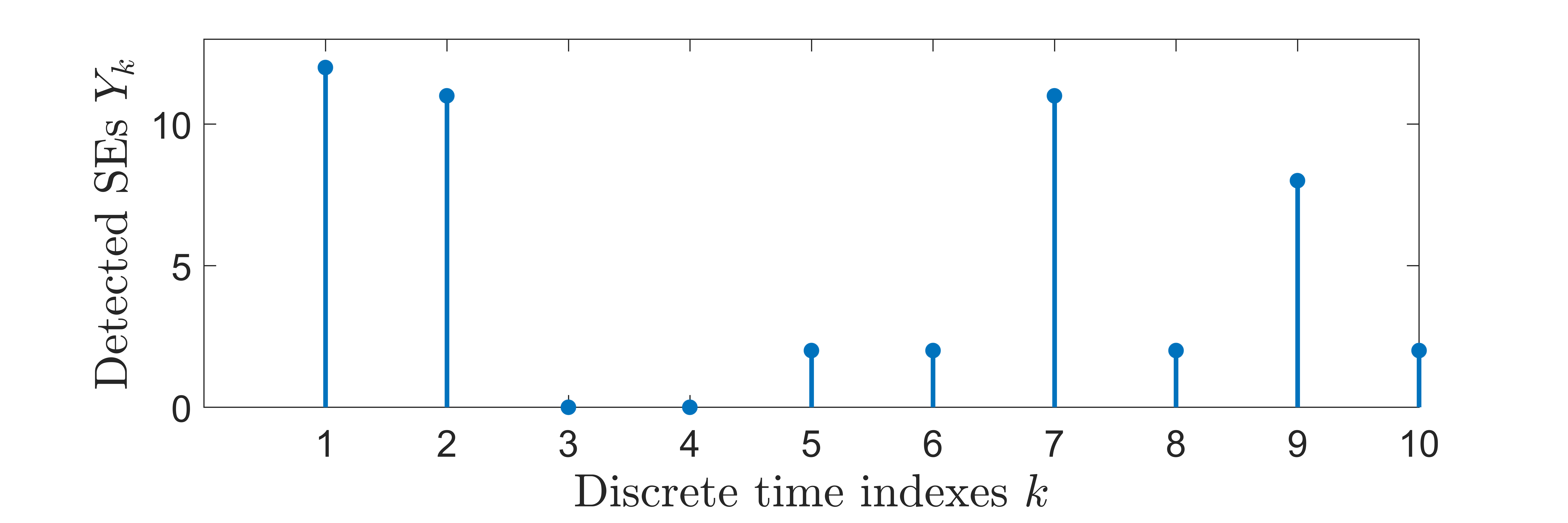}
      \subcaption{Discrete-time time-resolved measurement, $n = 10$.}
      \label{subfig:DT-measurements}
    \end{subfigure}
  \end{center}
    \caption{Illustration of the random processes generated in the abstraction of FIB microscopy through one possible realization.
    (a) The underlying marked Poisson process
    $\{(T_1,X_1),\, (T_2,X_2),\, \ldots \}$,
    with ion incident at times $T_1,\,T_2,\,\ldots$
    generating detected SE counts $X_1,\,X_2,\,\ldots$.
    (b) The marked Poisson process
    $\{(\Ttilde_1,\Xtilde_1),\, (\Ttilde_2,\Xtilde_2),\, \ldots \}$,
    produced by discarding the ions for which no SEs are detected.
    (c) Illustration of dividing dwell time of $t = 20$\,s into $n = 10$ subintervals of equal length.
    (d) The resulting discrete-time SE count process.}
    \label{fig:marked-process-illustration}
\end{figure}

Note that a somewhat high 1 pA beam current corresponds to a
rate of $6.2 \times 10^{6}$ ions per second,
or a mean ion interarrival time of 160 ns.
The interaction between an incident ion and the sample
and the subsequent detection of SEs
occurs within a few femtoseconds~\cite{LiMD:19}.
With the SE detections happening so quickly,
we abstract the SE detections caused by an incident ion
to be simultaneous with the ion incidence.
Thus, the model
can be described as a marked Poisson process
$\{(T_1,X_1),\, (T_2,X_2),\, \ldots \}$,
where
$(T_1,\, T_2,\, \ldots)$ is the arrival time sequence of the ions.
The ion count $M$ is the largest $i$ such that $T_i \leq t$
(with $M=0$ when $T_1 > t$).
One realization on an interval $[0,t]$ is illustrated in Fig.~\ref{subfig:marked-CT}.
Note that the arrival times (horizontal) are arbitrary positive real numbers and
the marks (vertical) are nonnegative integers.

Since cases of $X_i=0$ produce no detected SEs,
the corresponding ion arrival time is not observable in practice.
Thus consider also the thinned process
$\{(\Ttilde_1,\Xtilde_1),\, (\Ttilde_2,\Xtilde_2),\, \ldots \}$,
where
$\Ttilde_i$ is the arrival time of the $i$th ion that produces a \emph{positive} number of detected SEs
and $\Xtilde_i$ is the corresponding number of detected SEs.
Define $\Mtilde$ to be the largest $i$ such that $\Ttilde_i \leq t$
(with $\Mtilde=0$ when $\Ttilde_1 > t$).
Note that the thinned process is also a marked Poisson process because the events of the form $\{X_i = 0\}$,
which determine whether an arrival in the original Poisson process is retained,
are independent of the arrival time process.
Fig.~\ref{subfig:thinned-process}
illustrates the thinned process for the realization of the underlying process in 
Fig.~\ref{subfig:marked-CT}.

Now suppose the observation time interval $[0,t]$ is evenly divided into $n$ subintervals of length $t/n$.
Counting the total number of SEs detected in each subinterval produces a discrete-time, discrete-valued
random process:
\begin{equation}
    \label{eq:Y_k-def}
    Y_k = \sum_{ \{ i \ : \ T_i \in [(k-1)t/n,\,kt/n) \} } X_i,
    \qquad
    k = 1,\,2,\,\ldots,\,n.
\end{equation}
We call an observation over a subinterval a \emph{subacquisition}.
Fig.~\ref{subfig:divide-time}
illustrates the partition of $[0,t]$ into subintervals and
Fig.~\ref{subfig:DT-measurements}
illustrates the resulting discrete-time process
for the realization of the underlying process in 
Fig.~\ref{subfig:marked-CT}.
Because of the independence of a Poisson process over disjoint intervals,
$\{Y_k\}_{k=1}^n$ is an independent and identically distributed (i.i.d.) process.
We can also view $\{Y_k\}_{k=1}^n$
as a marked Bernoulli process,
where $\{Y_k > 0\}$ indicates an arrival in discrete time slot $k$
and the ``mark'' is then the (nonzero) value $Y_k$.

The abstraction described here applies similarly to SEM and FIB microscopy.
The main difference is the typical values of the mean SE yield $\eta$.
In SEM,
neglecting topographical effects (which tend to increase yield),
for a sample with atomic number up to 83,
the SE yield at the maximizing electron energy
is typically 0.6 to 2~\cite{LinJ:05}.
In FIB microscopy, SE yield is typically between 1 and 8~\cite{notte2007introduction}.
The advantages of TR measurements that are established in this paper diminish at smaller $\eta$ values.
Furthermore, in FIB microscopy, sample damage is more of an impediment to improving image quality by increasing dose~\cite{livengood2009subsurface, castaldo2009influence, orloff1996fundamental}.
Thus, we concentrate on FIB microscopy.

\subsection{Measurement Models}
We consider four measurement models for the probabilistic experiment described in Section~\ref{sec:physical-abstraction}:
\begin{itemize}
\item \emph{Oracle}:
    Observe
    \begin{equation}
      \label{eq:oracle-observation}
      \{M,\,(T_1,X_1),\,(T_2,X_2),\,\ldots,\,(T_M,X_M)\}.  
    \end{equation}
    Though no current instrument provides this information, this measurement model provides a useful benchmark.
\item \emph{Conventional}:
    Observe only
    \begin{equation}
      \label{eq:Y-def}
      Y = \sum_{i=i}^M X_i.
    \end{equation}
    This would be the standard operation of a FIB microscope that has direct detection of SEs.
    Note that
    \begin{equation}
      \label{eq:Y-def-Xtilde}
      Y = \sum_{i=1}^{\Mtilde} \Xtilde_i
    \end{equation}
    and
    \begin{equation}
      \label{eq:Y-def-Yk}
      Y = \sum_{k=1}^n Y_k
    \end{equation}
    are equivalent to the definition of $Y$.
    As the sum of a $\Poisson(\lambda)$ number of mutually independent $\Poisson(\eta)$ random variables,
    \eqref{eq:Y-def} is
    the simplest of the expressions.
\item \emph{Continuous-time time-resolved (CTTR)}:
    Observe
    \begin{equation}
      \label{eq:CTTR-observation}
      \{\Mtilde,\,(\Ttilde_1,\Xtilde_1),\,(\Ttilde_2,\Xtilde_2),\,\ldots\,\,(\Ttilde_{\Mtilde},\Xtilde_{\Mtilde})\}.
    \end{equation}
    This is an idealization of a FIB microscope with direct detection of SEs with perfect temporal precision.
\item \emph{Discrete-time time-resolved (DTTR)}:
    Observe
    \begin{equation}
      \label{eq:DTTR-observation}
      \{Y_1,\,Y_2,\,\ldots,\,Y_n\}.
    \end{equation}
    This is a model for the use of a FIB microscope to collect a set of low-dose subacquisitions.
\end{itemize}
Having established these abstractions, we can reiterate that our principal goal is to
demonstrate substantial improvements from time-resolved measurements.
In our previous work~\cite{peng2020source},
we introduced the concept of DTTR measurement along with estimators to apply with these measurements, and we showed empirical improvements over the trivial estimator that is routinely applied with conventional measurements.
The analysis of estimators in that work is limited,
and certain theoretical assertions are made without proofs.
The CTTR model introduced here is easier to analyze and represents a bound for what can be done with DTTR measurements.
We also provide new analyses of estimators for the DTTR model.
Through these results and Monte Carlo simulations,
the convergence of DTTR estimators to CTTR estimators as
$n \rightarrow \infty$ can be understood precisely
(see Sections~\ref{ssec:DTFI_ratio} and~\ref{ssec:DTQM_analysis}).

\subsection{Analyses for Oracle Measurement}
We initially assume $M = m > 0$.
Then the oracle measurement \eqref{eq:oracle-observation} includes nonempty sets
of ion arrival times $\{T_1,\,T_2,\,\ldots,\,T_m\}$ and
of SE counts $\{X_1,\,X_2,\,\ldots,\,X_m\}$.
The arrival times have beta distributions, with no dependence on parameter of interest $\eta$.
Thus, arrival times are immaterial to estimation of $\eta$ (i.e., redundant with knowing $M$).
The SE counts are i.i.d.\ observations with the $\Poisson(\eta)$ distribution,
and it is elementary to show that $Y = X_1+X_2+\cdots+X_m$ is a sufficient
statistic and that $Y/m$ is an efficient estimator of $\eta$ from the available data.
In the case of $M=0$, there is no basis for any estimate of $\eta$, so we must assign some arbitrarily chosen number $\eta_0$.

From the arguments above, we define the oracle estimator
\begin{equation}
  \label{eq:oracle-estimator}
  \etaOracle(M,X_1,X_2,\ldots,X_M) = \openCaserl
                          \eta_0, & M = 0; \\
    \Frac{Y}{M}, & M > 0.
    \closeCase
\end{equation}
Conditioned on $M = m > 0$, the mean-squared error (MSE) of this estimator is $\eta/m$
because $Y$ has mean $m\eta$ and variance $m\eta$.
Using the Poisson distribution for $M$ and the total expectation theorem,
\begin{equation}
\label{eq:MSE-oracle-exact}
\mathrm{MSE}(\etaOracle) = e^{-\lambda}(\eta-\eta_0)^2 + \sum_{m=1}^\infty \frac{\eta}{m} \frac{\lambda^m}{m!} e^{-\lambda}.
\end{equation}
For large enough $\lambda$, the arbitrary guess of $\eta_0$ when $M=0$ has little impact on the MSE\@.
Approximating the second term
using \eqref{eq:g-asymptotic} from Appendix~\ref{app:oracle}
gives
\begin{equation}
\label{eq:MSE-oracle-approx}
\mathrm{MSE}(\etaOracle) \approx \frac{\eta}{\lambda}
\qquad \mbox{for large $\lambda$}.
\end{equation}

\subsection{Analyses for Conventional Measurement}
\label{sec:conventional}
It is straightforward to show that the conventional measurement $Y$ has \emph{Neyman Type A} probability mass function (PMF)
\begin{equation}
  \label{equ:neyman}
    \mathrm{P}_{Y}(y \sMid \eta, \lambda) = \frac{e^{-\lambda} \eta^y}{y!} \sum_{m = 0}^{\infty} \frac{(\lambda e^{-\eta}) ^m m^y}{m!},
    \quad
    y = 0,\,1,\,\ldots,
\end{equation}
mean
\begin{equation}
  \label{eq:Y-mean}
    \E{Y} = \lambda \eta,
\end{equation}
and variance
\begin{equation}
  \label{eq:Y-var}
  \var{Y} = \lambda \eta + \lambda \eta^2.
\end{equation}
Starting from \eqref{eq:Y-def},
\eqref{equ:neyman} follows from the law of total probability,
\eqref{eq:Y-mean} from iterated expectation with conditioning on $M$, and
\eqref{eq:Y-var} from the law of total variance with conditioning on $M$~\cite{peng2020source}.
Reaching the same conclusions starting from \eqref{eq:Y-def-Xtilde} or \eqref{eq:Y-def-Yk} involves more complicated computations.

From~\eqref{eq:Y-mean}, the baseline estimator
\begin{equation}
  \etaBaseline(Y) = \frac{Y}{\lambda}
  \label{eq:eta-baseline}
\end{equation}
is unbiased, and from \eqref{eq:Y-var} its MSE is
\begin{equation}
  \mathrm{MSE}(\etaBaseline) = \frac{\eta(1+\eta)}{\lambda}.
   \label{eq:MSE-baseline}
\end{equation}
If $\lambda$ ions were deterministically incident upon the sample,
$Y$ would be a $\Poisson(\lambda \eta)$ random variable,
with variance $\lambda \eta$,
and the MSE of the baseline estimator would be $\eta/\lambda$.
The excess variance in \eqref{eq:Y-var},
consistent with experimental observations~\cite{Frank2005},
and excess MSE in \eqref{eq:MSE-baseline}
are due to the random variation in the number of incident ions
or the \emph{source shot noise}.
Our line of work mitigates this noise.

Since estimation under a Neyman Type A observation model is not well known,
the potential efficiency of the baseline estimator is not evident.
To this end, it is natural to evaluate the Fisher information (FI)
about $\eta$ in $Y$ with $\lambda$ as a known parameter,
which we denote by
$\mathcal{I}_Y(\eta \sMid \lambda)$.

The FI is defined as
\begin{equation} 
\label{equ:Fisher_info_equation_PP_define}
\mathcal{I}_Y(\eta \sMid \lambda)
  = \E{ \left(\frac{\partial \log \mathrm{P}_Y(y \sMid \eta, \lambda)}{\partial \eta} \right)^2 \sMid \eta },
\end{equation}
where a known non-random parameter in the expectation is emphasized by putting it after a semicolon.
From \eqref{equ:neyman},
\begin{align*}
\log&\mathrm{P}_Y(y \sMid \eta, \lambda) \\
    &= -\lambda + y\log \eta - \log y! + \log\Bigg( \sum_{m = 0}^{\infty}\frac{(\lambda e^{-\eta})^m m^y}{m!} \Bigg).
\end{align*}
Then taking the derivative with respect to $\eta$, we find that
\begin{eqnarray*}
\frac{\partial \log \mathrm{P}_Y(y \sMid \eta, \lambda)}{\partial \eta} 
&\!=\!& \frac{y}{\eta} - \frac{\sum_{m = 0}^{\infty} \dfrac{m(\lambda e^{-\eta})^m m^y}{m!}}{\sum_{m = 0}^{\infty} \dfrac{(\lambda e^{-\eta})^m m^y}{m!}} \\
&\!\eqlabel{a}\!& \frac{y}{\eta}
    - \frac{{\mathrm{P}_Y(y + 1 \sMid \eta, \lambda)}\bigg/{\dfrac{e^{-\lambda} \eta^{y + 1}}{(y + 1)!}}}
           {{\mathrm{P}_Y(y \sMid \eta, \lambda)}\bigg/{\dfrac{e^{-\lambda} \eta^{y}}{y!}}} \\
&\!=\!& \frac{y}{\eta} - \frac{\mathrm{P}_Y(y + 1 \sMid \eta, \lambda)}{\mathrm{P}_Y(y \sMid \eta, \lambda)} \frac{y + 1}{\eta}, \\
\end{eqnarray*}
where (\emph{a}) follows from \eqref{equ:neyman}.
The FI is the second moment of the above expression:
\begin{equation} 
\mathcal{I}_Y(\eta \sMid \lambda)
   = \sum_{y = 0}^{\infty} \left(
     \frac{y}{\eta} {-} \frac{\mathrm{P}_Y(y + 1 \sMid \eta, \lambda)}{\mathrm{P}_Y(y \sMid \eta, \lambda)} \frac{y+1}{\eta}
     \right)^{\!2} \mathrm{P}_Y(y \sMid \eta, \lambda).
\label{equ:Fisher_info_equation_PP}
\end{equation}

\begin{figure}
  \begin{center}
    \includegraphics[width=0.8\linewidth]{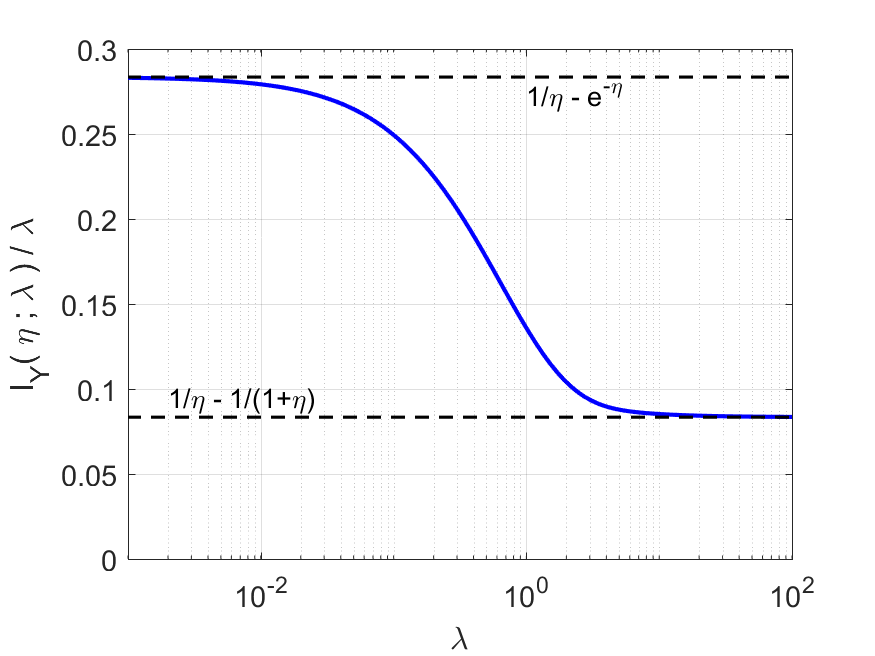}
  \end{center}
  \caption{Normalized Fisher information $\mathcal{I}_Y(\eta \sMid \lambda)/\lambda$
        as a function of $\lambda$ for $\eta = 3$.
        The marked asymptotes are derived in Appendix~\ref{app:FI_norm}.}
  \label{fig:normalized_Fisher_info}
\end{figure}

While \eqref{equ:Fisher_info_equation_PP} is not readily comprehensible,
it can be used to numerically evaluate $\mathcal{I}_Y(\eta \sMid \lambda)$
and to derive certain useful asymptotic approximations and limits.
One can interpret the ratio $\mathcal{I}_Y(\eta \sMid \lambda)/\lambda$ as the information gain per incident ion.
As illustrated in Fig.~\ref{fig:normalized_Fisher_info},
this \emph{normalized Fisher information}
is a decreasing function of $\lambda$,
with
\begin{equation}\label{eq:NFI-low-lambda}
  \lim_{\lambda \rightarrow 0}
    \frac{\mathcal{I}_Y(\eta \sMid \lambda)}{\lambda}
    = \frac{1}{\eta} - e^{-\eta}
\end{equation}
and
\begin{equation}\label{eq:NFI-high-lambda}
  \lim_{\lambda \rightarrow \infty}
    \frac{\mathcal{I}_Y(\eta \sMid \lambda)}{\lambda}
    = \frac{1}{\eta(1+\eta)}
    = \frac{1}{\eta} - \frac{1}{1+\eta}.
\end{equation}
Detailed derivations
are provided in Appendix~\ref{app:FI_norm}.

Using \eqref{eq:NFI-high-lambda} to write
\begin{equation}\label{eq:FI-high-lambda}
    \mathcal{I}_Y(\eta \sMid \lambda) \approx \frac{\lambda}{\eta(1+\eta)}
    \qquad
    \mbox{for large $\lambda$},
\end{equation}
we have a match to the reciprocal of the MSE in \eqref{eq:MSE-baseline}.
Thus, the baseline estimator achieves the {\CR} bound (CRB) asympotically as $\lambda \rightarrow \infty$,
but not otherwise (since the normalized FI is a decreasing function of $\lambda$).

One could seek improved estimators for low $\lambda$
or improved lower bounds to demonstrate that substantially better estimators do not exist.
We do not pursue those goals here.
In practice, improved estimators for low $\lambda$ may be of limited
interest---even if they improve significantly upon the baseline estimator.
For example, with reference to Fig.~\ref{fig:normalized_Fisher_info},
FI suggests that one may be able to improve upon the baseline estimator by
a factor of 3 at dose $\lambda = 10^{-2}$.
However, even the lowest possible MSE would be quite high at such a low dose.

The key observation from Fig.~\ref{fig:normalized_Fisher_info} and the limits
in \eqref{eq:NFI-low-lambda} and \eqref{eq:NFI-high-lambda} is that
low-dose measurements are more informative per incident ion than high-dose measurements.
The remainder of the paper studies methods to realize improvements related to this
gap while operating at any dose level---not only low dose.
Furthermore, notice that the MSE in \eqref{eq:MSE-baseline} has a simple
inversely proportional relationship with $\lambda$.
Most performance bounds and empirical performances in this paper share this
simple $1/\lambda$ behavior, so we place little emphasis on the performance
as $\lambda$ is varied.
Instead, we concentrate on comparisons among different methods and
the performance dependence on $\eta$.

\section{Continuous-Time Time-Resolved Measurement}
\label{sec:CTTR_measurement}
\subsection{Measurement Distributions}
\label{ssec:PP_model_continuous}
The CTTR measurement \eqref{eq:CTTR-observation}
contains the number of incident ions that result in positive detected SEs $\Mtilde$,
the arrival times of these ions $\{\Ttilde_1,\,\Ttilde_2,\,\ldots,\,\Ttilde_{\Mtilde}\}$,
and the corresponding SE counts $\{\Xtilde_1,\,\Xtilde_2,\,\ldots,\,\Xtilde_{\Mtilde}\}$.
As noted in Section~\ref{sec:physical-abstraction},
the mutual independence of the arrival times in the underlying process $\{T_1,\,T_2,\,\ldots\}$
and all events of the form $\{X_i = 0\}$ cause the Poisson process property to be preserved.

Since $\iP{X_i = 0} = e^{-\eta}$,
the rate $\Lambda$ of the underlying process is reduced to $\Lambda(1-e^{-\eta})$ for the thinned process.
For the thinned ion count over dwell time $t$, we have
$\Mtilde \sim \Poisson(\lambda(1 - e ^{-\eta}))$, or more explicitly the PMF
\begin{equation}
    \label{eq:Mtilde-PMF}
    \mathrm{P}_{\Mtilde}(\mtilde \sMid \eta,\lambda) = \exp(-\lambda(1 - e^{-\eta}))\frac{(\lambda(1 - e^{-\eta}))^{\mtilde}}{{\mtilde}!},
\end{equation}
for $\mtilde = 0,\,1,\,\ldots$.

The distribution of the $\Xtilde_i$ variables is simply the zero-truncation of the $\Poisson(\eta)$ distribution:
\begin{equation}
    \label{eq:PMF_Xtilde}
    \mathrm{P}_{\Xtilde_i}(j \sMid \eta) = \frac{e^{-\eta}}{1 - e^{-\eta}} \cdot \frac{\eta^{j}}{j!},
    \qquad j = 1,\,2,\,\ldots.
\end{equation}

While the interarrival times of the thinned process have a simple exponential distribution,
this is not relevant to our estimation tasks:
Under the CTTR measurement model, we have $\Mtilde$ available, and conditioned on $\Mtilde$,
the thinned arrival times have beta distributions with no dependence on the parameter of interest $\eta$.

\subsection{Fisher Information}
\label{ssec:CTFI_analysis}
We would like to evaluate the FI about $\eta$ in the CTTR measurement \eqref{eq:CTTR-observation} with $\lambda$ as a known parameter:
\begin{align}\label{eq:FI_TR_part1}
\mathcal{I}_{\rm CTTR}(\eta \sMid \lambda)
&\eqlabel{a} \mathcal{I}_{(\Ttilde_1,\Xtilde_1),\ldots,(\Ttilde_{\Mtilde},\Xtilde_{\Mtilde})|\Mtilde}(\eta \sMid \lambda) 
   + \mathcal{I}_{\Mtilde}(\eta \sMid \lambda) \nonumber\\
&\eqlabel{b} \mathcal{I}_{\Xtilde_1,\ldots,\Xtilde_{\Mtilde}|\Mtilde}(\eta \sMid \lambda) 
   + \mathcal{I}_{\Mtilde}(\eta \sMid \lambda) \nonumber\\
&\eqlabel{c} \E{\Mtilde}\mathcal{I}_{\Xtilde_i}(\eta \sMid \lambda) + \mathcal{I}_{\Mtilde}(\eta \sMid \lambda) \nonumber \\
&\eqlabel{d} \lambda(1-e^{-\eta})\mathcal{I}_{\Xtilde_i}(\eta \sMid \lambda) + \mathcal{I}_{\Mtilde}(\eta \sMid \lambda),
\end{align}
where
(\emph{a}) follows from the chain rule for FI~\cite{VanTrees:01a};
(\emph{b}) from the conditional distribution of each $\Ttilde_i$ given $\Mtilde$ having no dependence on $\eta$;
(\emph{c}) from additivity of FI and the independence of $\{\Mtilde,\,\Xtilde_1,\,\Xtilde_2,\,\ldots,\,\Xtilde_{\Mtilde}\}$; and
(\emph{d}) from substitution of the mean of $\Mtilde$.
Thus,
we need to evaluate $\mathcal{I}_{\Xtilde_i}(\eta \sMid \lambda)$
and $\mathcal{I}_{\Mtilde}(\eta \sMid \lambda)$.
The former, which represents the FI for estimating $\eta$ from $\Xtilde_i$ is
\begin{align}
\mathcal{I}_{\Xtilde_i}(\eta \sMid \lambda)
&= \E{ \left(\frac{\partial \log \mathrm{P}_{\Xtilde_i}(\Xtilde_i \sMid \eta, \lambda)}{\partial \eta} \right)^{\!2} \sMid \eta } \nonumber\\
&\eqlabel{a} \sum_{j = 1}^{\infty} \left( \frac{j}{\eta} - \frac{1}{1-e^{-\eta}} \right)^{\!2} \frac{e^{-\eta}}{1-e^{-\eta}}\frac{\eta^j}{j!} \nonumber\\
&= \frac{1}{1-e^{-\eta}} \frac{\eta + 1}{\eta} - \frac{1}{(1-e^{-\eta})^2},
\label{eq:FI_Xtilde}
\end{align}
where (\emph{a}) uses the PMF in \eqref{eq:PMF_Xtilde}.
Similarly, $\mathcal{I}_{\Mtilde}(\eta \sMid \lambda)$, which represents the FI for estimating $\eta$ from $\Mtilde$, is
\begin{align}
&\mathcal{I}_{\Mtilde}(\eta \sMid \lambda)
  = \E{\left(\frac{\partial \log P_{\Mtilde}( \Mtilde \sMid \eta , \lambda)}{\partial \eta}\right)^{\!2} \sMid \eta}\nonumber\\
&\eqlabel{a} \sum_{j = 0}^{\infty}\left( \frac{e^{-\eta}}{1-e^{-\eta}}j - \lambda e^{-\eta} \right)^{\!2}
       \frac{e^{-\lambda (1-e^{-\eta})} [\lambda (1-e^{-\eta})]^j}{j!}\nonumber\\
&=\frac{\lambda e^{-\eta}}{e^{\eta}-1},
\label{eq:FI_Mtilde}
\end{align}
where (\emph{a}) uses the PMF in \eqref{eq:Mtilde-PMF}.
Finally, substituting~\eqref{eq:FI_Xtilde} and~\eqref{eq:FI_Mtilde} into~\eqref{eq:FI_TR_part1} gives
\begin{align}\label{eq:FI_CT}
\mathcal{I}_{\rm CTTR}(\eta \sMid \lambda)
&= \lambda\left(\frac{1}{\eta} - {e^{-\eta}}\right).
\end{align}

Notice that the FI for CTTR measurement matches the low-dose asymptote given in \eqref{eq:NFI-low-lambda}.
It is exact and holds for all values of $\lambda$.
The greater FI for CTTR measurement than for conventional measurement is suggestive of being able to
improve upon the baseline estimator \eqref{eq:eta-baseline}.
In the following sections, we define new estimators and demonstrate their improvements.

\subsection{Estimators}
In this section, we introduce estimators applicable to CTTR measurement \eqref{eq:CTTR-observation}.
From \eqref{eq:Y-def-Xtilde}, it is clear that the total number of detected SEs $Y$ is available.
As we have explained in the derivation of $\mathcal{I}_{\rm CTTR}(\eta \sMid \lambda)$ in Section~\ref{ssec:CTFI_analysis},
with $\Mtilde$ available, there is no additional information about $\eta$ in the thinned ion incidence times
$\{\Ttilde_1,\,\Ttilde_2,\,\ldots,\,\Ttilde_{\Mtilde}\}$.
Furthermore, with $Y$ and $\Mtilde$ available, there is no additional information about $\eta$ in the positive SE counts
$\{\Xtilde_1,\,\Xtilde_2,\,\ldots,\,\Xtilde_{\Mtilde}\}$.
To see this, consider any observation vector
$(\Xtilde_1,\,\Xtilde_2,\,\ldots,\,\Xtilde_{\Mtilde}) = (j_1,\,j_2,\,\ldots,\,j_m)$
conditioned on $\Mtilde = \mtilde$.
Using independence and \eqref{eq:PMF_Xtilde}, the likelihood of the observation
(conditioned on $\Mtilde = \mtilde$)
is
\begin{equation}
    \label{eq:Xtilde-likelihood}
  \prod_{i=1}^{\mtilde}
    \mathrm{P}_{\Xtilde_i}(j_i \sMid \eta)
    = \left(\frac{e^{-\eta}}{1 - e^{-\eta}}\right)^{\!\mtilde} \frac{\eta^{j_1+j_2+\cdots+j_{\mtilde}}}{j_1! \, j_2! \, \cdots \, j_{\mtilde}!}.
\end{equation}
As a function of $\eta$, the dependence on SE counts is only through their sum.
Hence, all the estimators in the section depend only on $Y$ and $\Mtilde$.

\subsubsection{Continuous-Time Quotient Mode Estimator}
\label{ssec:QM_estimators}
Recall that the oracle estimator \eqref{eq:oracle-estimator} divides the total SE count $Y$ by the number of incident ions $M$.
Using $\Mtilde$ as a proxy for the number of incident ions
yields the \emph{continuous-time quotient mode} (CTQM) estimator
\begin{equation}
  \label{eq:eta_QM_conti}
  \etaCTQM(\Mtilde,Y) = \openCaserl
                                0, & \Mtilde = 0; \\
                \Frac{Y}{\Mtilde}, & \Mtilde > 0.
    \closeCase
\end{equation}
Note that the 0 estimate for $\Mtilde = 0$ is not arbitrary, it is the
ML estimate of $\eta$ for this case.
The name ``quotient mode'' is to acknowledge a similar concept in a presentation by John Notte of Zeiss~\cite{Notte2013}
and in a patent application~\cite{Zeiss_QM_patent}.

\subsubsection{Continuous-Time Lambert Quotient Mode Estimator}
The CTTR measurement observes a thinned version of the underlying ion incidence process,
so $\Mtilde \leq M$.  We can do better than to use $\Mtilde$ as a proxy for $M$.
Since
\[
  \E{ \Mtilde \smid M } = (1-e^{-\eta})M,
\]
$(1 - e^{-\eta})^{-1}\Mtilde$ would be an unbiased proxy for $M$.
Unfortunately, this has dependence on $\eta$, which is not known.
In the spirit of the oracle and CTQM estimators,
we may seek an estimate $\widehat{\eta}$ that satisfies
\[
  \widehat{\eta} = \frac{Y}{(1-e^{-\widehat{\eta}})^{-1}\Mtilde}.
\]
The solution of this equation gives the
\emph{continuous-time Lambert quotient mode} (CTLQM) estimator:
\begin{equation}\label{eq:eta_LQM_conti}
    \etaCTLQM = W(-\etaCTQM e^{-\etaCTQM}) + \etaCTQM,
\end{equation}
where $W(\cdot)$ is the Lambert W function~\cite{CorlessGHJK:96}.

\subsubsection{Continuous-Time Maximum Likelihood Estimator}
\label{ssec:CTTRML_estimator}
Rather than use a heuristic approximation for the number of incident ions $M$, 
one could instead use the statistically principled ML estimation approach as follows.
The ML estimate is the value of $\eta$ that maximizes the joint likelihood of the full CTTR observation.

We have already seen that we can drop the times
$\{\Ttilde_1,\,\Ttilde_2,\,\ldots\,\Ttilde_{\Mtilde}\}$,
and the conditional likelihood of
$\{\Xtilde_1,\,\Xtilde_2,\,\ldots\,\Xtilde_{\Mtilde}\}$ given $\Mtilde$ was given in \eqref{eq:Xtilde-likelihood}.
Thus, we must maximize the product of \eqref{eq:Mtilde-PMF} and \eqref{eq:Xtilde-likelihood} over $\eta$.
For observation
$(\Mtilde,\,\Xtilde_1,\,\Xtilde_2,\,\ldots,\,\Xtilde_{\Mtilde}) = (\mtilde,\,j_1,\,j_2,\,\ldots,\,j_{\mtilde})$,
by dropping factors that do not depend on $\eta$, we obtain
\emph{continuous-time ML} (CTML) estimator
\[
  \etaCTML = \argmax_\eta e^{-\lambda(1-e^{-\eta})} e^{-{\mtilde}\eta} \eta^y,
\]
where $y = j_1 + j_2 + \cdots + j_{\mtilde}$.
The unique maximizer satisfies
\begin{align}\label{eq:eta_trml_conti}
\etaCTML = \frac{Y}{\Mtilde + \lambda e^{-\etaCTML}},
\end{align}
which can be solved by using an appropriate root-finding algorithm.

\subsection{Analyzing the Continuous-Time Quotient Mode Estimator}
\label{sec:ctqm-analysis}

By computing the MSE of $\etaCTQM$, we can evaluate the efficacy of the quotient mode estimator.
We begin by noting that
\begin{align}
\mathrm{MSE}(\etaCTQM) = \bias(\etaCTQM)^2 + \var{\etaCTQM}.
\label{eq:MSE_bias_variance}
\end{align}
As detailed in Appendix~\ref{app:mse_CTQM_deri},
the bias is given by
\begin{align}\label{eq:eta_QM_cont_bias}
\bias(\etaCTQM) 
   &= \E{\etaCTQM} - \eta \nonumber\\
   &= \left( \frac{e^{-\eta} - e^{-\lambda(1-e^{-\eta})}}
                  {1-e^{-\eta}} \right) \eta.
\end{align}
For fixed $\eta$, this is a nonzero bias even as dose $\lambda \rightarrow \infty$,
consistent with the motivation for defining the CTLQM estimator to improve upon the CTQM estimator.
If $\eta \rightarrow \infty$ as well, the bias vanishes,
which is consistent with the convergence in distribution of $\Mtilde$ to $M$.

As also detailed in Appendix~\ref{app:mse_CTQM_deri},
the variance is given by
\begin{align}
\var{\etaCTQM}
&= \frac{\eta^2}{\PposM^2} e^{-\lambda \PposM} (1 - e^{-\lambda \PposM})
 \nonumber \\
& \quad +
     \frac{\eta(\PposM - \eta e^{-\eta})} {\PposM^2}
       \sum_{j=1}^{\infty} \frac{1}{j}\frac{e^{-\lambda\PposM} (\lambda \PposM)^j}{j!},
\label{eq:eta_QM_cont_total_var}
\end{align}
where we introduce
\[
  \PposM = \iP{X_i > 0} = \PposMfull
\]
as a shorthand to make certain expressions more compact.
Expression \eqref{eq:eta_QM_cont_total_var} can be
combined with \eqref{eq:g-asymptotic} from Appendix~\ref{app:oracle}
to show that the variance vanishes as $\lambda \rightarrow \infty$,
decaying asymptotically as $\sim 1/\lambda$.
However, because of nonzero bias, the MSE is not inversely proportion to $\lambda$,
and the MSE of $\etaCTQM$ relative to other estimates depends on $\lambda$.
Substituting \eqref{eq:eta_QM_cont_bias} and \eqref{eq:eta_QM_cont_total_var} into
\eqref{eq:MSE_bias_variance} gives an expression for the MSE of the CTQM estimator:
\begin{align}
\mathrm{MSE}(\etaCTQM)
 =& \ \frac{\eta^2}{\PposM^2} \left( e^{-\eta} - e^{-\lambda \PposM} \right)^{\!2} 
     + \frac{\eta^2}{\PposM^2} e^{-\lambda \PposM} (1- e^{-\lambda \PposM}) \nonumber\\
  & +
  \frac{\eta(\PposM - \eta e^{-\eta})} {\PposM^2}
  \sum_{j=1}^{\infty} \frac{1}{j}\frac{e^{-\lambda \PposM} (\lambda \PposM)^j}{j!}.
\label{eq:eta_QM_cont_MSE}
\end{align}
Substituting the upper bound
\eqref{eq:g-bound} from Appendix~\ref{app:oracle}
for the series in \eqref{eq:eta_QM_cont_MSE} gives
\begin{align}
\mathrm{MSE}(\etaCTQM)
   < & \ \frac{\eta^2}{\PposM^2} \left( e^{-\eta} - e^{-\lambda \PposM} \right)^{\!2} 
        + \frac{\eta^2}{\PposM^2} e^{-\lambda \PposM} (1- e^{-\lambda \PposM}) \nonumber\\
   & + 0.518 \frac{\eta(\PposM - \eta e^{-\eta})} {\PposM^2}.
\label{eq:eta_QM_cont_MSE_bound}
\end{align}

\begin{figure*}
    \hfill
    \begin{subfigure}{\threewidewidth}
      \includegraphics[width=\linewidth]{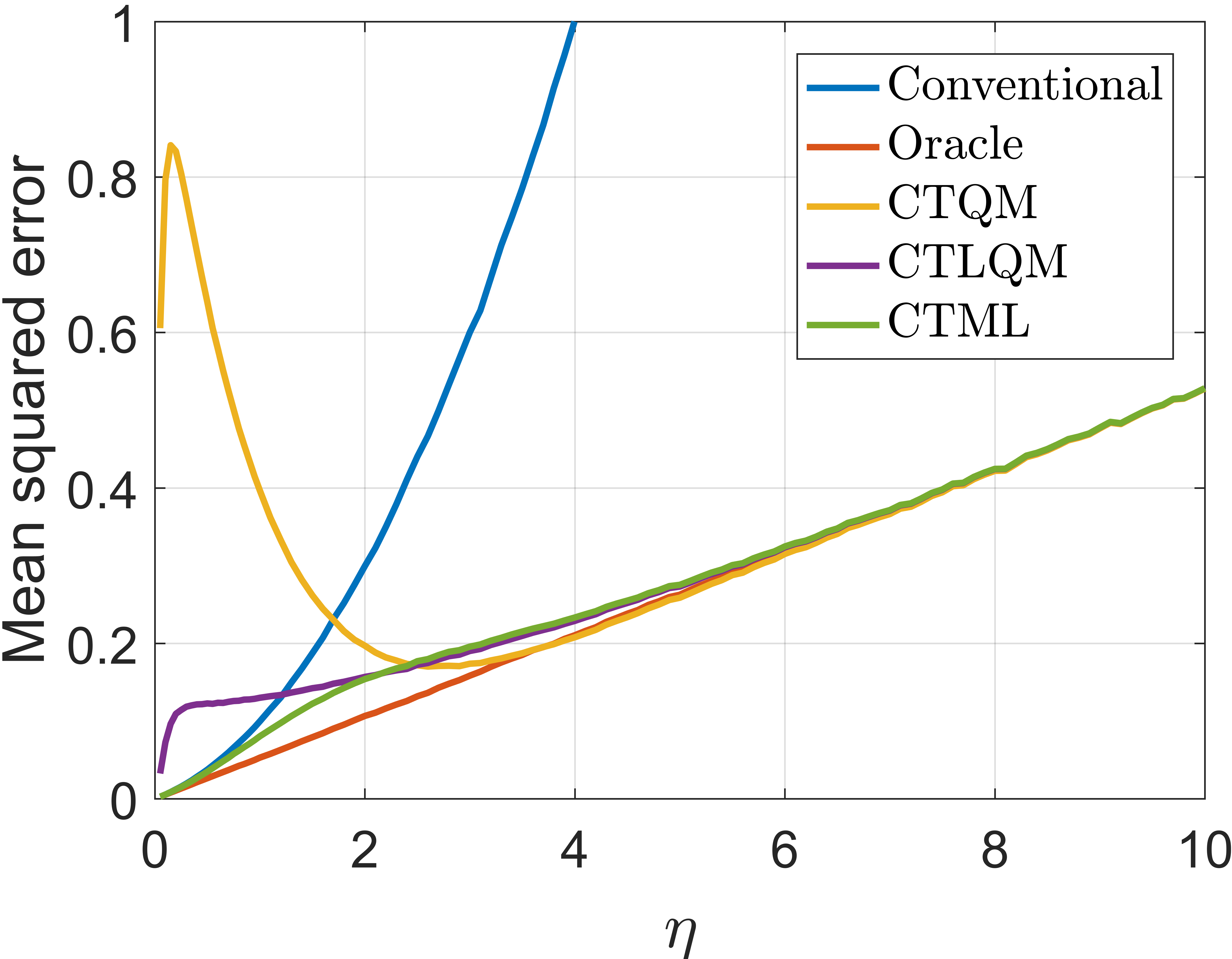}
      \subcaption{MSE across $\eta$.}
      \label{subfig:CT_MSE}
    \end{subfigure}
    \hfill
    \begin{subfigure}{\threewidewidth}
      \includegraphics[width=\linewidth]{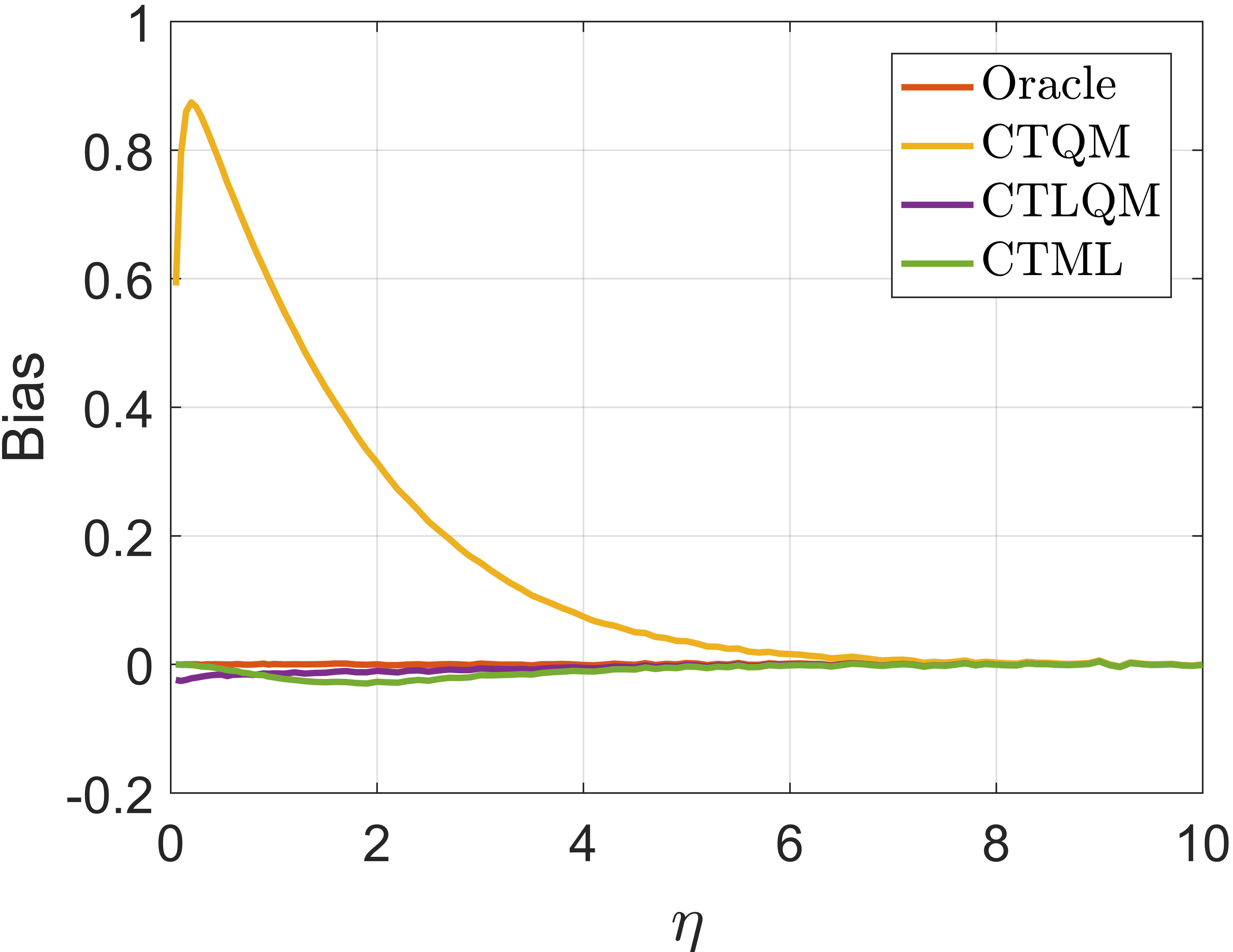}
      \subcaption{Bias across $\eta$.}
      \label{subfig:CT_bias}
    \end{subfigure}
    \hfill
    \begin{subfigure}{\threewidewidth}
      \includegraphics[width=\linewidth]{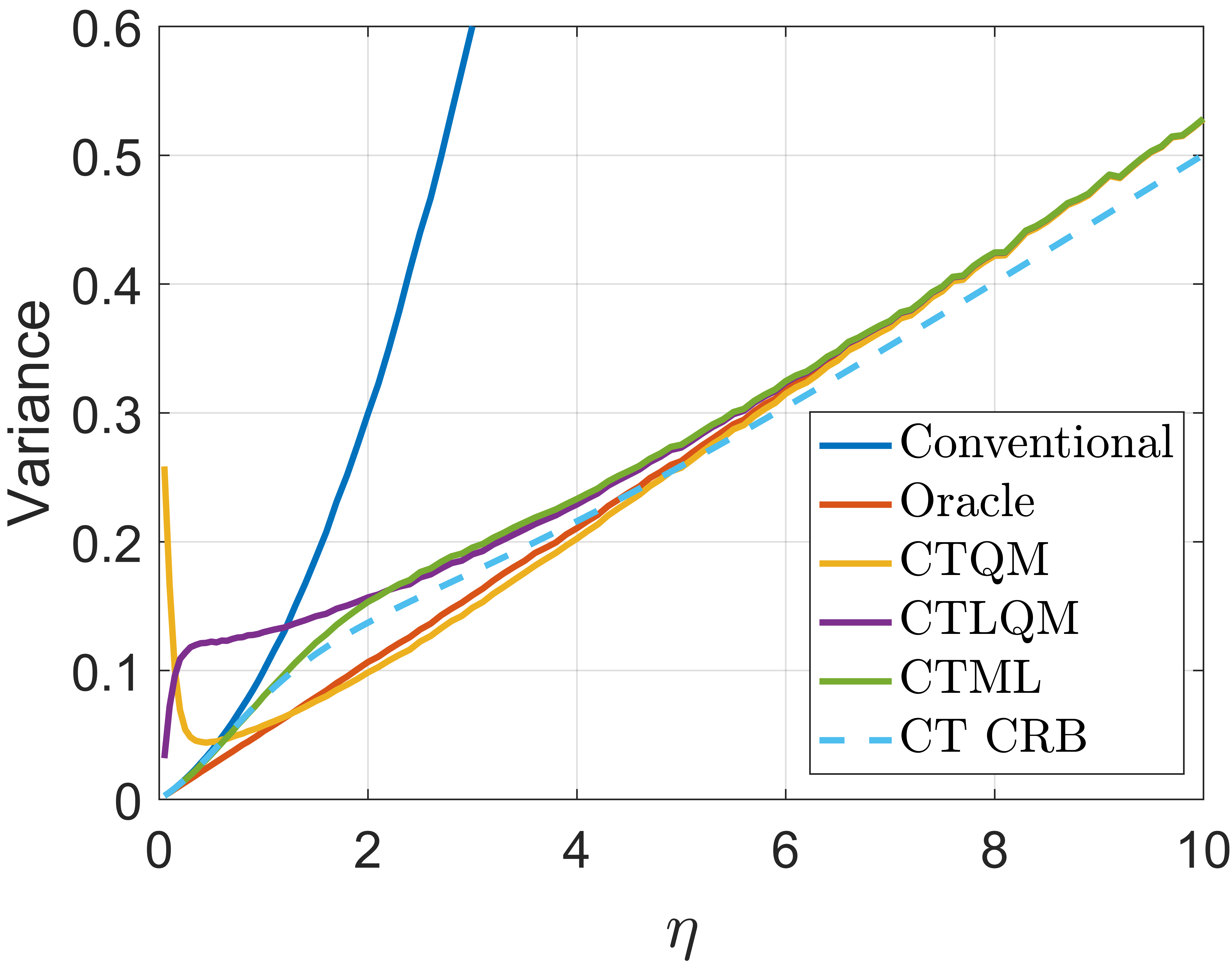}
      \subcaption{Variance across $\eta$.}
      \label{subfig:CT_var}
    \end{subfigure}
    \hfill
  \caption{Comparison of continuous-time time-resolved measurement estimators
    as a function of $\eta$.
    Conventional, oracle,
    continuous-time quotient mode, 
    continuous-time Lambert quotient mode, and
    continuous-time maximum likelihood estimators are simulated for
    dose rate $\Lambda = 1/1600$ ions per ns and 
    dwell time $t = 32\,000$~ns, hence total dose $\lambda = 20$ ions per pixel.
    (a) MSE\@.
    (b) Bias.  (Conventional estimator omitted because its bias is zero.)
    (c) Variance.
  }
  \label{fig:CTTR_comparison}
\end{figure*}

\subsection{Numerical Comparisons of Estimators}
To demonstrate the benefits afforded by CTTR measurement,
in Fig.~\ref{fig:CTTR_comparison}
we compare the conventional, oracle, CTQM, CTLQM and CTML estimators across ground truth $\eta \in [0,10]$.
The MSE values in Fig.~\ref{subfig:CT_MSE} are computed from $150\,000$ independent Monte Carlo trials, using a dose rate $\Lambda = 1/1600$ 
ions per~ns 
and dwell time $t = 32\,000$~ns, for a total dose $\lambda = 20$ 
ions per pixel.
The curve for $\etaBaseline$ matches the theoretical MSE expression~\eqref{eq:MSE-baseline}. 
Although unimplementable, the curve for $\etaOracle$ also matches the theoretical MSE in~\eqref{eq:MSE-oracle-exact}.%
\footnote{We did not need to choose a value for $\eta_0$ because the event $\{M=0\}$,
which has probability $e^{-20} \approx 2 \cdot 10^{-9}$,
did not occur in any of the trials.}
We can observe that $\etaCTQM$ has a large MSE for small $\eta$, caused largely by $\Mtilde$ severely underestimating $M$ in these cases.
Predictably, $\etaCTLQM$ reduces MSE tremendously for small values of $\eta$ because the role of $\Mtilde$ is modified the most in these cases.
For low $\eta$ (about 0 to 2.5),
$\etaCTML$ achieves lowest MSE amongst all implementable estimators;
for moderate $\eta$ (about 2.5 to 5.5),
$\etaCTQM$ is slightly better than the others; and
for large $\eta$ (about 5.5 and above),
$\etaCTQM$, $\etaCTLQM$ and $\etaCTML$ all give nearly identical performance.
It is noteworthy that CTQM converges with the oracle at $\eta$ above about 3.0,
though the oracle is unimplementable in practice.

The MSE trends and comparisons can be better understood through the
biases in Fig.~\ref{subfig:CT_bias} and variances in Fig.~\ref{subfig:CT_var}.
Curves for $\etaCTQM$ coincide with the bias and variance expressions derived in~\eqref{eq:eta_QM_cont_bias} and~\eqref{eq:eta_QM_cont_total_var}.
As previously noted, the large bias of $\etaCTQM$ for small values of $\eta$
is caused by the number of incident ions $M$ 
being severely underestimated by $\Mtilde$.
The bias of $\etaCTQM$ can be corrected by the use of $\etaCTLQM$.
The dashed cyan-colored curve in Fig.~\ref{fig:CTTR_comparison}(c) is the CRB
for any unbiased estimator, 
which is the reciprocal of~\eqref{eq:FI_CT}.
The variances of the implementable and approximately unbiased $\etaCTLQM$ and $\etaCTML$
estimators approximately coincide with the CRB\@.

\section{Discrete-Time Time-Resolved Measurement}
\label{sec:DTTR_measurement}
\subsection{Measurement Distributions}
The DTTR measurement \eqref{eq:DTTR-observation} is a length-$n$ vector of SE counts collected over
subacquisition dwell times of $t/n$.
Thus, some modeling and analysis for DTTR measurement follows from scaling of $\lambda$
in expressions from Section~\ref{sec:conventional}.
Since $\{Y_1,\,Y_2,\,\ldots,\,Y_n\}$
are independent, their joint PMF is simply
\begin{equation}
    \label{eq:DTTR-distribution}
    \mathrm{P}_{Y_1,Y_2,\ldots,Y_n}(y_1,y_2,\ldots,y_n \sMid \eta,\lambda )
    = \prod_{k=1}^n \mathrm{P}_Y( y_k \sMid \eta,\lambda/n ),
\end{equation}
written in terms of the PMF in \eqref{equ:neyman}.

\subsection{Fisher Information}
\label{ssec:DTFI_ratio}
To evaluate the FI about $\eta$ in the DTTR measurement \eqref{eq:DTTR-observation} with $\lambda$ as a known parameter
is also quite simple.
Because FI is additive over independent observations,
\begin{align}
  \label{eq:Fisher_info_TR}
    \mathcal{I}_{\rm DTTR}(\eta \sMid \lambda, n)
     &= n \, \mathcal{I}_{Y}(\eta \sMid \lambda/n),
\end{align}
expressed in terms of the FI in \eqref{equ:Fisher_info_equation_PP}.
While this FI inherits the complexity and lack of interpretability of \eqref{equ:Fisher_info_equation_PP},
the distinction is that the relevant dose parameter in $\mathcal{I}_Y$ has been reduced from $\lambda$ to $\lambda/n$,
so it is more reasonable to approximate with the low-dose asymptote \eqref{eq:NFI-low-lambda}.
Specifically, we can write
\begin{align}
\mathcal{I}_{\rm DTTR}(\eta \sMid \lambda, n)
 &= \lambda \, \frac{\mathcal{I}_Y (\eta \sMid \lambda/n)}{\lambda/n} \nonumber \\
 &\approx \lambda\left( \frac{1}{\eta} - e^{-\eta} \right),
\label{equ:Fisher_info_approx_TR}
\end{align}
where the approximation holds for large enough $n$ because of \eqref{eq:NFI-low-lambda}.
Note that~\eqref{equ:Fisher_info_approx_TR}
has the same expression as~\eqref{eq:FI_CT},
so as $n \rightarrow \infty$, the FI of DTTR measurement converges from below to the FI of CTTR measurement.

\subsection{Estimators}
In this section, we present estimators applicable to DTTR measurement \eqref{eq:DTTR-observation}
that were first introduced in \cite{peng2020source}.
The subsequent analyses are new.

\subsubsection{Discrete-Time Quotient Mode Estimator}
\label{ssec:DTQM_estimator}
Similar to the principle behind the CTQM estimator~\eqref{eq:eta_QM_conti},
the number of subacquisitions with positive SEs,
\begin{equation}
    \label{eq:L-def}
    L = \sum_{k=1}^n \mathbbm{1}_{\{ Y_k > 0\}}, 
\end{equation}
can be a proxy for the number of incident ions $M$.
The \emph{discrete-time quotient mode} (DTQM) estimator
is then defined as
\begin{equation}
  \label{eq:eta_DTQM}
  \etaDTQM(L,Y) = \openCaserl
                          0, & L = 0; \\
                \Frac{Y}{L}, & L > 0.
    \closeCase
\end{equation}

\subsubsection{Discrete-Time Lambert Quotient Mode Estimator}
\label{ssec:DTLQM_estimator}
At low $\eta$, $M$ is severely underestimated by $L$.
Correspondingly, the \emph{discrete-time Lambert quotient mode} (DTLQM) estimator
is obtained by incorporating the correction factor $(1 - e^{-\eta})^{-1}$,
\[
  \widehat{\eta} = \frac{Y}{(1-e^{-\widehat{\eta}})^{-1}L}.
\]
The solution of this equation gives the estimator:
\begin{equation}\label{eq:DTLQM_expression}
    \etaDTLQM = W(-\etaDTQM e^{-\etaDTQM}) + \etaDTQM.
\end{equation}

\subsubsection{Discrete-Time Maximum Likelihood Estimator}
\label{ssec:DTML_estimator}
The \emph{discrete-time ML} (DTML) estimate
is the value of $\eta$ that maximizes the joint likelihood in \eqref{eq:DTTR-distribution}:
\begin{align}
\label{eq:eta_DTML}
\etaDTML = \argmax_{\eta} \, \prod_{k=1}^{n} \mathrm{P}_{Y}(y_k \sMid \eta, \lambda/n ),  
\end{align}
where 
$\mathrm{P}_Y(\cdot \sMid \cdot, \cdot)$ is given by~\eqref{equ:neyman}. 
Unlike in the CTTR case, we have no fixed-point form for the estimator.
Instead, it can be computed by direct numerical optimization.
Since the decision variable is scalar, even a simple grid search is not impractical.

\subsection{Analyzing the Discrete-Time Quotient Mode Estimator}
\label{ssec:DTQM_analysis}
Like in Section~\ref{sec:ctqm-analysis},
we analyze the MSE of $\etaDTQM$
by finding expressions for its bias and variance.
In both calculations, we make use of a zero-truncated modification of the
Neyman Type A distribution at ion incidence parameter $\lambda/n$.
Let $p = \iP{Y_k > 0}$.
Then using \eqref{equ:neyman}, we find
\begin{equation}
    \label{eq:p-def}
  p = 1 - \mathrm{P}_{Y}( 0 \sMid \eta, \lambda/n ) = 1 - \exp\!\left(-\frac{\lambda}{n}(1-e^{-\eta})\right)\!.
\end{equation}
If $\Ytilde_k$ is the zero-truncated version of $Y_k$, then its PMF is
\begin{equation}
    \label{eq:ytilde-pmf}
  \mathrm{P}_{\Ytilde_k}(y_k) 
    = \frac{1}{p}\frac{e^{-\Frac{\lambda}{n}} \eta^{y_k}}{y_k!}
      \sum_{m = 0}^{\infty} \frac{(\frac{\lambda}{n} e^{-\eta})^m m^{y_k}}{m!},
  \quad
  y_k = 1,2,\ldots,
\end{equation}
its mean is
\begin{equation}
    \label{eq:ytilde-mean}
    \iE{ \Ytilde_k } = \frac{1}{p} \frac{\lambda \eta}{n},
\end{equation}
and its variance is
\begin{equation}
    \label{eq:ytilde-var}
    \ivar{ \Ytilde_k }
    =    \frac{1}{p}   \left(\frac{\lambda}{n}\eta + \frac{\lambda}{n}\eta^2 + \left(\frac{\lambda}{n}\eta\right)^2\right)
       - \frac{1}{p^2} \left(\frac{\lambda}{n}\eta\right)^2.
\end{equation}
Furthermore, as a sum of independent indicator random variables,
$L$ is a binomial random variable with $n$ trials and
success probability $p$ for each trial.

For $\ell > 0$,
\begin{align}
\E{\etaDTQM \smid L = \ell} 
&= \E{\frac{Y}{L} \smid L = \ell}
 = \frac{1}{\ell} \E{Y \smid L = \ell} \nonumber \\
&\eqlabel{a} \frac{1}{\ell}\E{\sum_{j = 1}^{\ell}\Ytilde_j}
 = \E{ \Ytilde_k },
\label{equ:QM_condi_mean}
\end{align}
where (\emph{a}) follows from using $\Ytilde_j$ to denote the $j$th positive subacquistion $Y_k$.
Trivially,
\begin{equation}
\label{equ:QM_condi_mean0}
\E{\etaDTQM \smid L = 0} = 0.
\end{equation}
Using $\iP{L>0} = 1-(1-p)^n$ and the total expectation theorem to combine
\eqref{equ:QM_condi_mean} and \eqref{equ:QM_condi_mean0} gives
\[
  \E{\etaDTQM} = \iE{\Ytilde_k} (1 - (1-p)^n)).
\]
The bias of $\etaDTQM$ is thus given by
\begin{align}\label{eq:DTQM_bias}
\bias(\etaDTQM)
&= \E{\etaDTQM} - \eta \nonumber \\
&= \frac{\lambda \eta}{n p} \left[1 - (1 - p)^n\right] - \eta,
\end{align}
where \eqref{eq:ytilde-mean} has been substituted.

Similar arguments, detailed in Appendix~\ref{app:mse_DTQM_deri}, yield
\begin{align}\label{eq:DTQM_var}
\var{\etaDTQM}
&= \left( \iE{ \Ytilde_k } \right)^2 \left[1 - (1 - p)^n\right] (1 - p)^n  \nonumber\\
&  \quad +
    \ivar{\Ytilde_k}
    \sum_{\ell = 1}^{n}
    \frac{1}{\ell}\binom{n}{\ell}p^\ell(1 - p)^{n-\ell},
\end{align}
where $\iE{ \Ytilde_k }$ is given in \eqref{eq:ytilde-mean} and
$\ivar{ \Ytilde_k }$ is given in \eqref{eq:ytilde-var}.
Since the behavior of the series in \eqref{eq:DTQM_var} for large $n$ is not evident,
we also derive in Appendix~\ref{app:mse_DTQM_deri} the lower bound
\begin{align}\label{eq:DTQM_var_lower_bound}
\var{\etaDTQM}
&\geq \left( \iE{ \Ytilde_k } \right)^2 \left[1 - (1 - p)^n\right] (1 - p)^n  \nonumber\\
& \quad +
    \ivar{\Ytilde_k} \frac{[1 - (1 - p)^n]^2}{np}.
\end{align}
The expressions in \eqref{eq:DTQM_var} and \eqref{eq:DTQM_var_lower_bound}
can be added to the square of the bias from
\eqref{eq:DTQM_bias} to obtain the MSE of $\etaDTQM$
and a lower bound for this MSE.

\begin{figure*}
  \hfill
  \begin{subfigure}{\threewidewidth}
    \includegraphics[width=\linewidth]{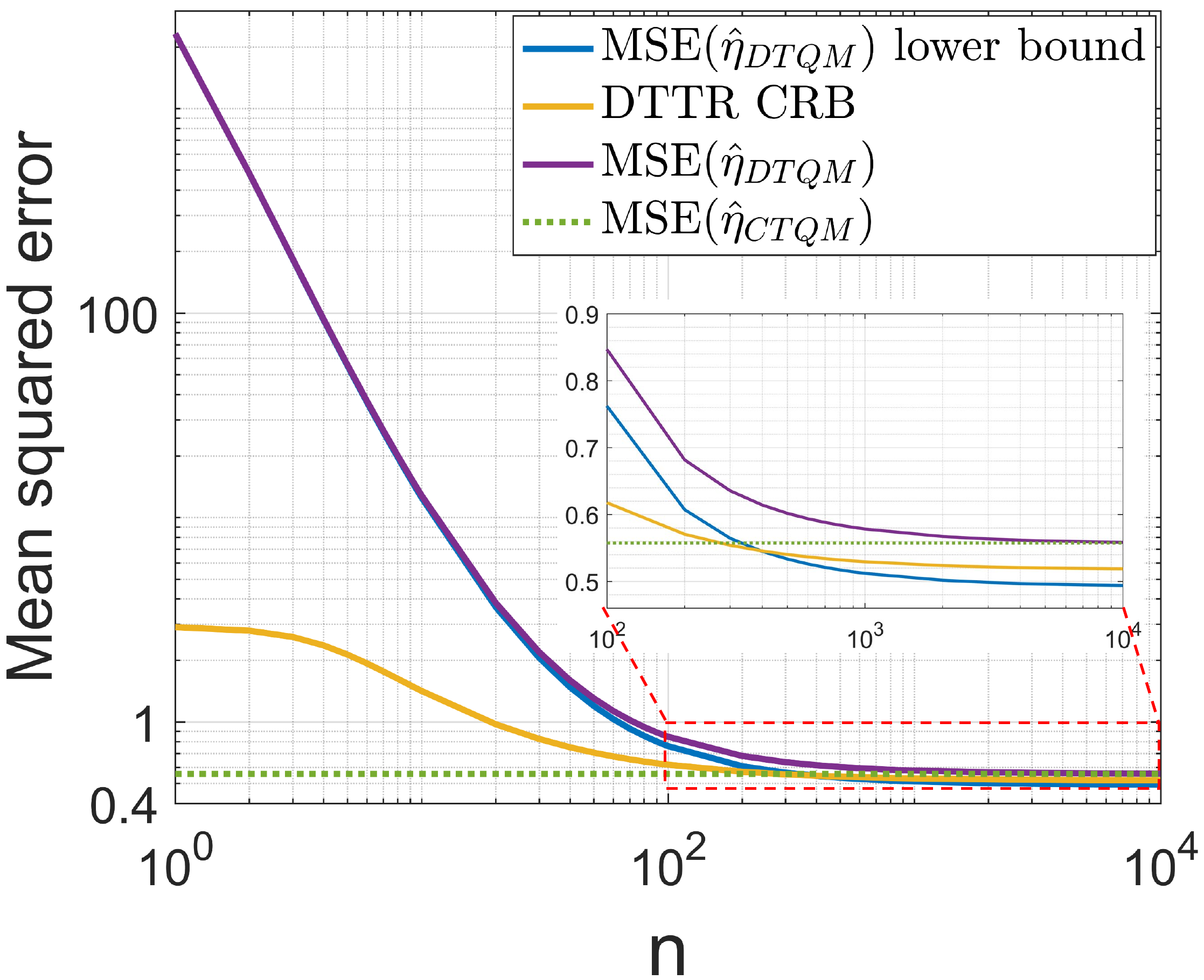}
    \subcaption{MSE across $n$}
    \label{fig:QM_mse}
  \end{subfigure}
  \hfill
  \begin{subfigure}{\threewidewidth}
    \includegraphics[width=\linewidth]{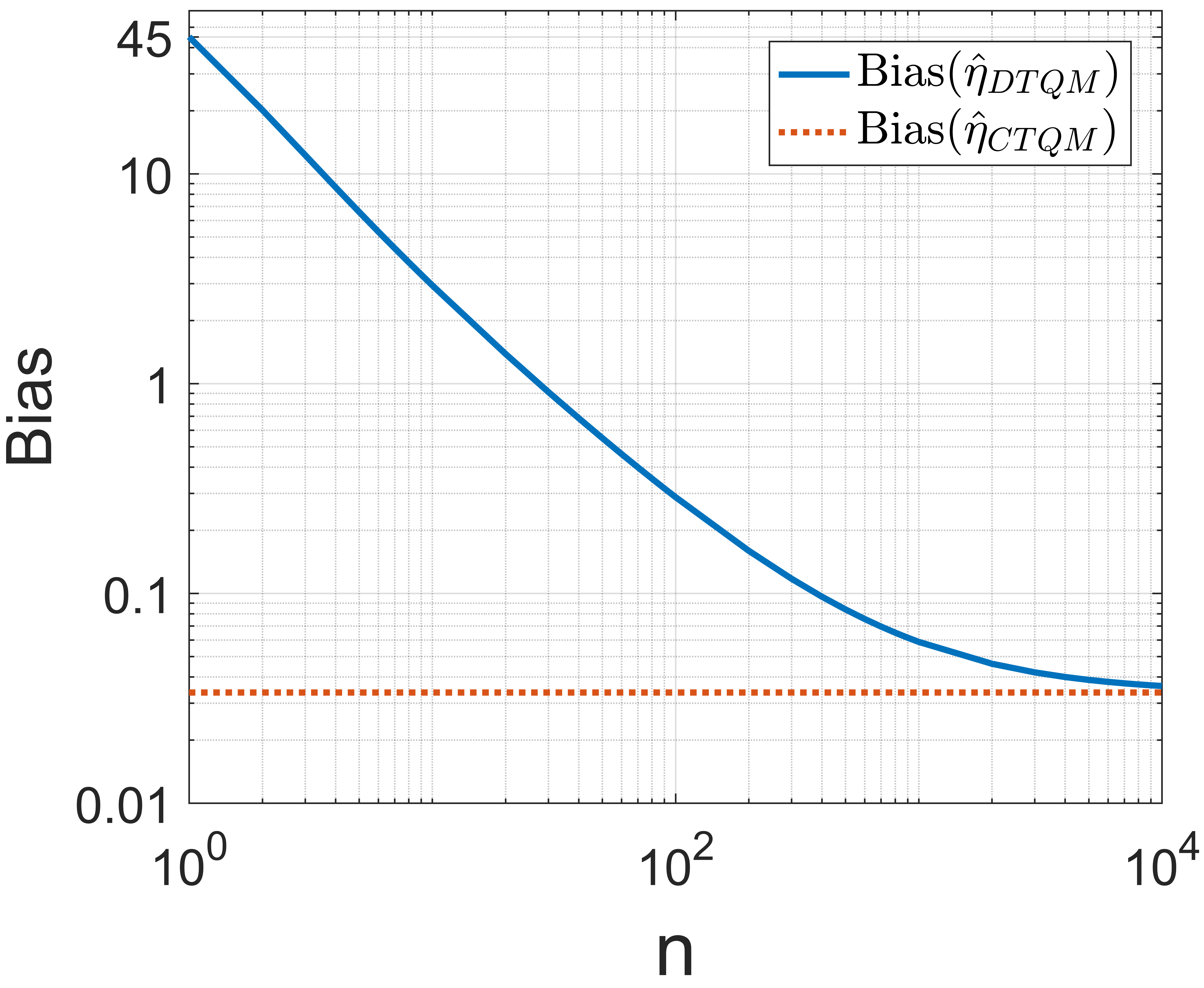}
    \subcaption{Bias across $n$}
    \label{fig:QM_mean}
  \end{subfigure}
  \hfill
  \begin{subfigure}{\threewidewidth}
    \includegraphics[width=\linewidth]{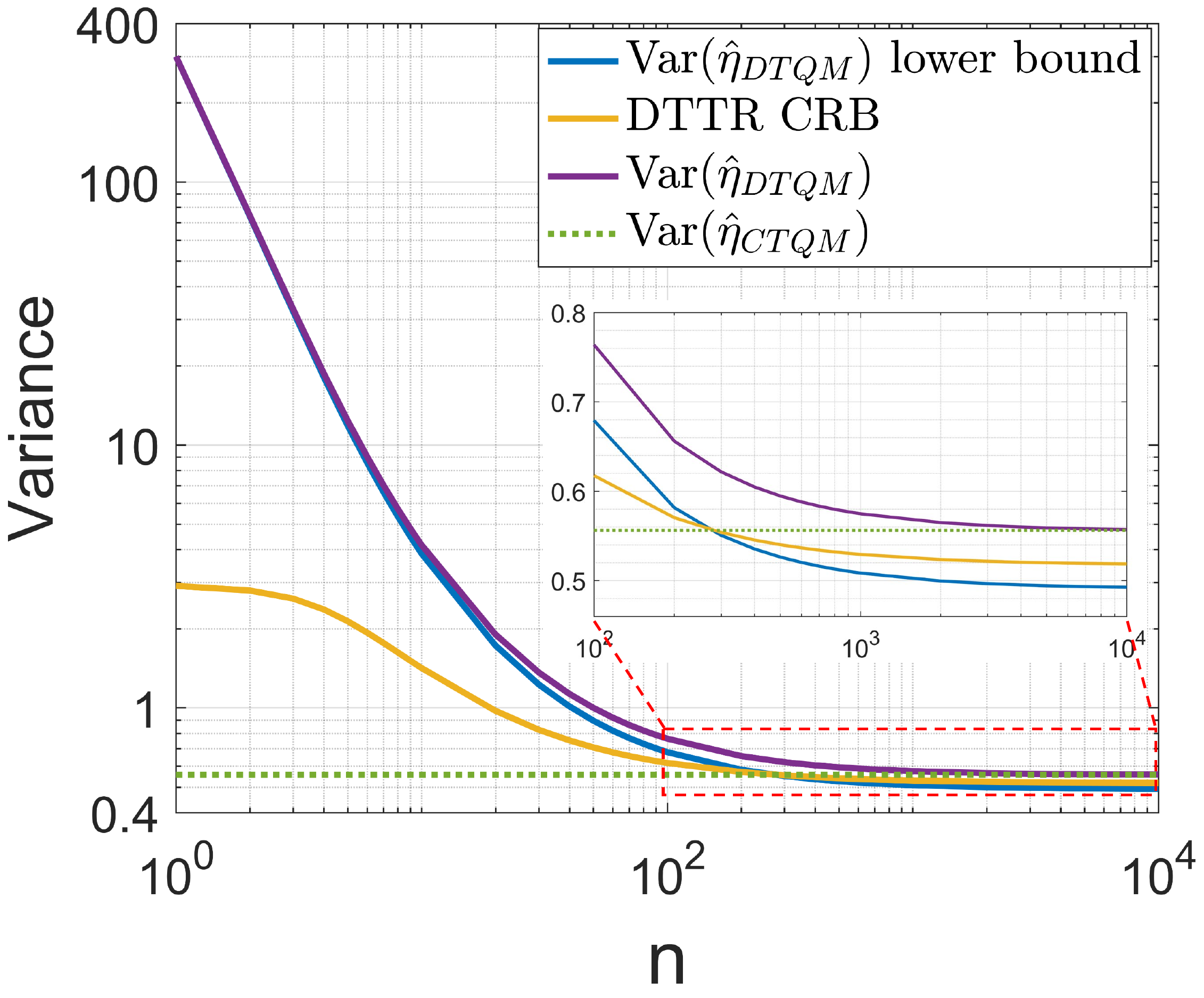}
    \subcaption{Variance across $n$}
    \label{fig:QM_var}
  \end{subfigure}
  \hfill
  \caption{MSE, bias, and variance of $\etaDTQM$ as functions of the number of subacquisitions $n$
    and
    those of $\etaCTQM$ (dashed lines) for $\lambda = 10$ and $\eta = 5$.
    The {\CR} lower bound for time-resolved measurements (yellow, see~\eqref{eq:Fisher_info_TR}) is plotted as well.} 
  \label{fig:QM_MSE_vs_n}
\end{figure*}

\subsubsection*{High $n$ Limits}
One of the themes of this paper is that
the CTTR measurement model is easier to analyze than
the DTTR measurement model,
yet it gives expressions relevant to understanding the practical DTTR measurement setting.
We would like to examine properties of $\etaDTQM$ at the high $n$ limit,
in part to demonstrate that we obtain matches to $\etaCTQM$ behavior.

For the limit of the bias,
we will use two facts proven in Appendix~\ref{app:high_n}:
\begin{align}
  \label{eq:lim_np}
    \lim_{n\to\infty} np      &= \lambda (1 - e^{-\eta}), \\
  \label{eq:lim_1minusp_n}
    \lim_{n\to\infty} (1-p)^n &= e^{-\lambda(1-e^{-\eta})}.
\end{align}
By substituting \eqref{eq:lim_np} and \eqref{eq:lim_1minusp_n} into \eqref{eq:DTQM_bias},
\begin{align}\label{eq:DTQM_bias_limit}
\lim_{n \to \infty} \bias(\etaDTQM) 
&= \frac{\lambda \eta}{\lambda(1 - e^{-\eta})}(1 - e^{-\lambda(1-e^{-\eta})}) - \eta \nonumber \\
&= \left( \frac{e^{-\eta} - e^{-\lambda(1-e^{-\eta})}} {1-e^{-\eta}} \right) \eta,
\end{align}
an exact match to $\bias(\etaCTQM)$ in \eqref{eq:eta_QM_cont_bias}.

\begin{figure*}
  \hfill
  \begin{subfigure}{\threewidewidth}
    \includegraphics[width=\linewidth]{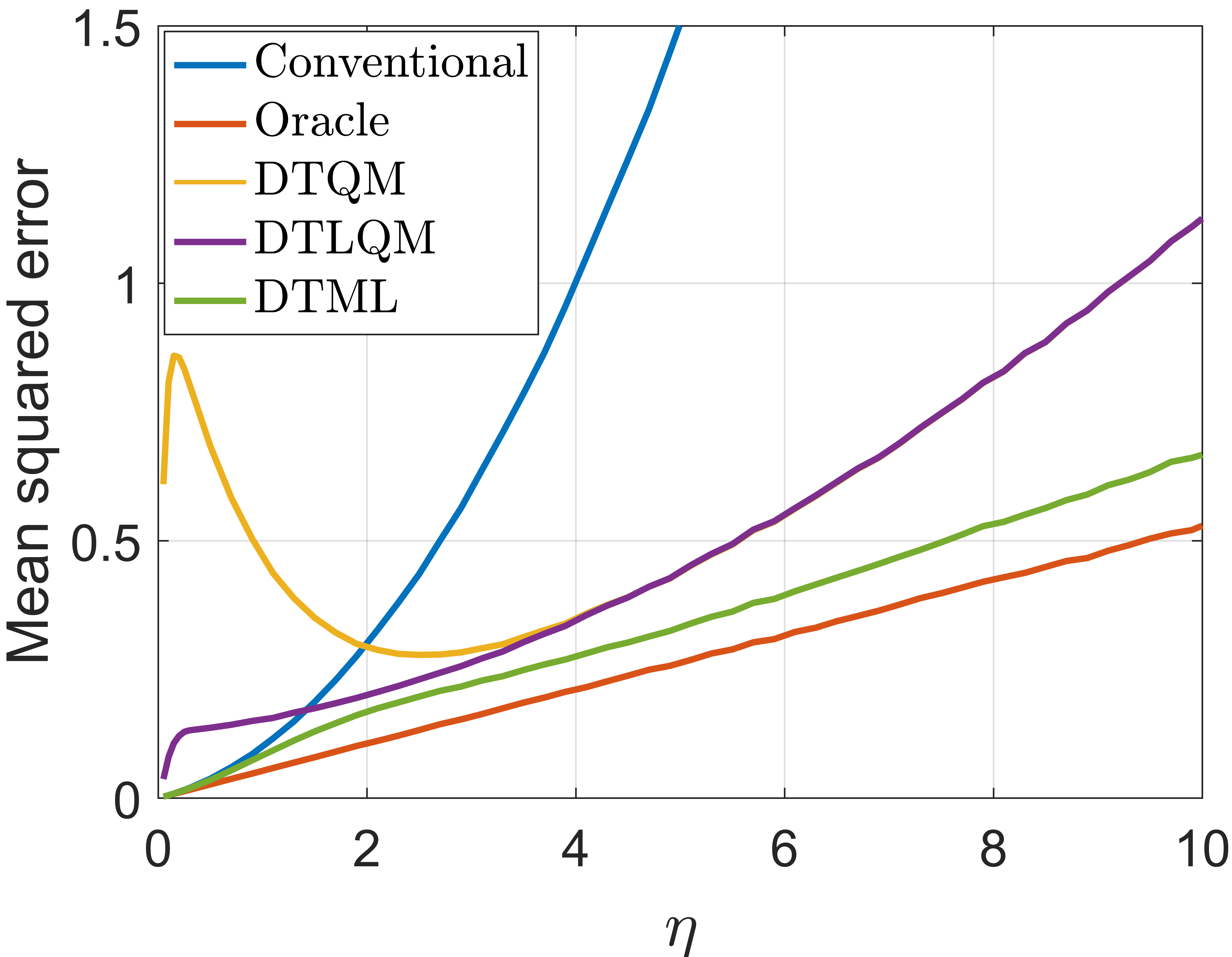}
    \subcaption{MSE across $\eta$}
  \end{subfigure}
  \hfill
  \begin{subfigure}{\threewidewidth}
    \includegraphics[width=\linewidth]{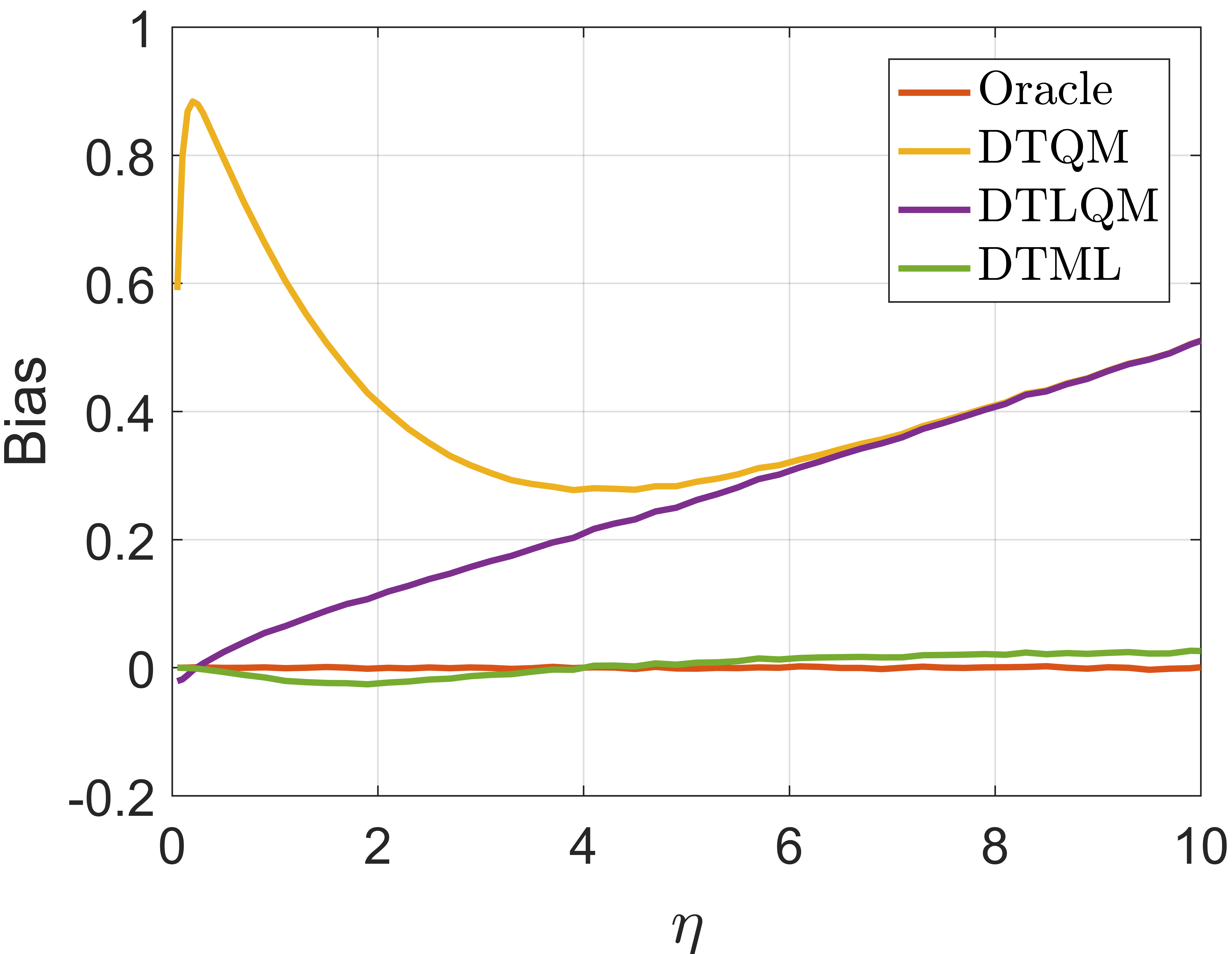}
    \subcaption{Bias across $\eta$}
  \end{subfigure}
  \hfill
  \begin{subfigure}{\threewidewidth}
    \includegraphics[width=\linewidth]{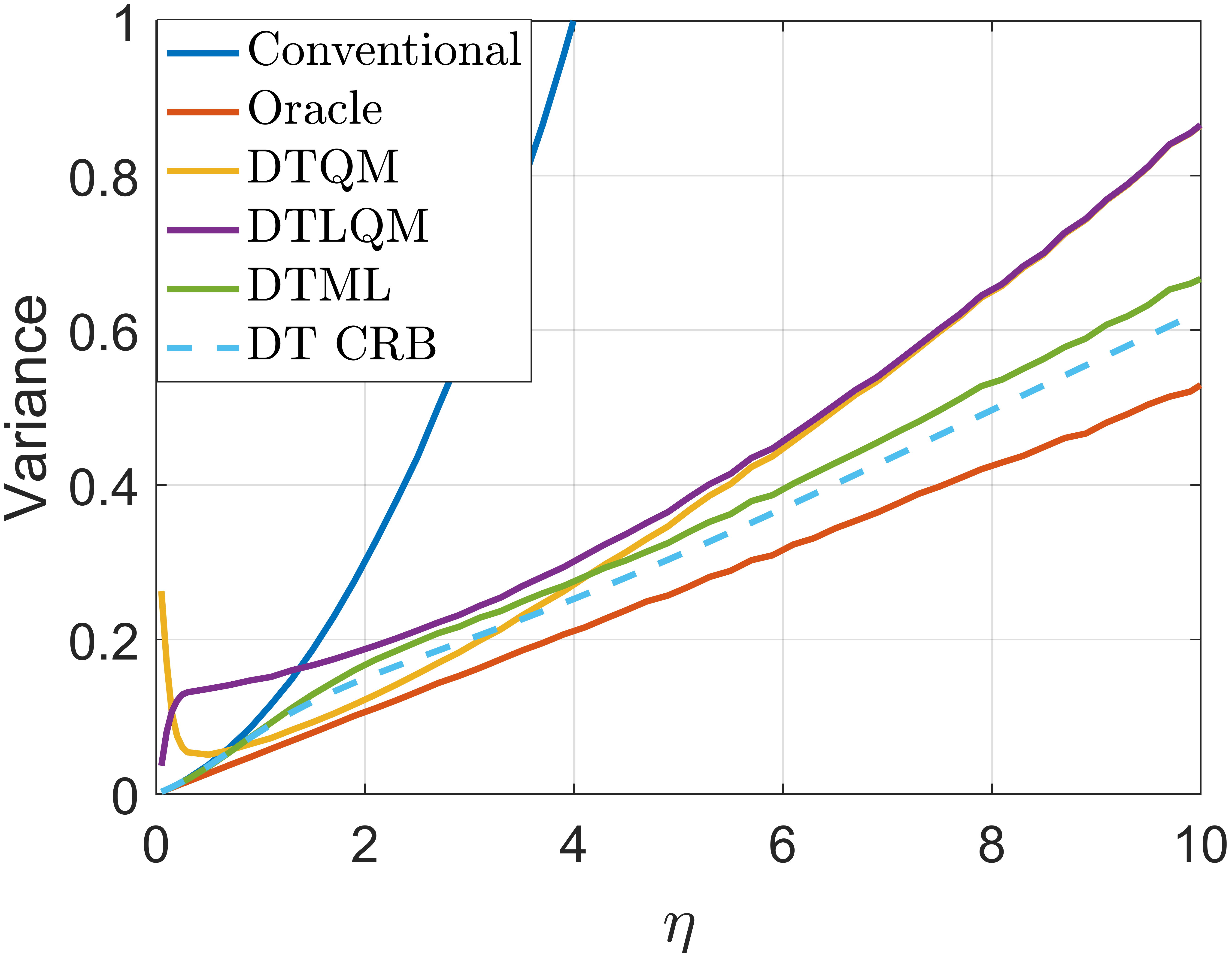}
    \subcaption{Variance across $\eta$}
  \end{subfigure}
  \hfill
  \caption{Comparison of discrete-time time-resolved estimators
    as a function of $\eta$.
    Conventional, oracle, discrete-time quotient mode, 
    discrete-time Lambert quotient mode, and discrete-time maximum likelihood estimators are simulated for total dose $\lambda = 20$ split over $n=200$ subacquisitions. (a) MSE\@. (b) Bias. (Conventional estimator excluded because its bias is zero.) (c) Variance.}
  \label{fig:PP_discrete_MSE_comparison}
\end{figure*}

For the limit of the variance,
we will use two additional facts proven in Appendix~\ref{app:high_n}:
\begin{align}
  \label{eq:lim_E_Ytilde}
    \lim_{n\to\infty} \iE{\Ytilde_k}   &= \frac{\eta}{1-e^{-\eta}} , \\
  \label{eq:lim_var_Ytilde}
    \lim_{n\to\infty} \ivar{\Ytilde_k} &= \frac{\eta - (\eta+\eta^2)e^{-\eta}}{(1-e^{-\eta})^2} .
\end{align}
Along with substitution of limits, we recognize that the summand in
\eqref{eq:DTQM_var}
includes a binomial probability for which there is a Poisson limit
because $np$ approaches a positive constant~\cite{Serfling:78}:
\begin{equation}
    \label{eq:binomial-Poisson}
    \binom{n}{\ell}p^\ell(1 - p)^{n-\ell}
    \rightarrow
    \frac{(
    \lambda (1 - e^{-\eta})
    )^\ell}{\ell!} e^{-
    \lambda (1 - e^{-\eta})
    }.
\end{equation}
It is now tedious but straightforward to substitute
\eqref{eq:lim_np}, \eqref{eq:lim_1minusp_n}, \eqref{eq:lim_E_Ytilde}, \eqref{eq:lim_var_Ytilde}, and \eqref{eq:binomial-Poisson}
into
\eqref{eq:DTQM_var} to obtain
\begin{align}\label{eq:DTQM_var_limit}
\lim_{n \to \infty} \var{\etaDTQM} 
&= \var{\etaCTQM},
\end{align}
where $\var{\etaCTQM}$ is given in \eqref{eq:eta_QM_cont_total_var}.

\subsection{Numerical Comparisons of Estimators}
Fig.~\ref{fig:QM_MSE_vs_n} shows the MSE, bias, and variance of $\etaDTQM$
as functions of the number of subacquisitions $n$ when $\lambda = 10$ and $\eta = 5$. 
Fig.~\ref{fig:QM_mean} shows the bias
approaching the asymptote given by
\eqref{eq:DTQM_bias_limit}.
Fig.~\ref{fig:QM_mse} and Fig.~\ref{fig:QM_var} show the MSE and variance
and their lower bounds based on \eqref{eq:DTQM_var_lower_bound}.
When $n$ is sufficiently large, these are small and close to the \CR \, bound.

Fig.~\ref{fig:PP_discrete_MSE_comparison} compares the
conventional, oracle, DTQM, DTLQM, and DTML estimators across
ground truth $\eta \in [0,10]$.
MSE, bias, and variance are computed by Monte Carlo simulation
using total dose $\lambda = 20$ split over $n=200$ subacquisitions. 
The conventional curves match those in the CT setting in Fig.~\ref{fig:CTTR_comparison}.
The DTQM estimator has a large bias for smaller $\eta$, 
which is absent from the DTLQM estimator.
Unlike in CT,
where the bias for all the studied estimators vanish at moderate and high $\eta$,
both the DTQM and DTLQM estimator have substantial bias that is approximately
linear in $\eta$ for moderate and high $\eta$.
The lack of bias of the DTML estimator explains its uniform superiority over the DTQM and DTLQM estimators;
this contrasts with the CT setting in which CTQM, CTLQM, and CTML estimators have nearly equal MSE at moderate and high $\eta$.

\newlength{\figHeight}
\setlength{\figHeight}{3.2cm}
\newlength{\firstcolwidth}
\setlength{\firstcolwidth}{0.17\textwidth}
\newlength{\restcolwidth}
\setlength{\restcolwidth}{0.15\textwidth}

\begin{figure*}
  \centering
  \begin{tabular}{@{}c|c@{\,\,}c@{\,\,}c@{\,\,}c@{\,\,}c@{\,\,}c@{}}
    &
    & \multicolumn{5}{c}{\small \bf Absolute errors of time-resolved estimators}
    \\
    {\small \bf Ground truth}
    &
    & {\small DT $n=50$}
    & {\small DT $n=100$}
    & {\small DT $n=200$}
    & {\small DT $n=500$}
    & {\small CT}
    \\
    \begin{subfigure}[t]{\firstcolwidth}
        \includegraphics[height=\figHeight]{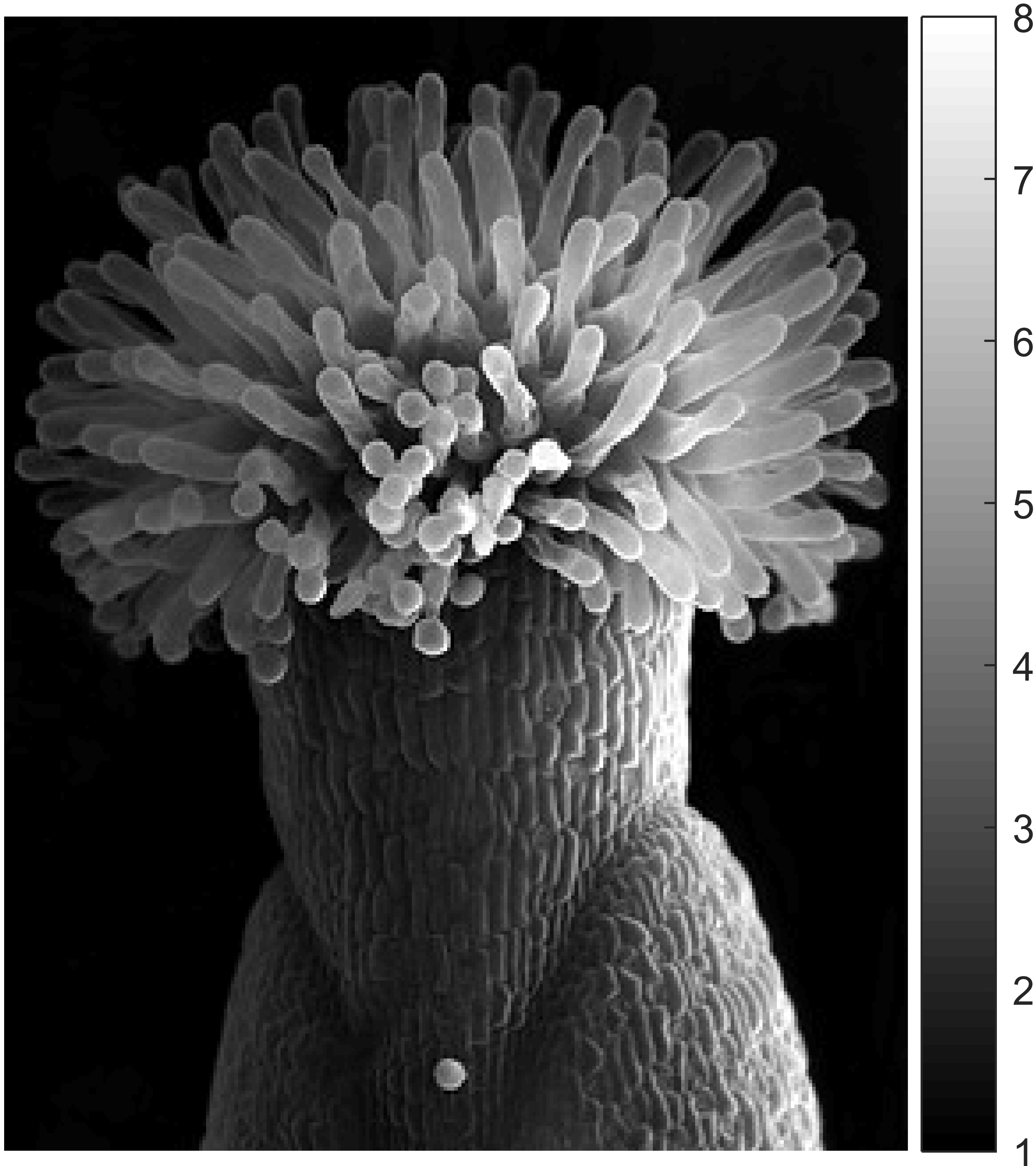}
    \end{subfigure} &
    \rotatebox[origin=l]{90}{\qquad \qquad \small QM}
    &
    \begin{subfigure}[t]{1.01\restcolwidth}
        \includegraphics[height=\figHeight]{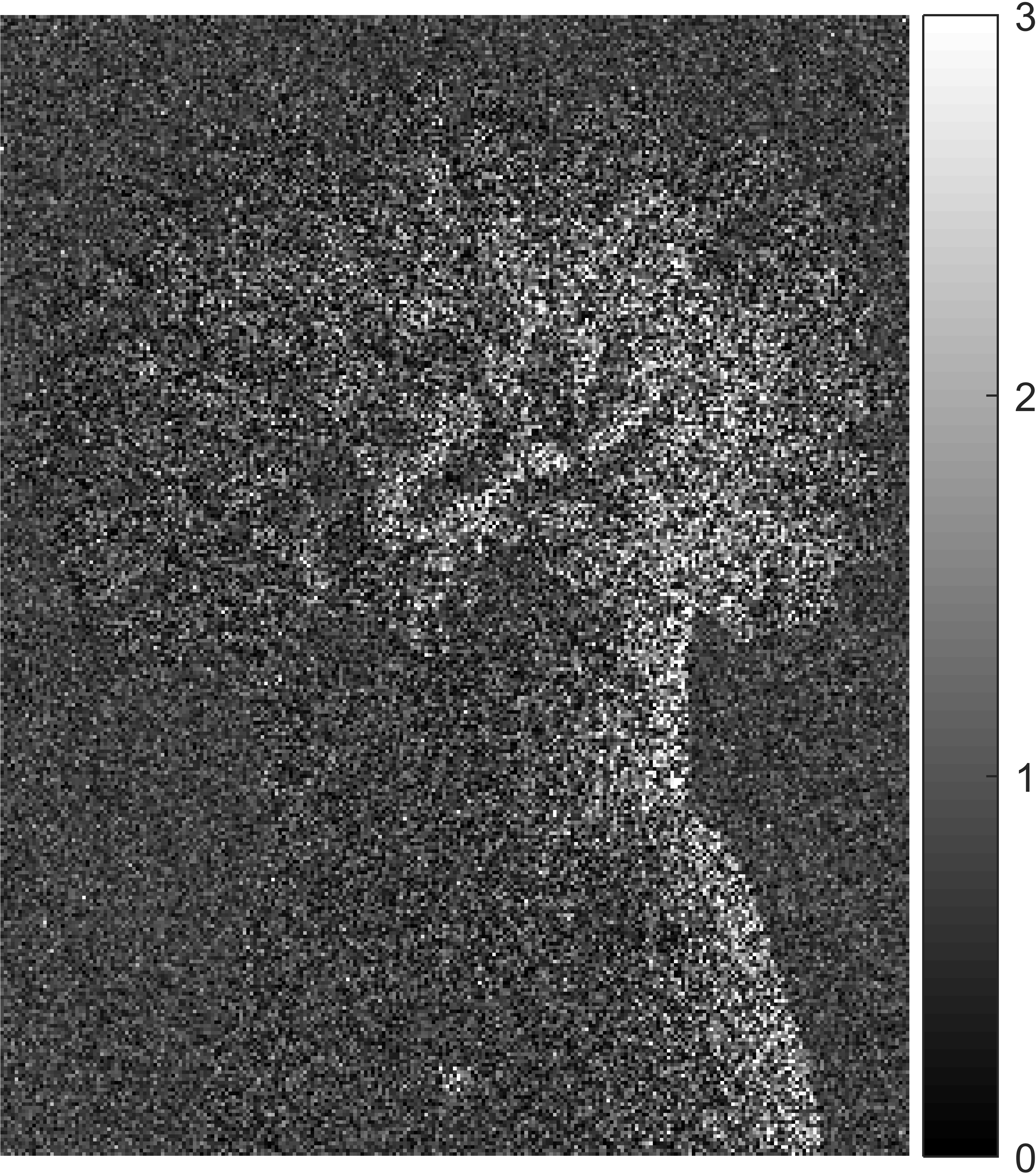}
        \caption*{$\mathrm{MSE} = 1.111$}
    \end{subfigure} &
    \begin{subfigure}[t]{\restcolwidth}
        \includegraphics[height=\figHeight]{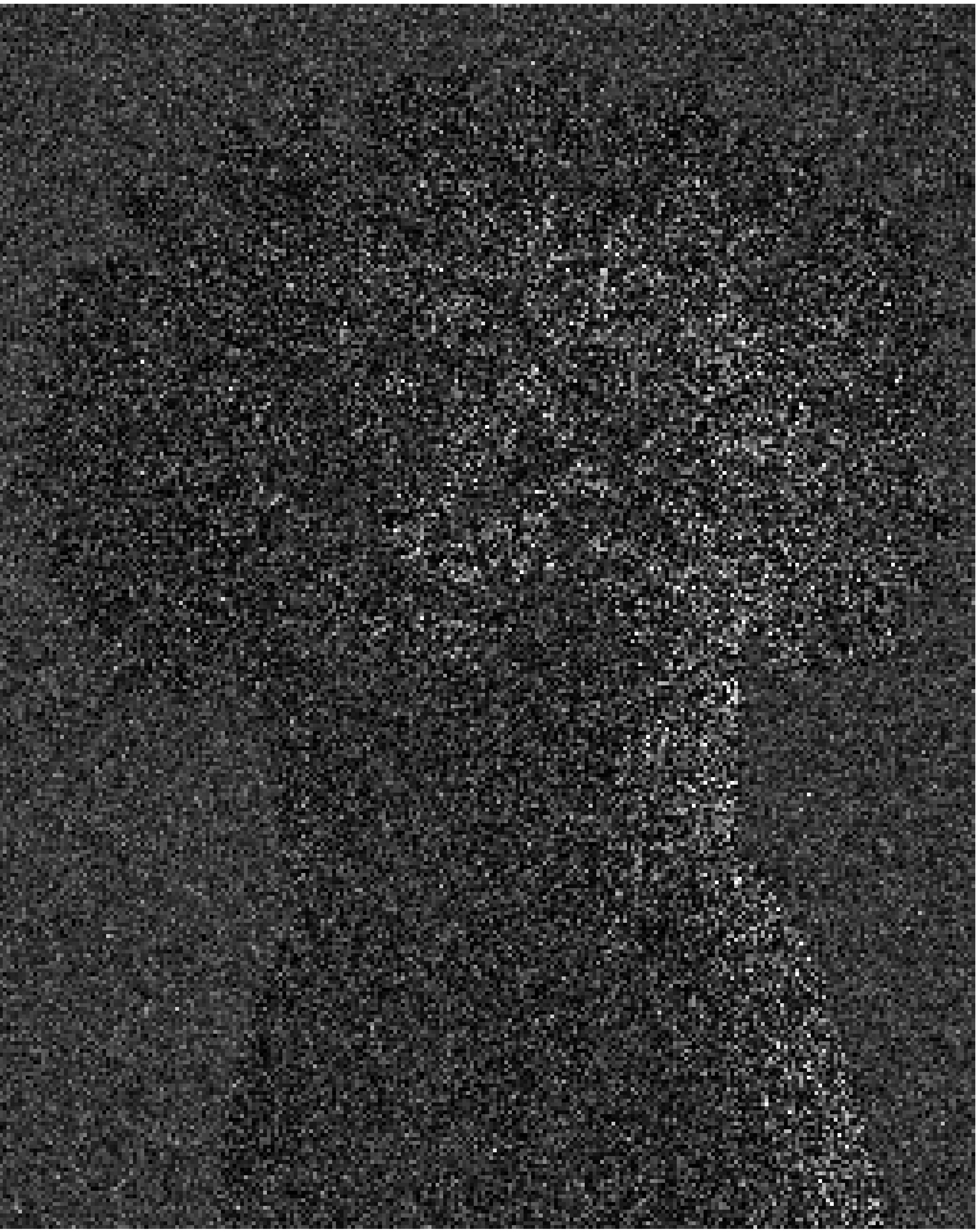}
        \caption*{$\mathrm{MSE} = 0.566$}
    \end{subfigure} &
    \begin{subfigure}[t]{\restcolwidth}
        \includegraphics[height=\figHeight]{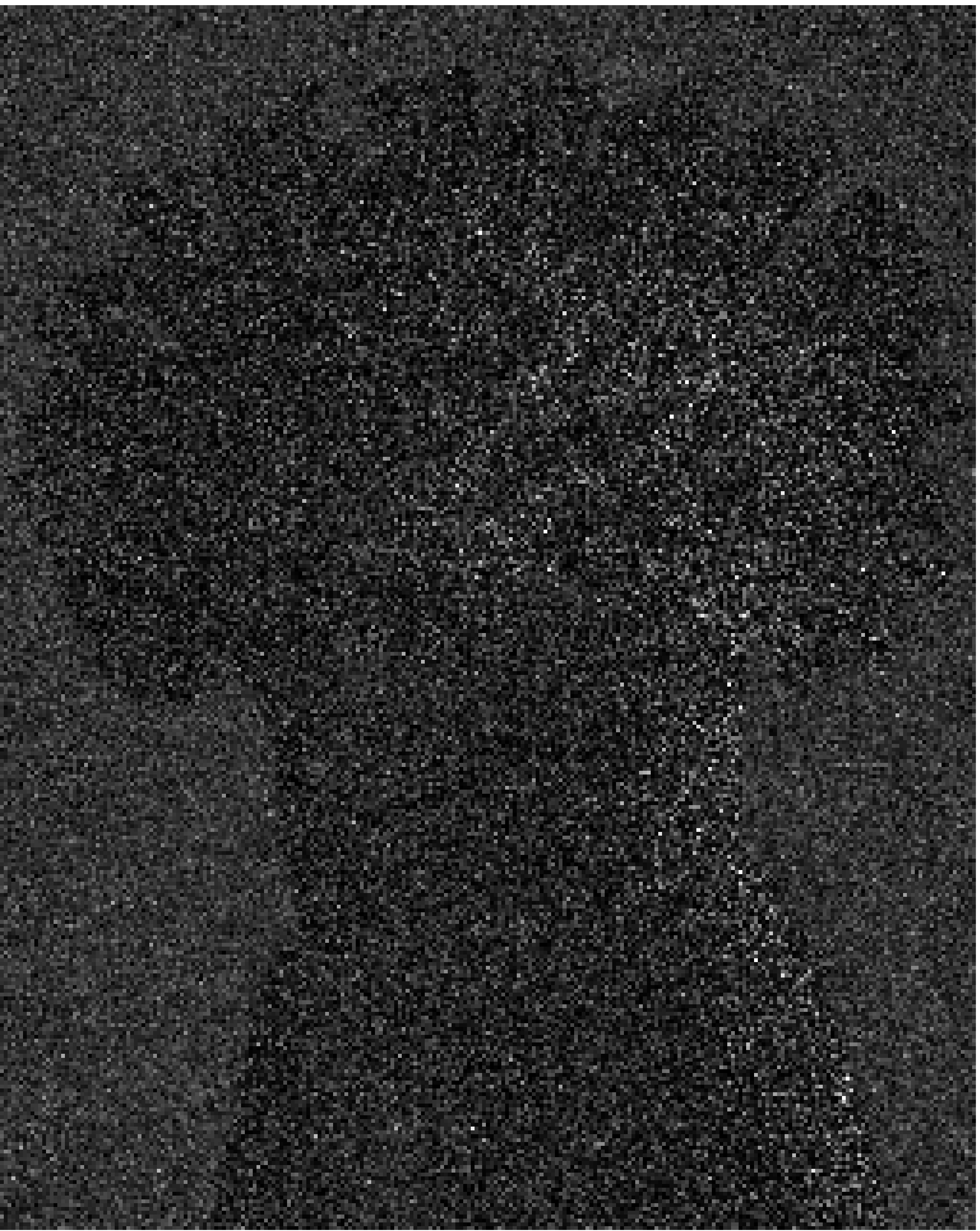}
        \caption*{$\mathrm{MSE} = 0.392$}
    \end{subfigure} &
    \begin{subfigure}[t]{\restcolwidth}
        \includegraphics[height=\figHeight]{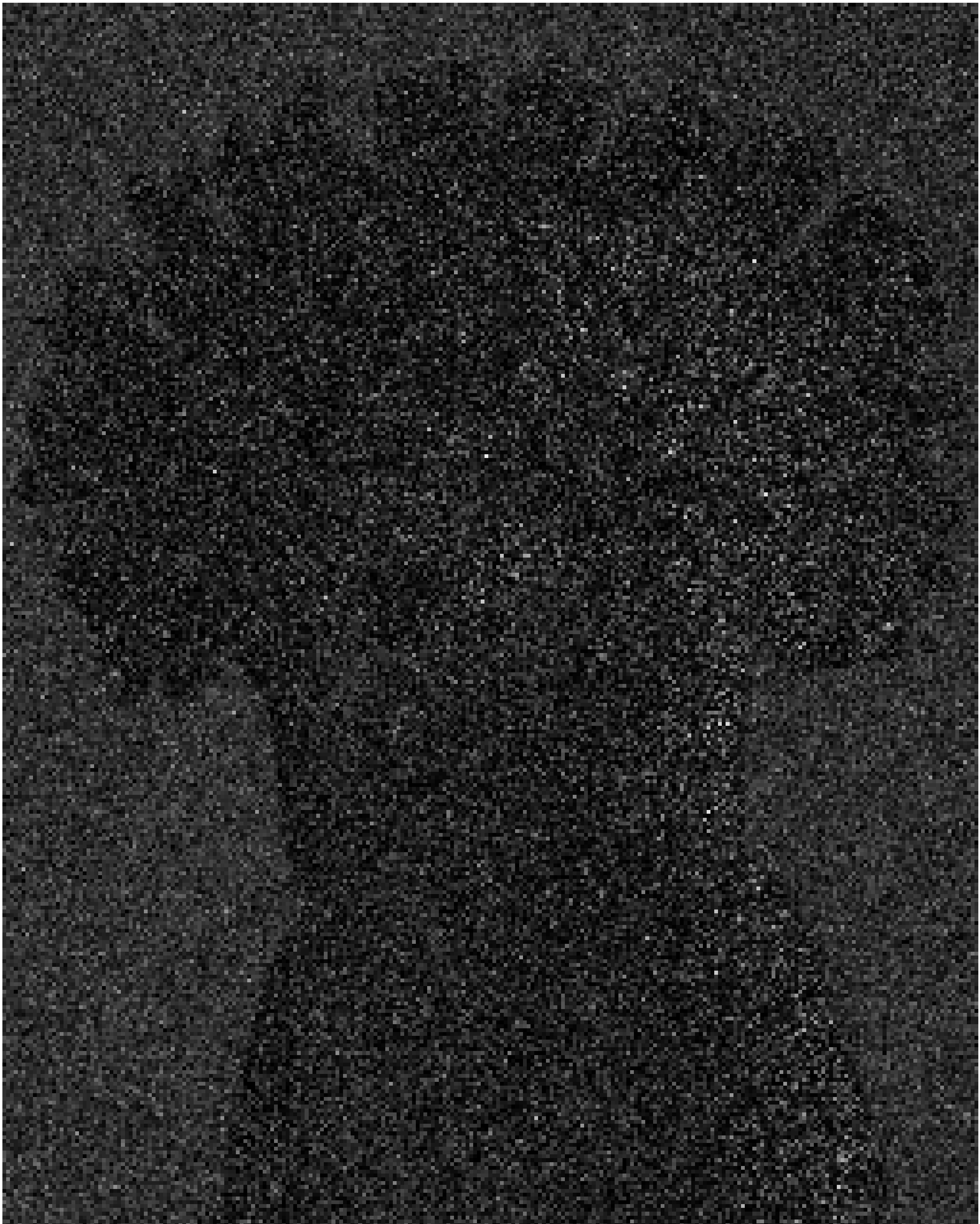}
        \caption*{$\mathrm{MSE} = 0.313$}
    \end{subfigure} &
    \begin{subfigure}[t]{\restcolwidth}
        \includegraphics[height=1.01\figHeight]{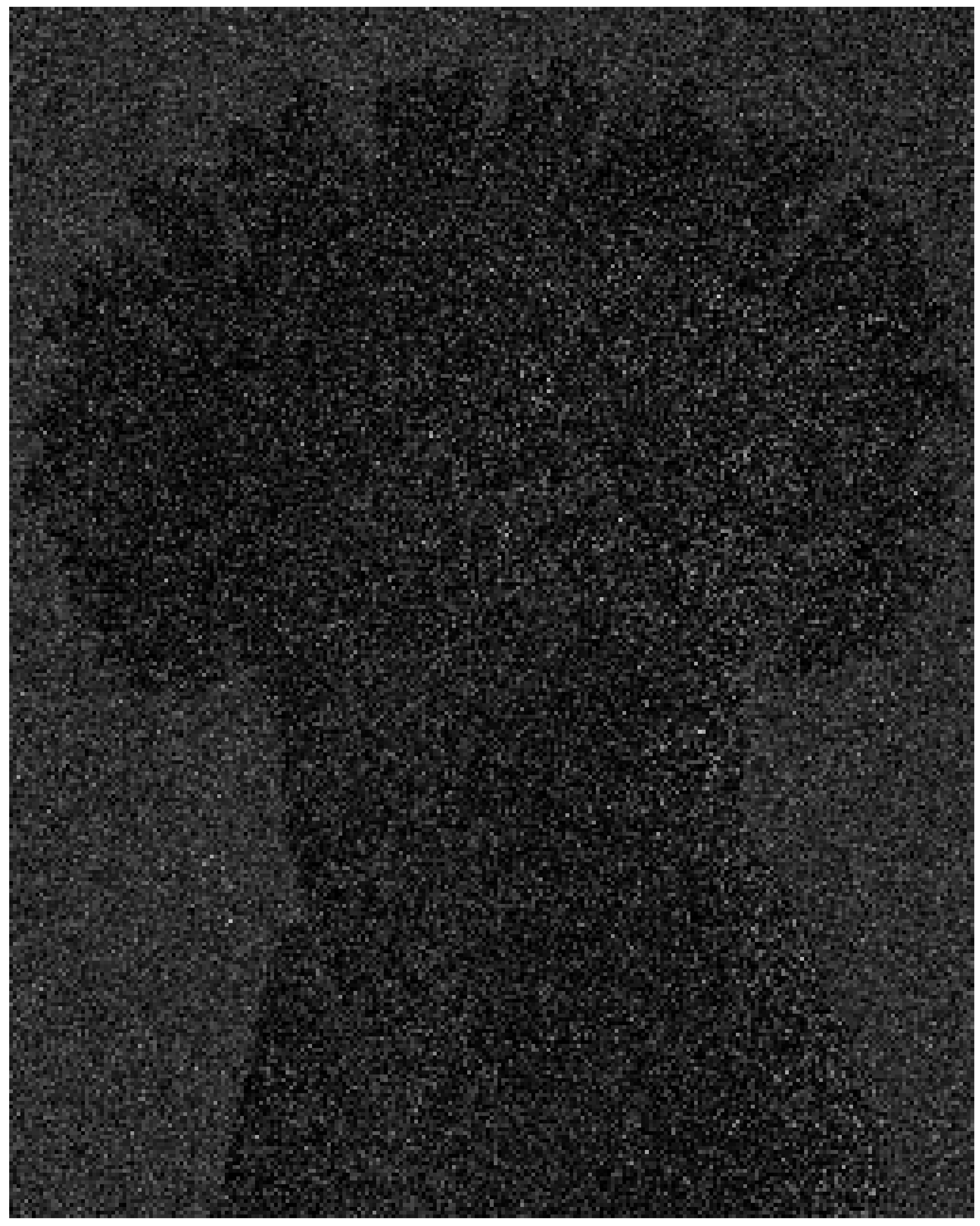}
        \caption*{$\mathrm{MSE} = 0.272$}
    \end{subfigure} \\
    {\small \bf Absolute error of} \\[-1mm]
    {\small \bf conventional estimator} \\
    \begin{subfigure}[t]{\firstcolwidth}
        \includegraphics[height=\figHeight]{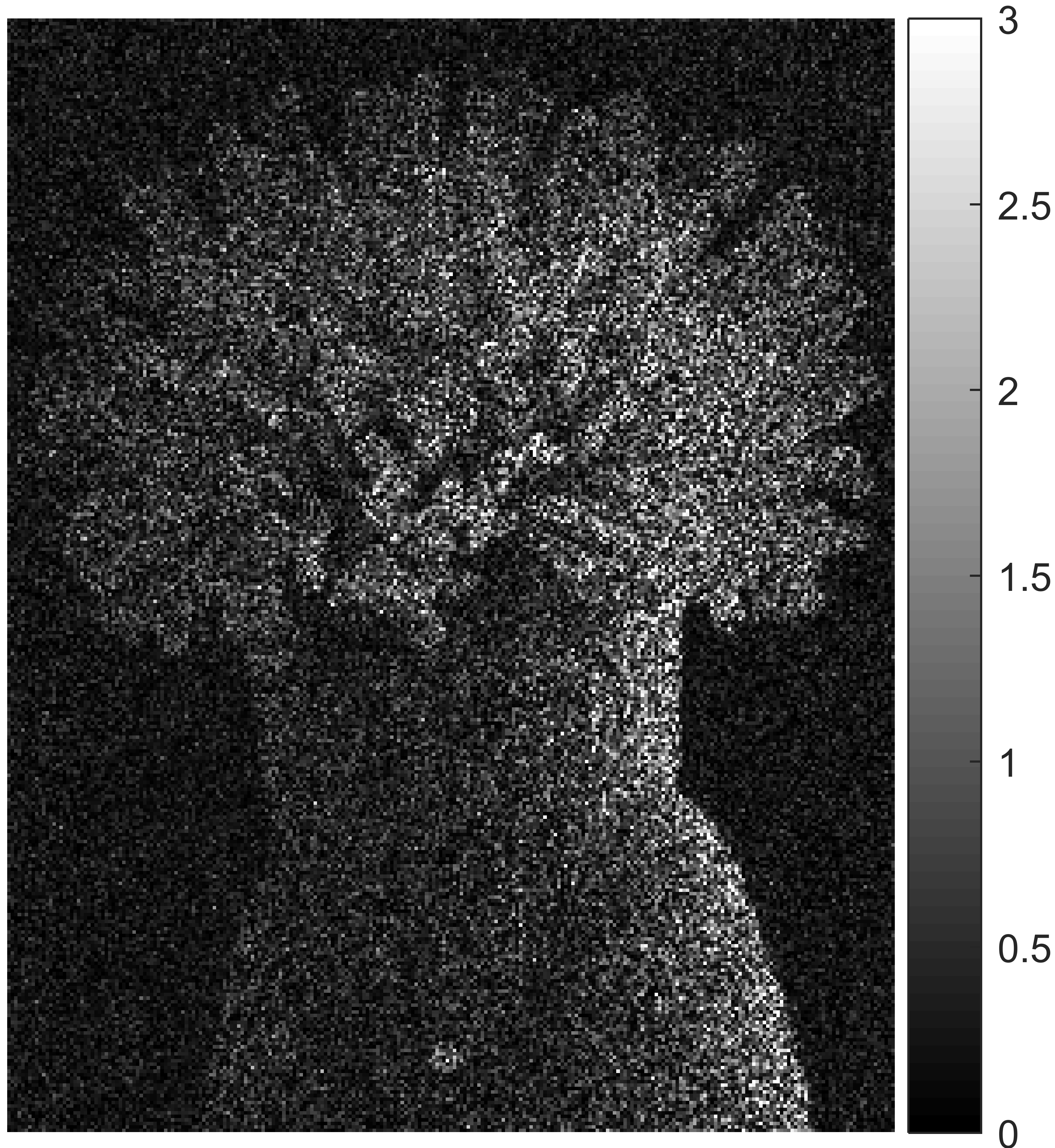}
        \caption*{$\mathrm{MSE} = 0.599$}
    \end{subfigure} & 
    \rotatebox[origin=l]{90}{\qquad \qquad \small LQM}
    &
    \begin{subfigure}[t]{\restcolwidth}
        \includegraphics[height=\figHeight]{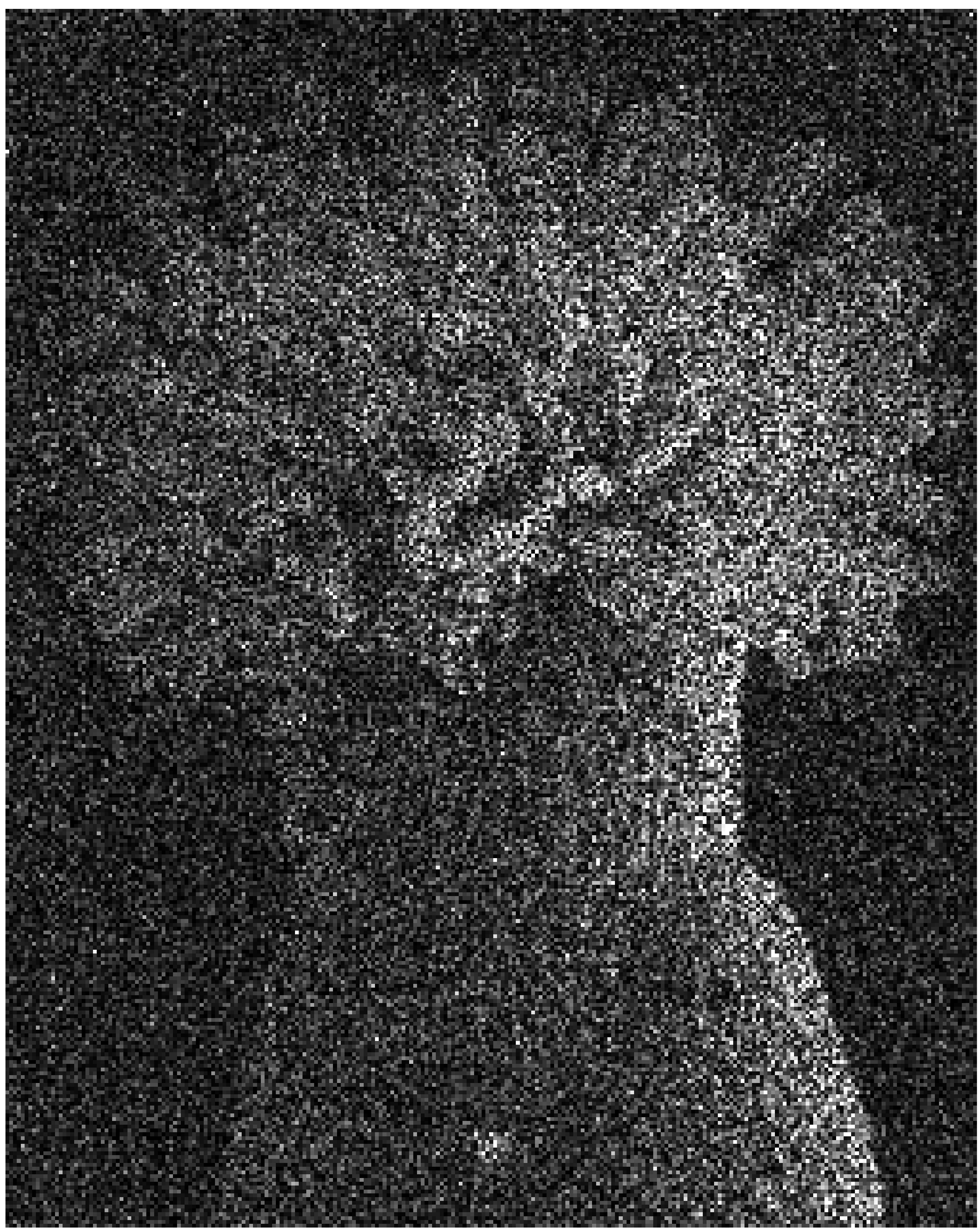}
        \caption*{$\mathrm{MSE} = 0.877$}
    \end{subfigure} & 
    \begin{subfigure}[t]{\restcolwidth}
        \includegraphics[height=\figHeight]{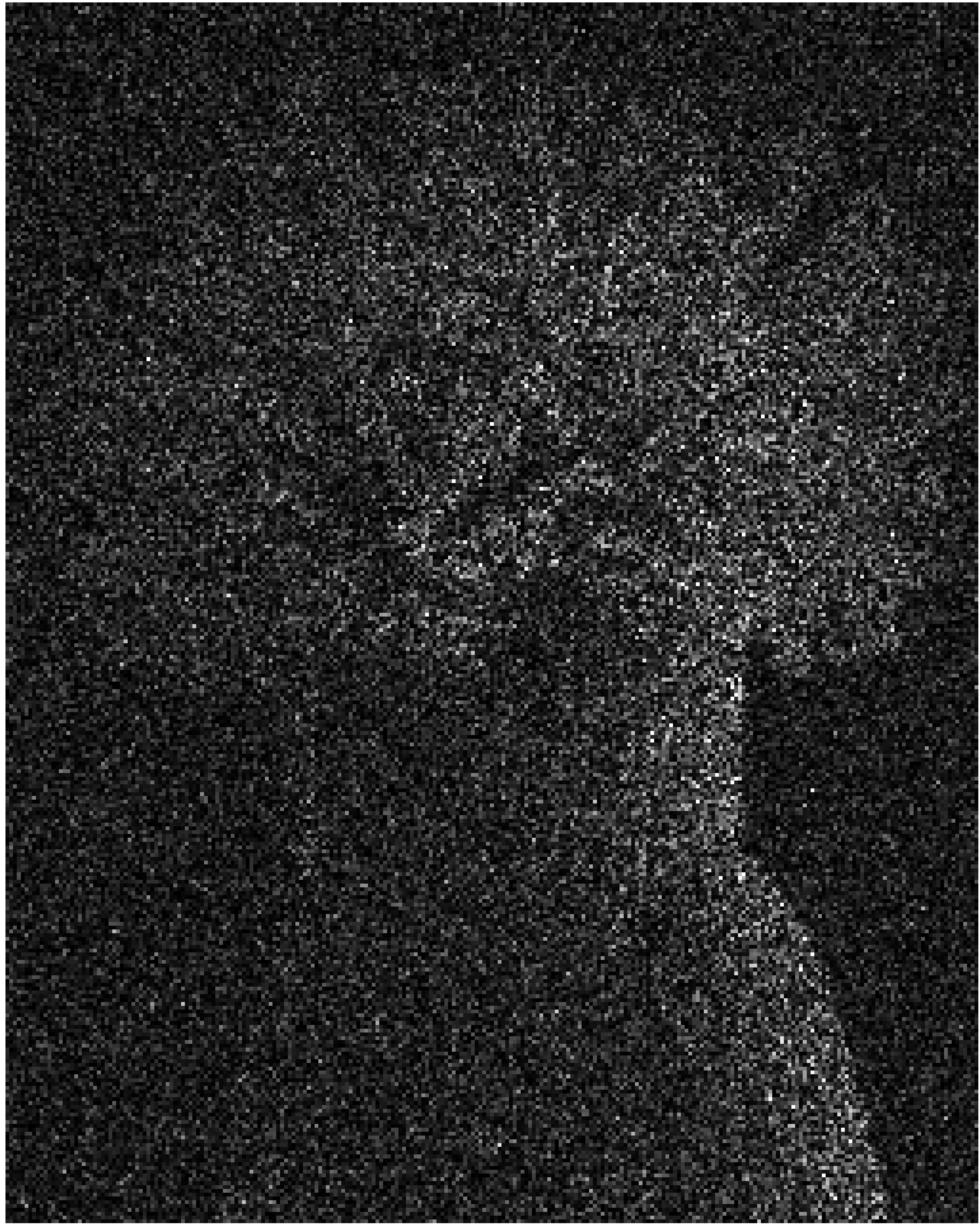}
        \caption*{$\mathrm{MSE} = 0.398$}
    \end{subfigure} &
    \begin{subfigure}[t]{\restcolwidth}
        \includegraphics[height=\figHeight]{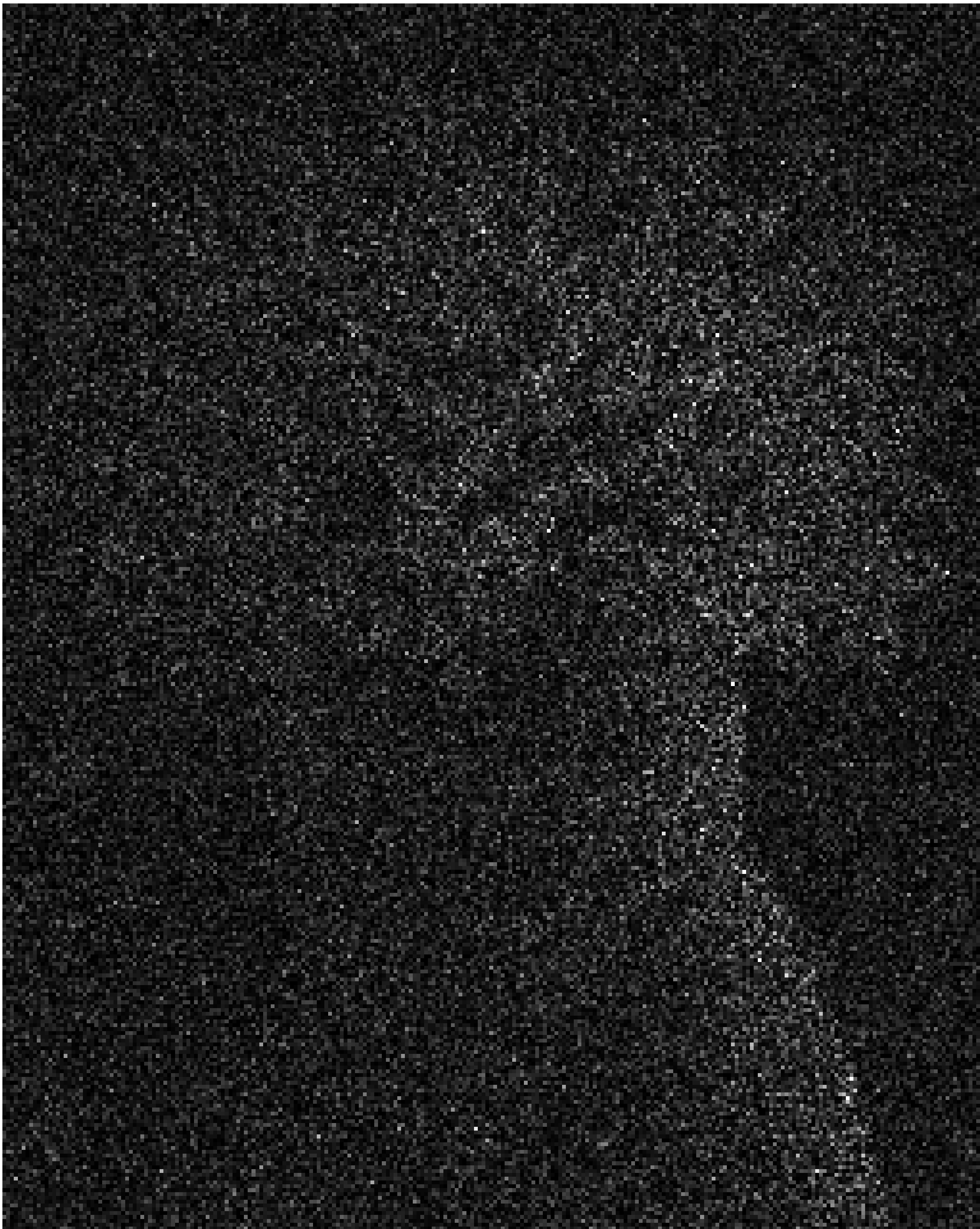}
        \caption*{$\mathrm{MSE} = 0.259$}
    \end{subfigure} & 
    \begin{subfigure}[t]{\restcolwidth}
        \includegraphics[height=\figHeight]{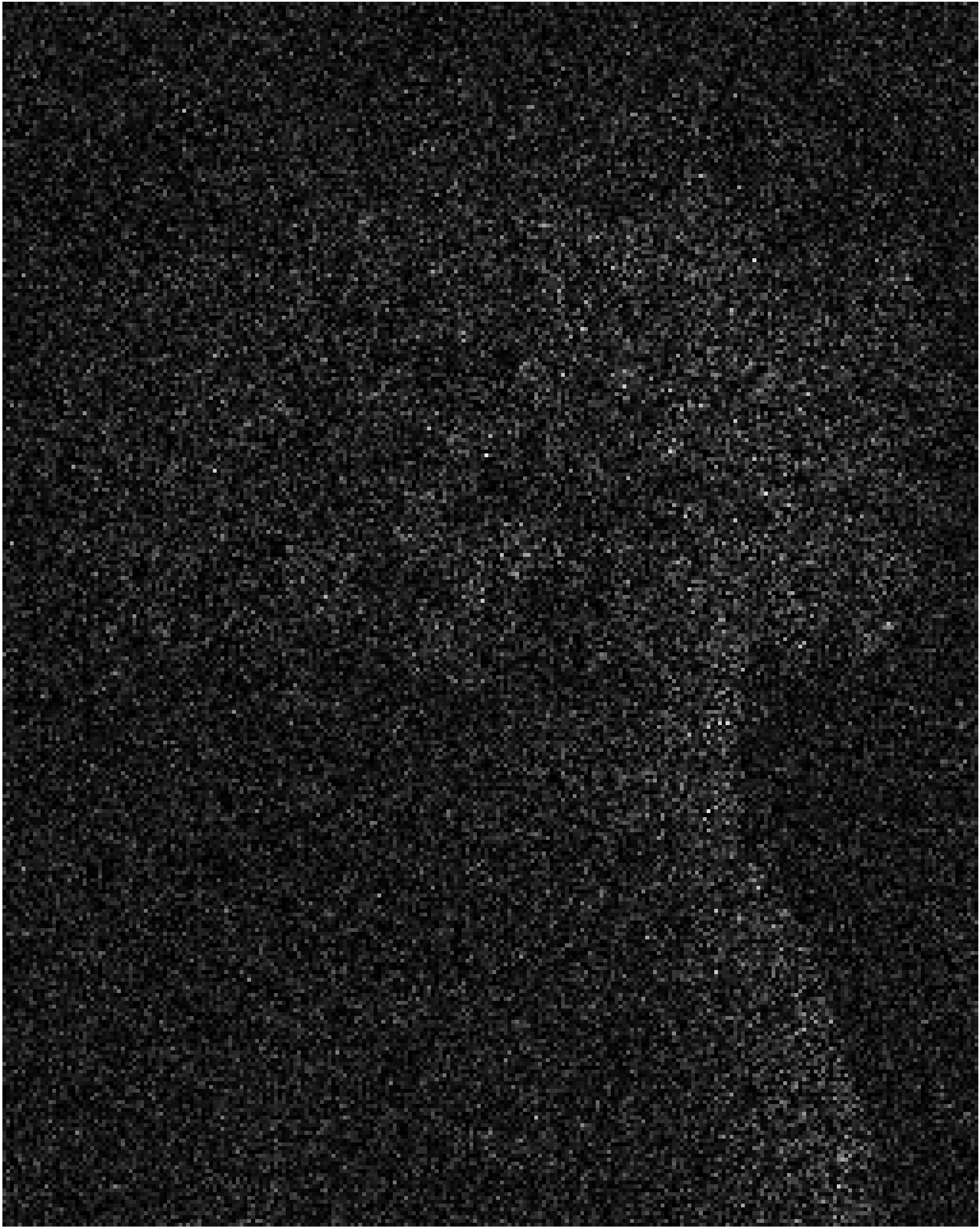}
        \caption*{$\mathrm{MSE} = 0.206$}
    \end{subfigure} & 
    \begin{subfigure}[t]{\restcolwidth}
        \includegraphics[height=1.01\figHeight]{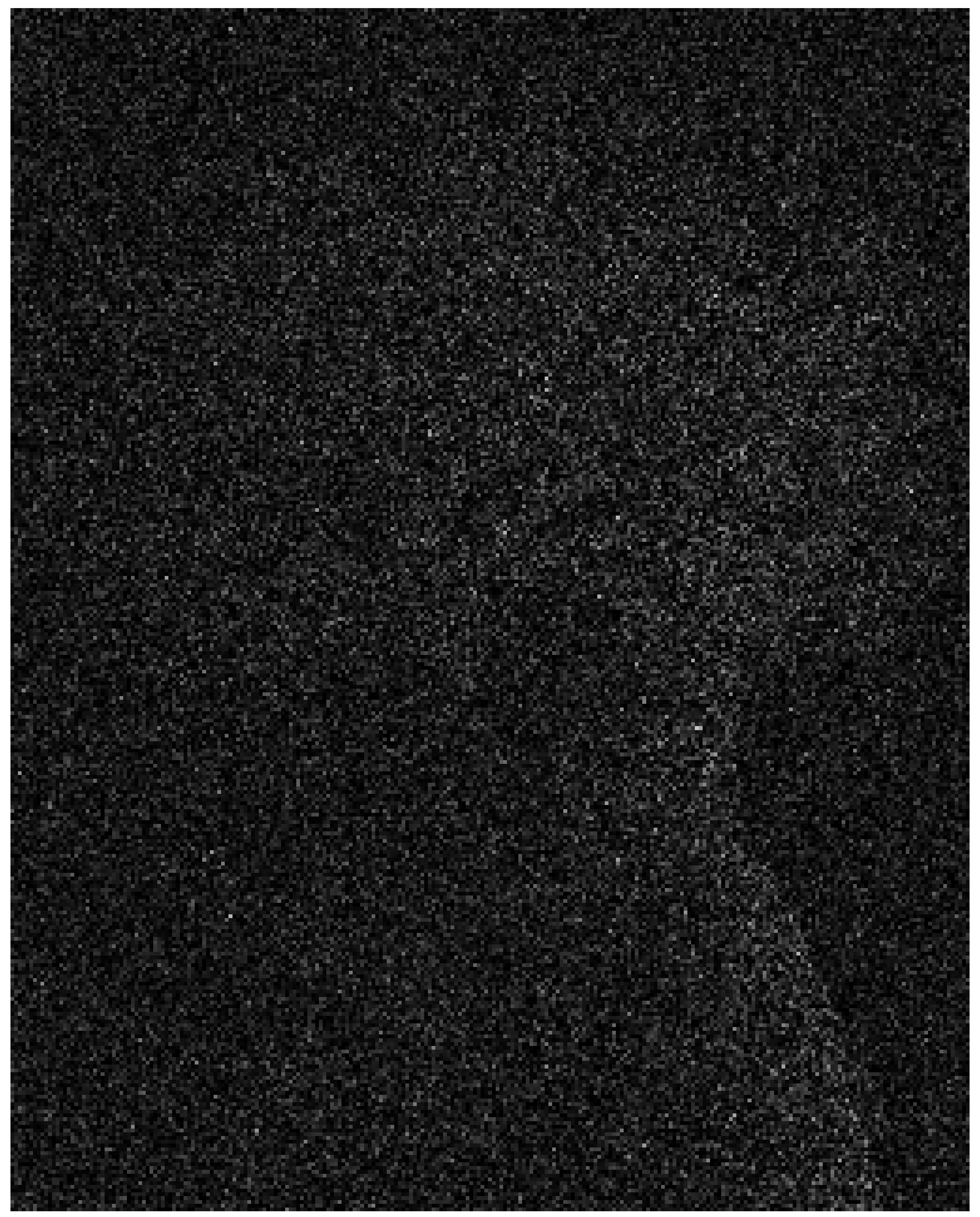}
        \caption*{$\mathrm{MSE} = 0.181$}
    \end{subfigure} \\
    {\small \bf Absolute error of} \\[-1mm]
    {\small \bf oracle estimator} \\
    \begin{subfigure}[t]{\firstcolwidth}
        \includegraphics[height=\figHeight]{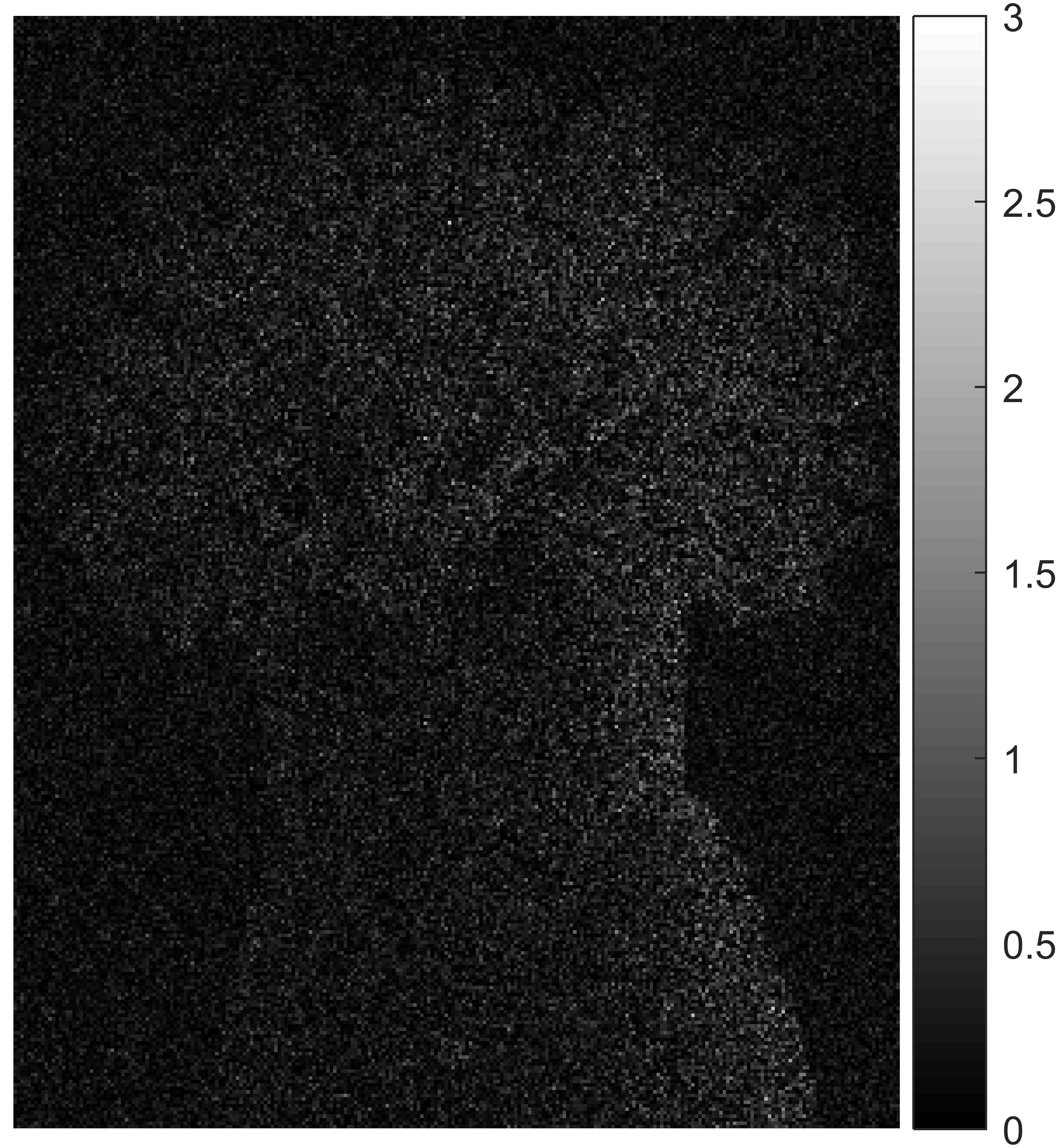}
        \caption*{$\mathrm{MSE} = 0.135$}
    \end{subfigure} & 
    \rotatebox[origin=l]{90}{\qquad \qquad \small ML}
    &
    \begin{subfigure}[t]{\restcolwidth}
        \includegraphics[height=\figHeight]{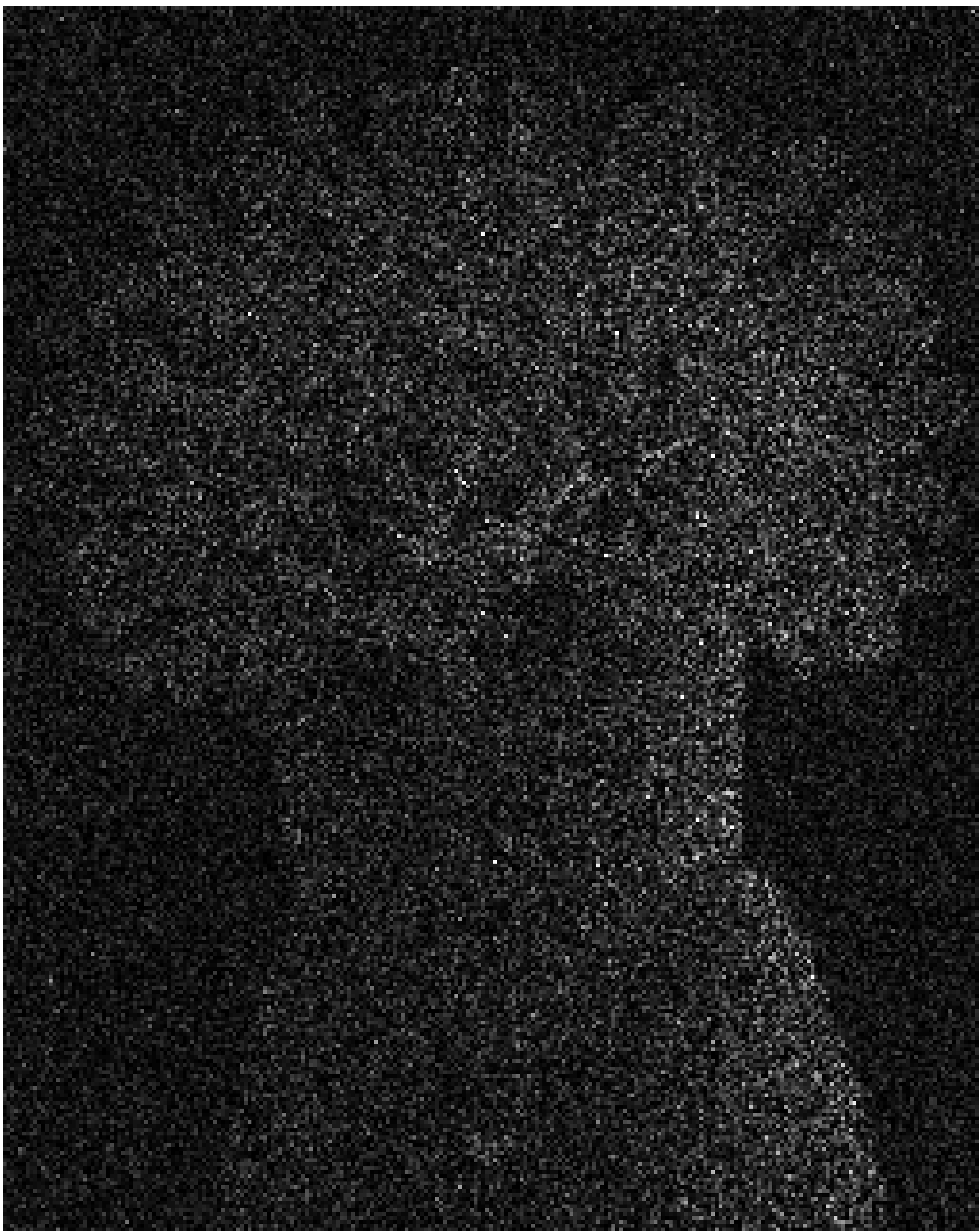}
        \caption*{$\mathrm{MSE} = 0.246$}
    \end{subfigure} & 
    \begin{subfigure}[t]{\restcolwidth}
        \includegraphics[height=\figHeight]{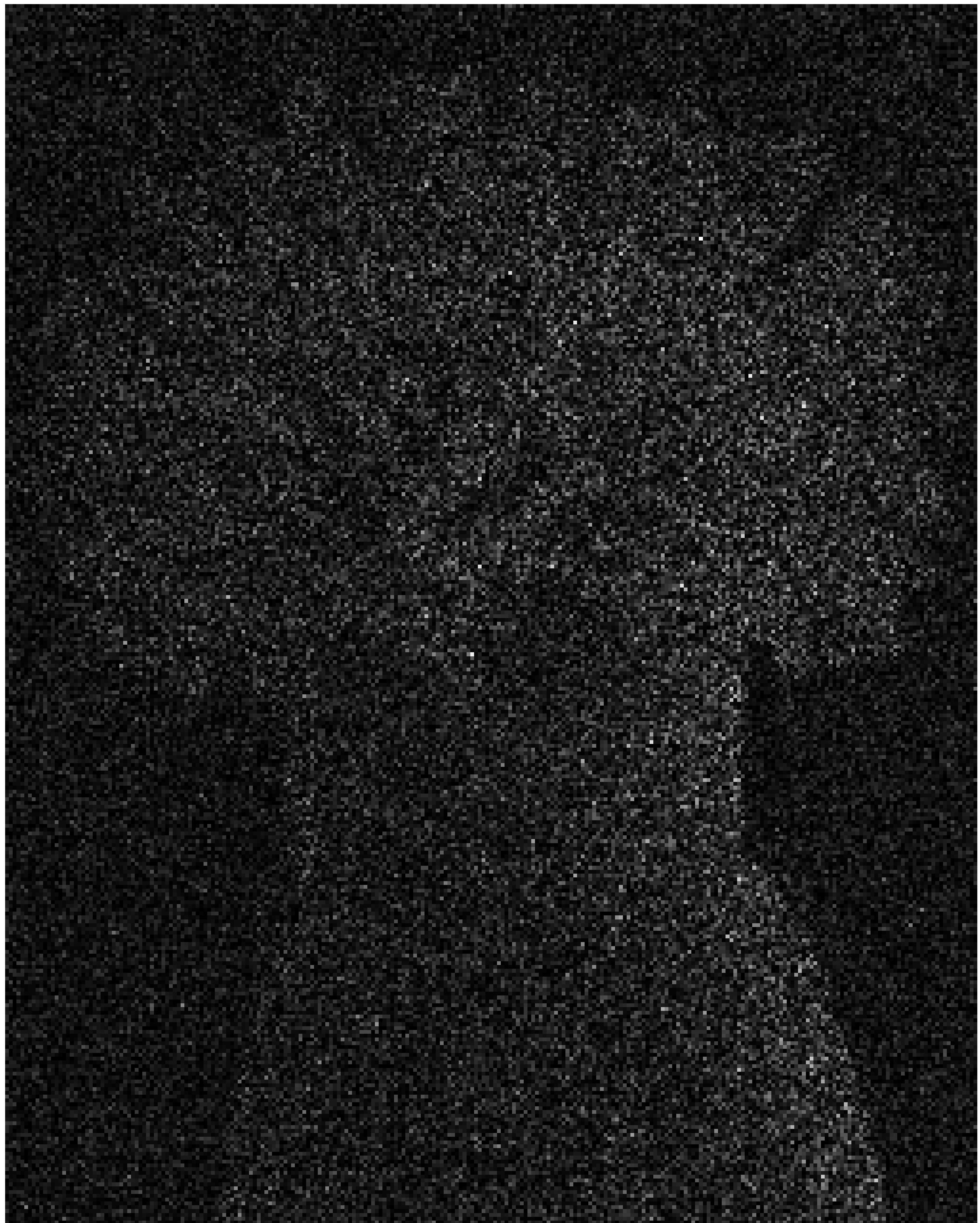}
        \caption*{$\mathrm{MSE} = 0.211$}
    \end{subfigure} & 
    \begin{subfigure}[t]{\restcolwidth}
        \includegraphics[height=\figHeight]{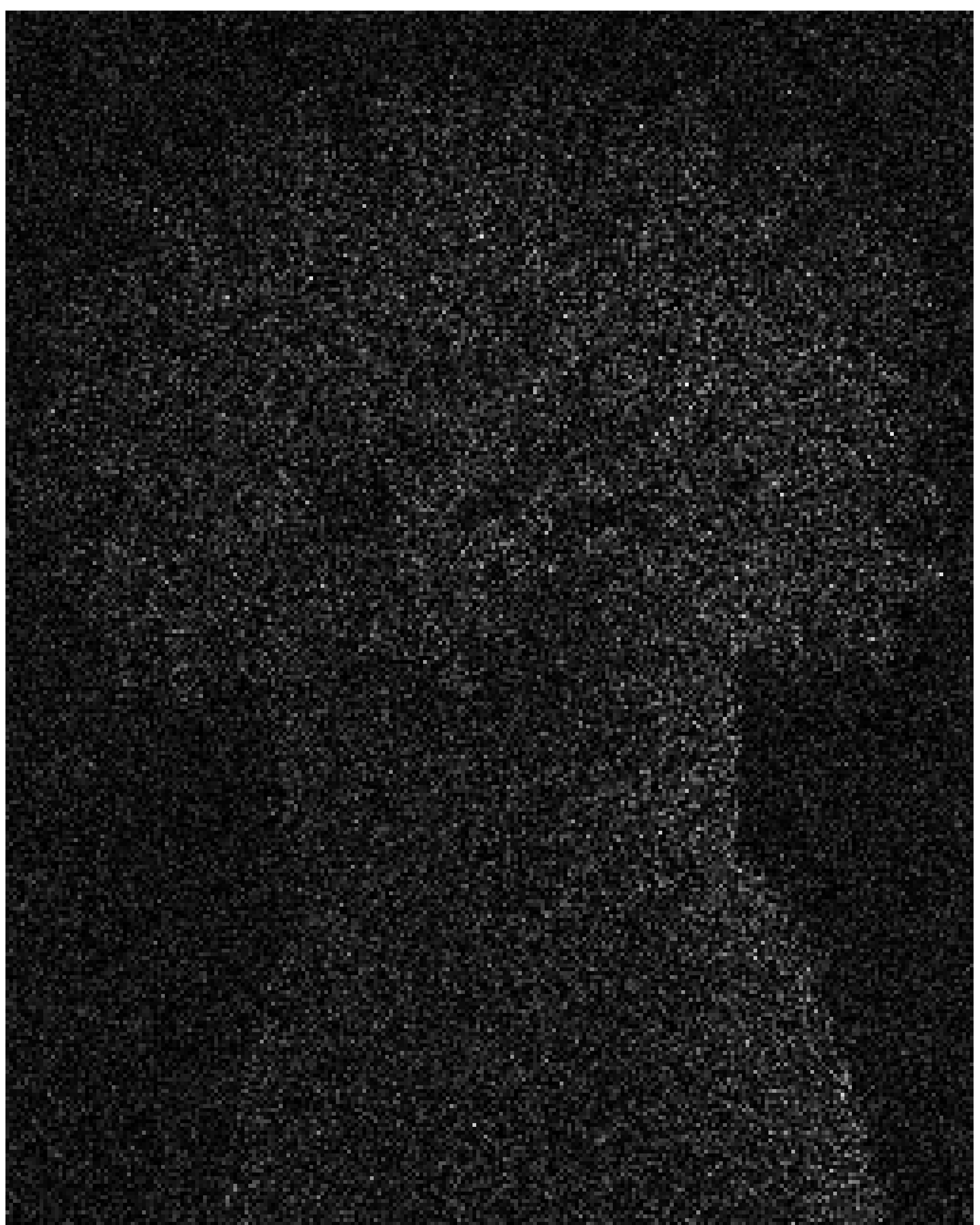}
        \caption*{$\mathrm{MSE} = 0.191$}
    \end{subfigure} & 
    \begin{subfigure}[t]{\restcolwidth}
        \includegraphics[height=\figHeight]{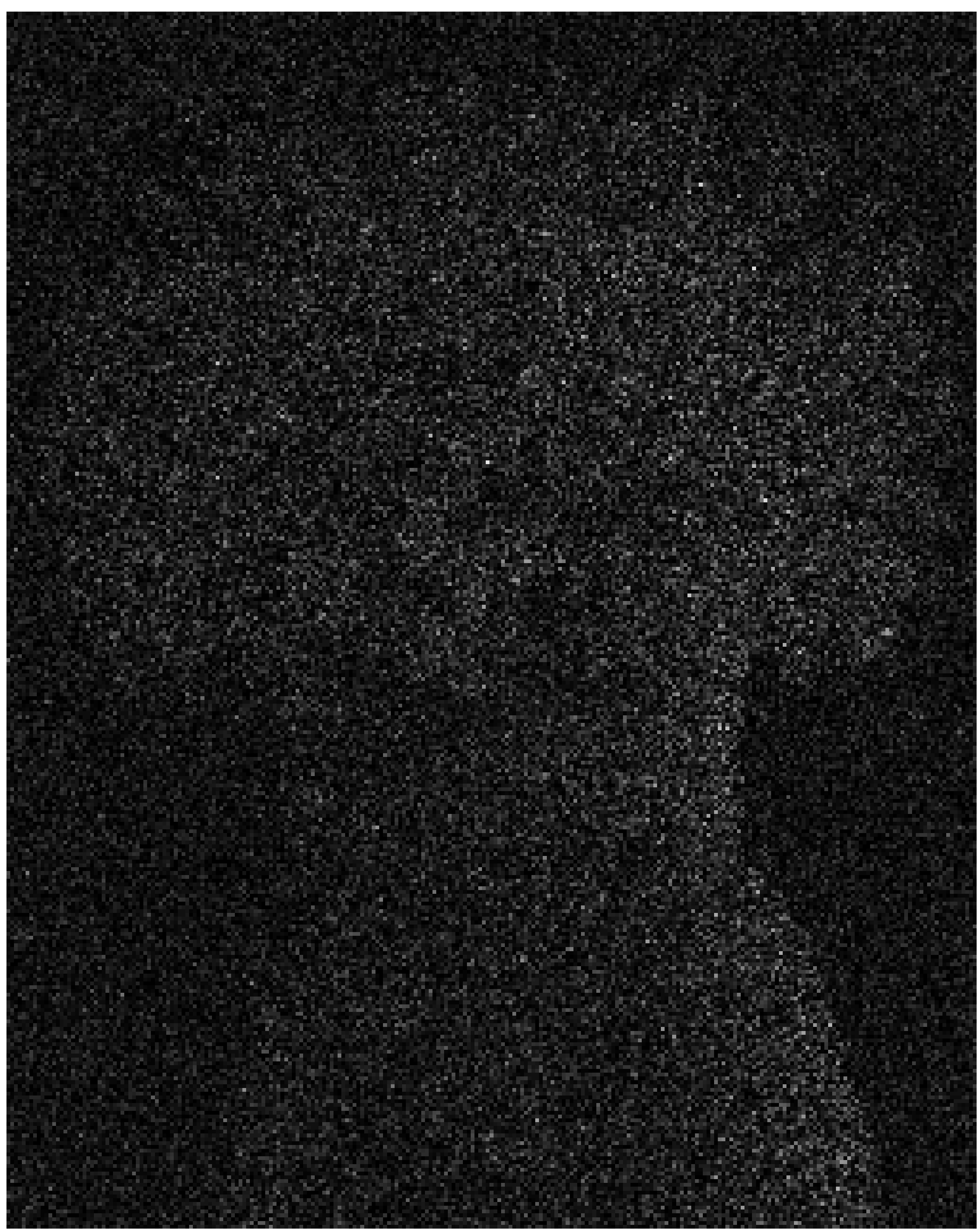}
        \caption*{$\mathrm{MSE} = 0.178$}
    \end{subfigure} & 
    \begin{subfigure}[t]{\restcolwidth}
        \includegraphics[height=\figHeight]{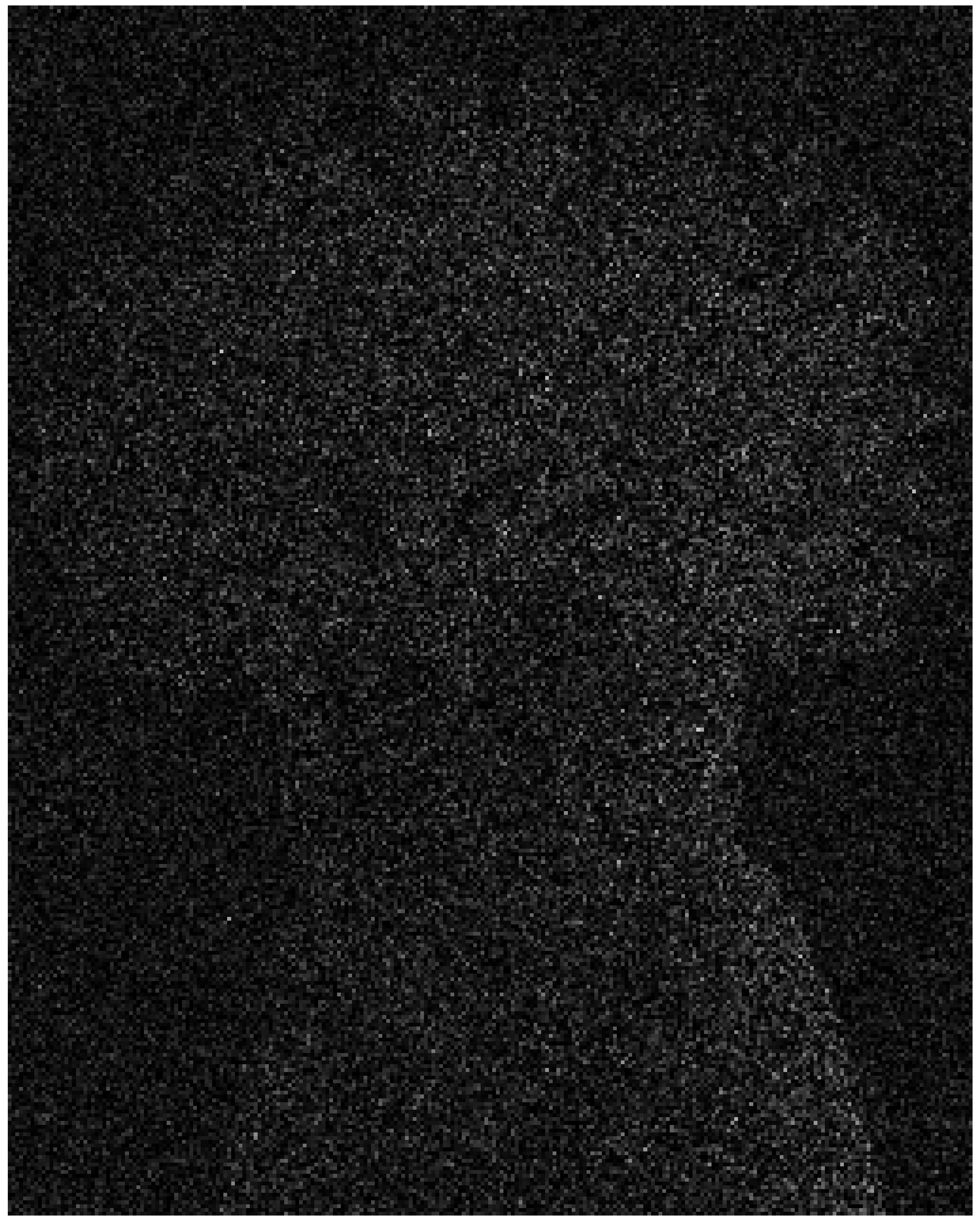}
        \caption*{$\mathrm{MSE} = 0.166$}
    \end{subfigure} 
  \end{tabular}
  \caption{Simulated FIB microscopy experiment with time-resolved estimators 
    under discrete- and continuous-time settings. 
    Aside from the ground truth in the upper-left corner,
    with $\eta \in [1,8]$,
    all images are of the absolute value of the error.
    All results are for total dose $\lambda = 20$ mean incident ions per pixel.
    The columns of time-resolved estimators are for increasing numbers of subacquisitions $n$, culminating in the limiting continuous-time case.
    Quotient mode, Lambert quotient mode, maximum likelihood estimators are compared,
    with the conventional estimator \eqref{eq:eta-baseline}
    and oracle estimator \eqref{eq:oracle-estimator} provided for context.
    None of these estimators include spatial regularization.
    }
  \label{fig:mushroom}
\end{figure*}

\section{Simulated Microscopy Results}
\label{sec:simulated-microscopy}
Figures~\ref{fig:CTTR_comparison} and
\ref{fig:PP_discrete_MSE_comparison}
show that time-resolved estimators improve upon the conventional
processing of abstracted FIB microscope data,
and Fig.~\ref{fig:QM_MSE_vs_n} shows that the performances of DTTR estimators improve with increasing numbers of subacquisitions, converging to the performances of corresponding CTTR estimators.
We conclude with visual results to demonstrate these properties
in simulated FIB microscopy experiments.

\begin{figure}
  \begin{center}
    \includegraphics[width=0.8\linewidth]{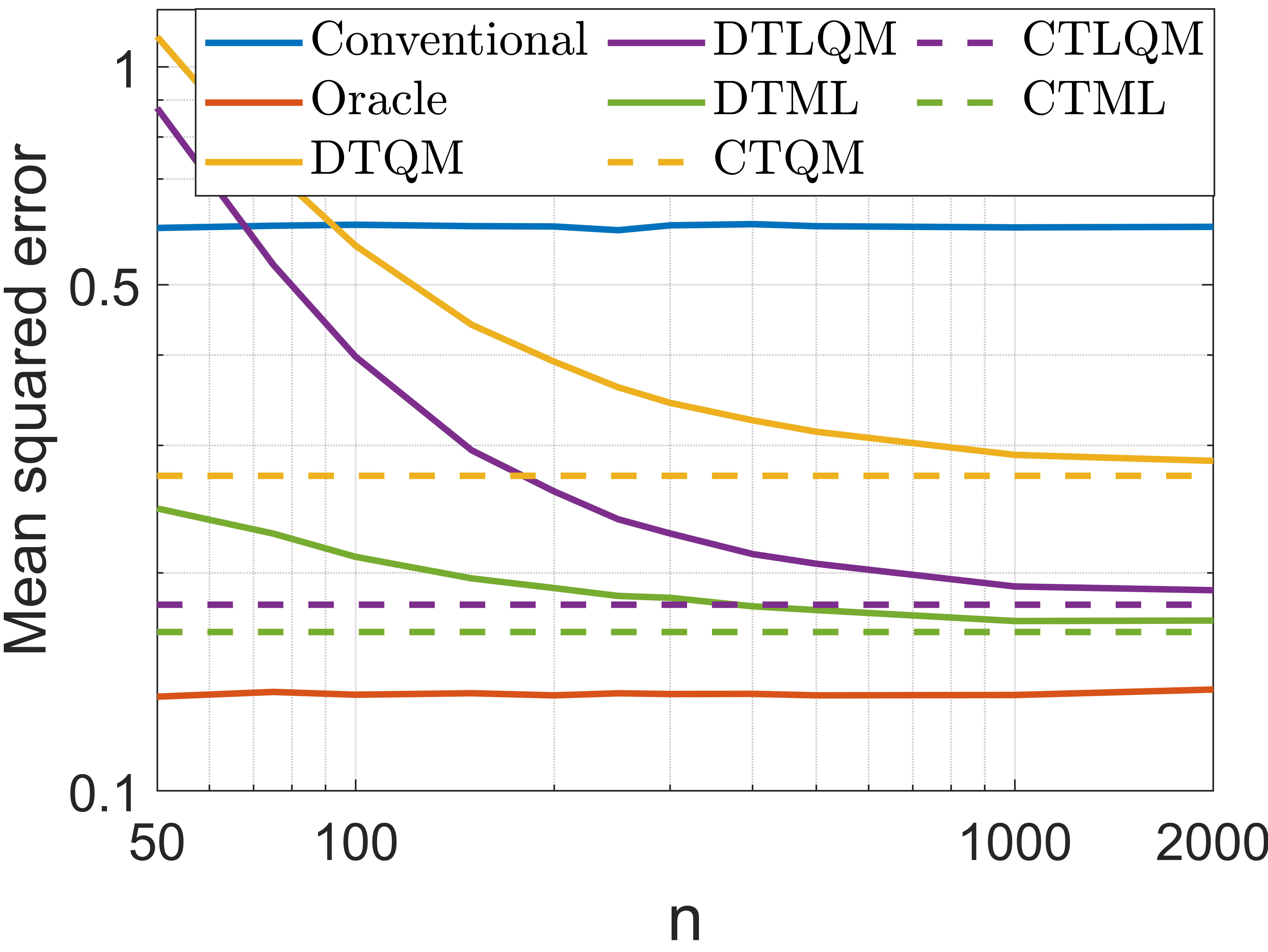}
  \end{center}
  \caption{Mean-squared errors
  as functions of number of subacquisitions $n$
  in simulated FIB microscopy of the image in Fig.~\ref{fig:mushroom} using
  conventional, oracle,
  quotient mode, Lambert quotient mode, and maximum-likelihood estimators.
  Total dose is $\lambda = 20$, and 
  mean secondary yield $\eta$ of the ground truth image is scaled to $[1, 8]$.}
  \label{fig:hairstyle_mse_vs_n}
\end{figure}

We use the \emph{Hairstyle}\footnote{A SEM image of the upper part of the style and stigma from an Arabidopsis flower,
\url{https://www.flickr.com/photos/fei_company/9316514268/in/set-72157634429801580/}}
image from ThermoFisher Scientific
(upper-left of Fig~\ref{fig:mushroom})
as the ground truth image, scaled to have SE yield $\eta \in [1,\,8]$.
All experiments use total dose $\lambda = 20$.
Fig.~\ref{fig:mushroom} displays absolute error images for
quotient mode, Lambert quotient mode and maximum likelihood estimators.
CTTR measurements are simulated and,
by division of the dwell time,
also interpreted as DTTR measurements for
$n \in \{50,\, 100,\, 200,\, 500\}$ subacquisitions.
Fig.~\ref{fig:hairstyle_mse_vs_n}
complements Fig.~\ref{fig:mushroom}
by plotting MSEs as functions of $n$ for these and additional values of $n$.

In Fig.~\ref{fig:mushroom},
each estimator shows improving performance as $n$ increases,
with $n=500$ coming close to the CT performance.
With $n$ increased to 2000 in Fig.~\ref{fig:hairstyle_mse_vs_n},
convergence to CT performance is more clearly indicated.
For each value of $n$, the ranking of estimators has
ML best, LQM second, and QM worst.
For $n \geq 100$, all the TR methods perform better than the conventional estimator.
One way to summarize is to see that with a factor of 4.4 separating the MSEs of the conventional and oracle estimators,
the CTML estimator achieves a 3.6 times lower MSE than the conventional estimator.

\section{Conclusion}
\label{sec:conclusion}
In this work, we establish an abstract framework for TR measurement in FIB microscopy with direct electron detection.
Through estimation-theoretic analyses, analyses of estimators, and Monte Carlo FIB imaging simulations,
we show the extent to which source shot noise can be mitigated by TR measurement methods.
The most easily interpreted conclusion comes from the Fisher information of continuous-time TR measurements $\lambda(1/\eta-e^{-\eta})$ in \eqref{eq:FI_CT};
when $\eta$ is not too small, this is only slightly smaller than the $\lambda/\eta$, which is equal to the FI that would be obtained with a deterministic incident particle beam.
The dependence on mean SE yield $\eta$ shows that TR methods have greater potential in FIB microscopy than in SEM\@.

Continuous-time measurement is not implementable in any foreseeable technology.
Instead, it is intended as a greatly simpler way to understand the limits of performance under very fine time resolution than to consider $n \rightarrow \infty$ limits for discrete-time results.
Through performance comparisons such as those in Figs.~\ref{fig:QM_MSE_vs_n}, \ref{fig:mushroom}, and \ref{fig:hairstyle_mse_vs_n},
one can predict the time resolution that is necessary to approach the CT limit within a desired margin.
Necessary time resolution can also be understood through the use of Fig.~\ref{fig:normalized_Fisher_info} to choose a sufficiently small value for $\lambda/n$.
For example, for the illustrated value of $\eta=3$, the normalized Fisher information plot suggests that when the time resolution is fine enough for $\lambda/n < 0.1$,
at least 83\% of the improvement created by time-resolved measurement will be attained.

We study three types of unconventional estimators for both continuous- and discrete-time TR measurements.
QM estimators are the simplest to implement and are similar to estimators proposed by Zeiss but not made commercially available.
LQM estimators greatly reduce a source of bias at low $\eta$ and merely require a table lookup to be applied to QM estimates.
ML estimators require root-finding or minimization of a non-convex function.
The relationships among the estimators are nontrivial:
though generally best, the ML estimator does not outperform the others uniformly over $\eta$;
and though far better at low $\eta$, the LQM estimator does not outperform the QM estimator uniformly over $\eta$.

The improvements presented here seem to be rooted entirely in making the number of incident ions estimable,
and this is potentially applicable even without direct SE detection.
Indirect electron detection creates uncertainty in the number of detected SEs, including uncertainty in whether any SEs were detected
and hence in whether an ion was incident.
However, the experimental results of~\cite{peng2020source} suggest that 
the degradation
in mitigating source shot noise
can be small,
since the improvements presented therein are similar to the results presented here.
The QM and LQM estimators may be extended to cases in which the probabilistic model relating
measurements to numbers of SEs is complicated or uncertain.  For an advantage from TR measurement,
it may be enough to have a mean instrument output that is monotonically increasing with the number of detected SEs.
Then a QM or LQM estimator can use $\Mtilde$ or $(1-e^{-\eta})^{-1} \Mtilde$
to normalize a sum of nonlinearly scaled TR measurements to mitigate source shot noise.
This is one of several lines of inquiry suggested by the results presented here.

\appendices
\section{High-Dose Performance of Oracle Estimator}
\label{app:oracle}
Let
\begin{equation}
    \label{eq:g-def}
  g(x) = e^{-x} \sum_{m=1}^\infty \frac{1}{m} \frac{x^m}{m!}.
\end{equation}
This function appears in the performance of the oracle estimator
\eqref{eq:MSE-oracle-exact}
and the variance
\eqref{eq:eta_QM_cont_total_var}
and MSE
\eqref{eq:eta_QM_cont_MSE}
of the CTQM estimator.
We are interested in approximating and bounding
$g(x)$ to better understand those expressions.

\begin{figure}
  \begin{center}
    \includegraphics[width=0.8\linewidth]{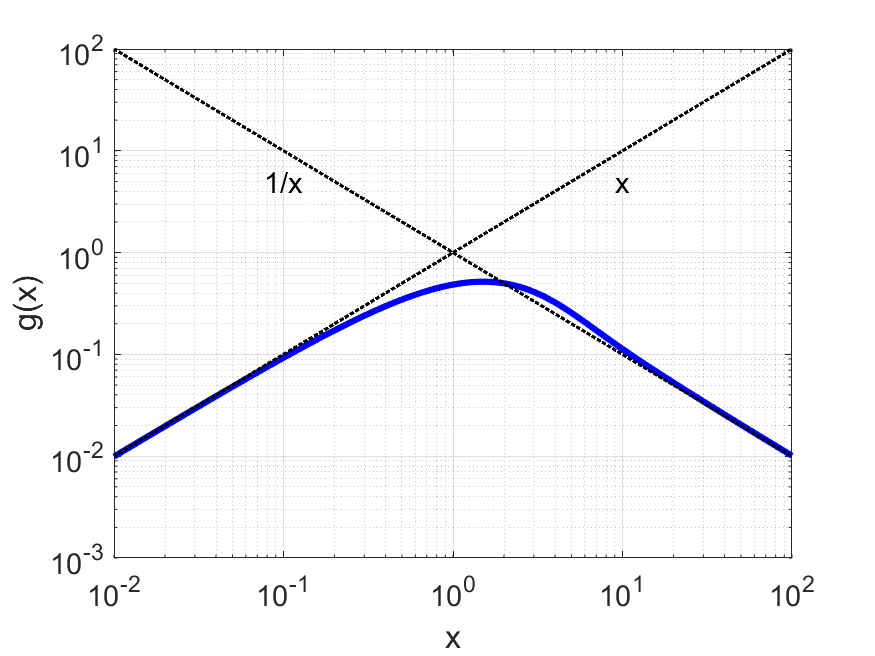}
  \end{center}
  \caption{Function $g(x)$ in \eqref{eq:g-def} along with low- and high-$x$ asymptotes.}
  \label{fig:g}
\end{figure}

For $x \ll 1$, the first term is dominant,
so $g(x) \approx x$.
The behavior at large $x$ is less obvious.
The series in \eqref{eq:g-def} converges to
$\mathrm{Ei}(x) -\gamma - \log x$,
where
$\mathrm{Ei}(x)$ is the exponential integral function and
$\gamma$ is the Euler--Mascheroni constant~\cite{Masina2019}.
Asympototically for $x \rightarrow \infty$,
$\mathrm{Ei}(x) \sim \Frac{e^x}{x}$,
meaning that the ratio of the expressions approaches 1\@.
Thus,
\begin{equation}
    \label{eq:g-asymptotic}
    g(x) \sim \frac{1}{x}.
\end{equation}
The log-log plot of $g(x)$ in Fig.~\ref{fig:g} 
shows the accuracies of the low- and high-$x$ approximations.
Being nonnegative, continuous, and vanishing as $x \rightarrow \infty$,
$g(x)$ has a finite upper bound:
\begin{equation}
    \label{eq:g-bound}
    g(x) \leq 0.518,
    \qquad
    \mbox{for all $x \in [0,\infty)$.}
\end{equation}

\section{Normalized Fisher Information Limits}
\label{app:FI_norm}
\subsection{Low-Dose Limit}
To evaluate
$\lim_{\lambda \rightarrow 0} \Frac{\mathcal{I}_Y(\eta \sMid \lambda)}{\lambda}$,
we first find $\lambda \rightarrow 0$ limits of expressions that appear in
\eqref{equ:Fisher_info_equation_PP}, including both the
PMF in \eqref{equ:neyman} and the probability ratio
$\Frac{\mathrm{P}_Y(y+1 \sMid \eta, \lambda )}
      {\mathrm{P}_Y(y \sMid \eta, \lambda )}$.

For $y = 0$,
\begin{align}
\mathrm{P}_Y(0 \sMid \eta, \lambda)
&= \frac{e^{-\lambda} \eta^0}{0!} \sum_{m = 0}^{\infty} \frac{(\lambda e^{-\eta})^m m^0}{m!} \nonumber \\
&\eqlabel{a} e^{-\lambda} \sum_{m = 0}^{\infty} \frac{(\lambda e^{-\eta})^m}{m!} 
 \eqlabel{b} e^{-\lambda}\exp(\lambda e^{-\eta}),
 \label{eq:P_Y_0}
\end{align}
where (\emph{a}) follows from $m^0 =1$; and
(\emph{b}) from identifying the series expansion of the exponential function.
Similarly, for $y = 1$,
\begin{align}
\mathrm{P}_Y(1 \sMid \eta, \lambda)
&= \frac{e^{-\lambda} \eta^1}{1!} \sum_{m = 0}^{\infty} \frac{(\lambda e^{-\eta})^m m^1}{m!} \nonumber \\
&= (e^{-\lambda} \eta)(\lambda e^{-\eta})\exp(\lambda e^{-\eta}),
 \label{eq:P_Y_1}
\end{align}
and for $y = 2$,
\begin{align}
\mathrm{P}_Y(2 \sMid \eta, \lambda)
&= \frac{e^{-\lambda} \eta^2}{2!} \sum_{m = 0}^{\infty} \frac{(\lambda e^{-\eta})^m m^2}{m!} \nonumber \\
&= \frac{e^{-\lambda} \eta^2}{2}
(\lambda e^{-\eta})
(1 + \lambda e^{-\eta})
\exp(\lambda e^{-\eta}).
 \label{eq:P_Y_2}
\end{align}
For general $y > 0$,
\begin{align}
\mathrm{P}_Y(y \sMid \eta, \lambda)
&= \frac{e^{-\lambda} \eta^y}{y!} \sum_{m = 0}^{\infty} \frac{(\lambda e^{-\eta})^m m^y}{m!} \nonumber \\
&= \frac{e^{-\lambda} \eta^y}{y!}
(\lambda e^{-\eta})\,
\mathrm{poly}_{y-1}(\lambda e^{-\eta})
\exp(\lambda e^{-\eta}),
 \label{eq:P_Y_y}
\end{align}
where $\mathrm{poly}_y(\lambda e^{-\eta})$
is a degree-$y$ polynomial in $\lambda e^{-\eta}$
with unit constant term.
This allows us to conclude, for any $y > 0$,
\begin{equation}
\label{eq:lim_P_y_normalized}
  \lim_{\lambda \rightarrow 0} \frac{\mathrm{P}_Y(y \sMid \eta, \lambda)}{\lambda}
   = \frac{\eta^y}{y!} e^{-\eta}.
\end{equation}

From \eqref{eq:P_Y_0} and \eqref{eq:P_Y_1}, we obtain, for $y = 0$,
\begin{equation}
\label{eq:P_ratio_0}
\frac{\mathrm{P}_Y(y + 1 \sMid \eta, \lambda)}{\mathrm{P}_Y(y \sMid \eta, \lambda)}
= \frac{\mathrm{P}_Y(1 \sMid \eta, \lambda)}{\mathrm{P}_Y(0 \sMid \eta, \lambda)}
= \eta \lambda e^{-\eta}.
\end{equation}
From \eqref{eq:P_Y_1} and \eqref{eq:P_Y_2}, we obtain, for $y = 1$,
\begin{equation}
\label{eq:P_ratio_1}
\frac{\mathrm{P}_Y(y + 1 \sMid \eta, \lambda)}{\mathrm{P}_Y(y \sMid \eta, \lambda)}
= \frac{\mathrm{P}_Y(2 \sMid \eta, \lambda)}{\mathrm{P}_Y(1 \sMid \eta, \lambda)}
= \frac{1}{2}\eta (1 + \lambda e^{-\eta}).
\end{equation}
For general $y > 0$, it follows from \eqref{eq:lim_P_y_normalized} that
\begin{equation}
\label{eq:lim_P_ratio}
\lim_{\lambda \rightarrow 0}
  \frac{\mathrm{P}_Y(y + 1 \sMid \eta, \lambda)}{\mathrm{P}_Y(y \sMid \eta, \lambda)}
  = \frac{\eta}{y + 1}.
\end{equation}

Now to evaluate
$\lim_{\lambda \rightarrow 0} \Frac{\mathcal{I}_Y(\eta \sMid \lambda)}{\lambda}$,
we can pass the limit through to each term in
\eqref{equ:Fisher_info_equation_PP}.
The first term is
\begin{align*}
&\lim_{\lambda \rightarrow 0}
   \left(\frac{0}{\eta} - \frac{\mathrm{P}_Y(1 \sMid \eta,\lambda)}{\mathrm{P}_Y(0 \sMid \eta,\lambda)} \frac{1}{\eta} \right)^2
   \frac{\mathrm{P}_Y(0 \sMid \eta,\lambda)}{\lambda} \\
 &\quad \eqlabel{a}  \lim_{\lambda \rightarrow 0}
   \left(\frac{0}{\eta} -
   \eta \lambda e^{-\eta}
   \frac{1}{\eta} \right)^2
   \frac{e^{-\lambda}\exp(\lambda e^{-\eta})}{\lambda} 
  = 0,
\end{align*}
where (\emph{a}) follows from \eqref{eq:P_Y_0} and \eqref{eq:P_ratio_0}.
By substituting \eqref{eq:lim_P_y_normalized} and \eqref{eq:lim_P_ratio}
in \eqref{equ:Fisher_info_equation_PP},
the remaining terms give
\begin{align*}
&\lim_{\lambda \rightarrow 0}
 \frac{\mathcal{I}_Y(\eta \sMid \lambda)}{\lambda}
  = \sum_{y=1}^\infty
        \left(\frac{y}{\eta}
              - \frac{\eta}{y+1} \frac{y+1}{\eta} \right)^2
              \frac{\eta^y}{y!}e^{-\eta} \\
 &\quad = \sum_{y=1}^\infty
        \left(\frac{y}{\eta}
              - 1 \right)^2
              \frac{\eta^y}{y!}e^{-\eta} 
  = \left(\frac{e^{\eta}}{\eta} - 1\right) e^{-\eta}
  = \frac{1}{\eta} - e^{-\eta} .
\end{align*}
This proves \eqref{eq:NFI-low-lambda}, as desired.

\subsection{High-Dose Limit}
Let us first compute the Fisher information for
the parameter $\eta$ when a Gaussian random variable
has mean $\eta$ and variance $f(\eta)$ for some
twice-differentiable function $f$.
Let $S \sim \mathcal{N}(\eta,\,f(\eta))$.
Then the log-likelihood of $S$ is
\begin{equation}
\log f_S(s \sMid \eta)
  = - \frac{1}{2}\log (2\pi)
    - \frac{1}{2}\log f(\eta)
    - \frac{(s-\eta)^2}{2f(\eta)}.    
\end{equation}
The derivative of $\log f_S(s \sMid \eta)$ with respect to $\eta$ is
\begin{align*}
\frac{\partial \log f_S(s \sMid \eta)}{\partial \eta}
&= - \frac{f'(\eta)}{2f(\eta)}
   - \frac{2(\eta - s)f(\eta) - (\eta-s)^2 f'(\eta)}{2f(\eta)^2}.
\end{align*}
The second derivative is then
\begin{align*}
\frac{\partial^2 \log f_S(s \sMid \eta)}{\partial \eta^2}
&= - \frac{f''(\eta)f(\eta) - f'(\eta)^2}{2f(\eta)^2}
   - \frac{1}{f(\eta)}\\
& \quad + \frac{2f'(\eta)}{f(\eta)^2}(\eta - s)\\
&  \quad - \frac{ 2[f'(\eta)]^2 - f''(\eta)f(\eta)}{2f(\eta)^3} (\eta-s)^2.
\end{align*}    
The Fisher information for the estimation of $\eta$ is
\begin{align}
\mathcal{I}_S(\eta)
&= \E{ -\frac{\partial^2 \log f_S(S \sMid \eta)}{\partial \eta^2} \sMid \eta } \nonumber\\
&= \frac{f''(\eta)f(\eta) - f'(\eta)^2}{2f(\eta)^2}
   + \frac{1}{f(\eta)}
   - \frac{2f'(\eta)}{f(\eta)^2}\E{\eta - S} \nonumber\\
&  \quad + \frac{2[f'(\eta)]^2 - f''(\eta)f(\eta)}{2f(\eta)^3} \E{(\eta - S)^2} \nonumber\\
&\eqlabel{a} \frac{f''(\eta)f(\eta) - f'(\eta)^2}{2f(\eta)^2}
   + \frac{1}{f(\eta)}
   - \frac{2f'(\eta)}{f(\eta)^2}\cdot 0 \nonumber\\
&  \quad + \frac{ 2[f'(\eta)]^2 - f''(\eta)f(\eta)}{2f(\eta)^3} \cdot f(n) \nonumber\\
&= \frac{1}{f(\eta)}
   + \frac{[f'(\eta)]^2}{2f(\eta)^2},
\label{eq:FI_S}
\end{align}
where (\emph{a}) follows from substituting
$\E{\eta-S} = 0$ and $\iE{(\eta-S)^2} = \var{S} = f(\eta)$.
(Note that this simplifies to the familiar
reciprocal of the variance when $f(\eta)$ is a constant.)

At high dose, $Y/\lambda$ is well-approximated as a
$\mathcal{N}(\eta,\,\eta(\eta+1)/\lambda)$ random variable~\cite[Sect.~IV]{Teich:81}.
Thus, define
$f(\eta) = \Frac{\eta(\eta+1)}{\lambda}$
so that $Y/\lambda$ is approximated well by $S$.
Substituting
$f'(\eta) = \Frac{(2\eta+1)}{\lambda}$
into \eqref{eq:FI_S} gives
\begin{align}
\mathcal{I}_S(\eta) 
&= \frac{\lambda}{\eta(\eta+1)}
   + \frac{(2\eta+1)^2}{2\eta^2(\eta+1)^2}.
\label{equ:Fisher_info_PP_large_lambda}
\end{align}
Since $Y \approx \lambda S$,
\begin{align}
\lim_{\lambda \rightarrow \infty}
\frac{\mathcal{I}_Y( \eta \sMid \lambda)}{\lambda}
&= \lim_{\lambda \rightarrow \infty}
    \left[ \frac{1}{\eta(\eta+1)}
    + \frac{(2\eta+1)^2}{2 \lambda \eta^2(\eta+1)^2} \right]\nonumber\\
&= \frac{1}{\eta(\eta+1)},\nonumber
\end{align}
as desired.

\section{Derivation of Mean-Squared Error for Continuous-Time Quotient Mode Estimator}
\label{app:mse_CTQM_deri}

\subsection{Bias of $\etaCTQM$}
For $m > 0$,
\begin{align}
    &\E{ \etaCTQM \smid \Mtilde = m }
     = \E{ \frac{Y}{\Mtilde} \smid \Mtilde = m} \nonumber \\
    & \quad = \frac{1}{m} \E{ Y \smid \Mtilde = m} 
     \eqlabel{a} \frac{1}{m} \E{ \sum_{i=1}^m \Xtilde_i } 
     = \E{ \Xtilde_i } \nonumber \\
    & \quad \eqlabel{b} \frac{\eta}{1-e^{-\eta}},
  \label{eq:E_etaCTQM_m}
\end{align}
where (\emph{a}) follows from using \eqref{eq:Y-def-Xtilde} as an expression for $Y$;
and (\emph{b}) from the effect of zero-truncation on the $\Poisson(\eta)$ distribution.
Trivially,
\begin{align}
  \label{eq:E_etaCTQM_0}
    \E{ \etaCTQM \smid \Mtilde = 0 } = 0.
\end{align}
Using $\iP{ \Mtilde > 0 } = 1 - e^{-\lambda(1-e^{-\eta})}$
from \eqref{eq:Mtilde-PMF}
and the total expectation theorem to combine
\eqref{eq:E_etaCTQM_m} and \eqref{eq:E_etaCTQM_0} gives
\begin{equation}
  \label{eq:E_etaCTQM}
    \E{ \etaCTQM }
       =  \left( 1 - e^{-\lambda(1-e^{-\eta})} \right)
                 \frac{\eta}{1-e^{-\eta}}.
\end{equation}
Subtracting $\eta$ gives \eqref{eq:eta_QM_cont_bias}.

\subsection{Variance of $\etaCTQM$}

From \eqref{eq:E_etaCTQM_m} and \eqref{eq:E_etaCTQM_0},
$\iE{ \etaCTQM \smid \Mtilde}$ is a two-valued random variable
equal to $\eta/(1-e^{-\eta})$ with probability $1 - e^{\lambda(1-e^{-\eta})}$
and equal to 0 otherwise.  Thus,
\begin{align}
    &\var{ \E{ \etaCTQM \smid \Mtilde } } \nonumber \\
    & \ = \frac{\eta^2}{(1-e^{-\eta})^2}
                     e^{-\lambda(1-e^{-\eta})} 
          \left( 1 - e^{-\lambda(1-e^{-\eta})} \right)
          \label{eq:app_var_cond_mean_CTQM}
\end{align}
by direct calculation.

Toward computing
$\iE{\ivar{ \etaCTQM \smid \Mtilde}}$,
let us first examine
$\ivar{ \etaCTQM \smid \Mtilde = m}$.
For $m > 0$,
\begin{align}
    &\var{ \etaCTQM \smid \Mtilde = m }
    = \var{ \frac{Y}{\Mtilde} \smid \Mtilde = m} \nonumber \\
    & \quad = \frac{1}{m^2} \var{ Y \smid \Mtilde = m} 
     \eqlabel{a} \frac{1}{m^2} \var{ \sum_{i=1}^m \Xtilde_i } \nonumber \\
    & \quad = \frac{1}{m} \var{ \Xtilde_i } 
     \eqlabel{b} \frac{1}{m}
                 \left( \frac{\eta + \eta^2}{1-e^{-\eta}}
                 -
                        \frac{\eta^2}{(1-e^{-\eta})^2} \right),
  \label{eq:var_etaCTQM_m}
\end{align}
where (\emph{a}) follows from using \eqref{eq:Y-def-Xtilde} as an expression for $Y$;
and (\emph{b}) from the effect of zero-truncation on the $\Poisson(\eta)$ distribution.
Trivially,
\begin{align}
  \label{eq:var_etaCTQM_0}
    \var{ \etaCTQM \smid \Mtilde = 0 } = 0.
\end{align}
Using the PMF of $\Mtilde$ from \eqref{eq:Mtilde-PMF} and the total expectation theorem
to combine \eqref{eq:var_etaCTQM_m} and \eqref{eq:var_etaCTQM_0} gives
\begin{align}
\label{eq:app_mean_cond_var_CTQM}
\E{\var{\etaCTQM \smid \Mtilde}}
&= \left( \frac{\eta + \eta^2}{1-e^{-\eta}} - \frac{\eta^2}{(1-e^{-\eta})^2} \right) \!
       e^{-\lambda(1-e^{-\eta})}
       \nonumber\\
&   \qquad \cdot \sum_{m = 1}^{\infty} \frac{1}{m}\frac{ [\lambda (1-e^{-\eta})]^m}{m!}.
\end{align}

By summing~\eqref{eq:app_var_cond_mean_CTQM} and~\eqref{eq:app_mean_cond_var_CTQM} we obtain the variance of $\etaCTQM$, which verifies~\eqref{eq:eta_QM_cont_total_var}.

\section{Derivation of
Mean-Squared Error for Discrete-Time Quotient Mode Estimator}
\label{app:mse_DTQM_deri}

\subsection{Variance of $\etaDTQM$}
From \eqref{equ:QM_condi_mean} and \eqref{equ:QM_condi_mean0},
$\iE{ \etaDTQM \smid L }$ is a two-valued random variable
equal to $\iE{ \Ytilde_k }$ with probability $1-(1-p)^n$
and equal to 0 otherwise.  Thus,
\begin{align}
    \var{ \E{ \etaDTQM \smid L } } 
    = \left( \E{\Ytilde_k} \right)^2 [1-(1-p)^n](1-p)^n,
          \label{eq:app_var_cond_mean_DTQM}
\end{align}
where $\iE{\Ytilde_k}$ is given in \eqref{eq:ytilde-mean}.

Toward computing
$\iE{\ivar{ \etaDTQM \smid L }}$,
let us first examine
$\ivar{ \etaDTQM \smid L = \ell }$.
For $\ell > 0$,
\begin{align}
    &\var{ \etaDTQM \smid L = \ell }
    = \var{ \frac{Y}{L} \smid L = \ell} 
    = \frac{1}{\ell^2} \var{ Y \smid L = \ell } \nonumber \\
    & \quad = \frac{1}{\ell^2} \var{ \sum_{j=1}^\ell \Ytilde_j } 
     = \frac{1}{\ell} \var{ \Ytilde_k },
  \label{eq:var_etaDTQM_ell}
\end{align}
where $\ivar{\Ytilde_k}$ is given in \eqref{eq:ytilde-var}.
Trivially,
\begin{align}
  \label{eq:var_etaDTQM_0}
    \var{ \etaDTQM \smid L = 0 } = 0.
\end{align}
Using the binomial PMF of $L$
and the total expectation theorem to combine
\eqref{eq:var_etaDTQM_ell} and \eqref{eq:var_etaDTQM_0} gives
\begin{align}
\label{eq:app_mean_cond_var_DTQM}
\E{\var{\etaDTQM \smid L}}
&= \var{\Ytilde_k} \sum_{\ell=1}^n \frac{1}{\ell} \binom{n}{\ell} p^\ell(1-p)^{n-\ell}.
\end{align}
By summing~\eqref{eq:app_var_cond_mean_DTQM} and~\eqref{eq:app_mean_cond_var_DTQM} we obtain the variance of $\etaDTQM$, which verifies~\eqref{eq:DTQM_var}.

\subsection{Variance Lower Bound}
Let $\Ltilde$ be the zero-truncated version of binomial random variable $L$,
with PMF
\[
  \mathrm{P}_{\Ltilde}(\ell)
  = \frac{1}{1-(1-p)^n} \binom{n}{\ell} p^{\ell}(1-p)^{n-\ell},
  \quad \ell = 1,2,\ldots,n.
\]
Then the series in \eqref{eq:DTQM_var} equals
$(1-(1-p)^n) \, \iE{1/\Ltilde}$.
Using Jensen's inequality, we can bound $\iE{1/\Ltilde}$ using $\iE{\Ltilde}$,
which has a simple closed-form expression:
\begin{equation}
\label{eq:mean-Ltilde}
    \E{\Ltilde} = \frac{1}{1-(1-p)^n} \E{L}
                = \frac{np}{1-(1-p)^n}.
\end{equation}
Specifically,
\begin{align}
  &\sum_{\ell = 1}^{n}
    \frac{1}{\ell}\binom{n}{\ell}p^\ell(1 - p)^{n-\ell}
  = (1-(1-p)^n) \, \E{\frac{1}{\Ltilde}} \nonumber \\
  & \quad \geqlabel{a} (1-(1-p)^n) \, \frac{1}{\E{\Ltilde}} 
   \eqlabel{b} \frac{(1-(1-p)^n)^2}{np},
  \label{eq:bound-from-Jensen}
\end{align}
where (\emph{a}) follows from Jensen's inequality; and
(\emph{b}) from \eqref{eq:mean-Ltilde}.
Substituting \eqref{eq:bound-from-Jensen} in
\eqref{eq:DTQM_var} gives \eqref{eq:DTQM_var_lower_bound}.

\section{Derivation of High $n$ Limits}
\label{app:high_n}

\newlength{\appEnegspace}
\setlength{\appEnegspace}{-3mm}

\subsection{Proof of \eqref{eq:lim_np}}

\vspace{\appEnegspace}

\begin{align*}
\lim_{n\to\infty} np
&= \lim_{n\to\infty}
       \frac{1 - e^{-{(\Frac{\lambda}{n})}(1 - e^{-\eta})}}
            {1/n} \\
&\eqlabel{a} \lim_{n\to\infty}
       \frac{-e^{-{(\Frac{\lambda}{n})}(1 - e^{-\eta})}(1 - e^{-\eta})\lambda /n^2}
            {-1/n^2} \\
&= \lim_{n\to\infty}
       e^{-{(\Frac{\lambda}{n})}(1 - e^{-\eta})}(1 - e^{-\eta})\lambda \\
&= \lambda (1 - e^{-\eta}),
\end{align*}
where (\emph{a}) follows from L'H\^{o}pital's rule.

\subsection{Proof of \eqref{eq:lim_1minusp_n}}

\vspace{\appEnegspace}

\begin{align*}
\lim_{n\to\infty} (1-p)^n
&\eqlabel{a} \lim_{n\to\infty} \left( \exp\!\left( -\frac{\lambda}{n} (1-e^{-\eta}) \right) \right)^n \\
&= \exp( -\lambda (1-e^{-\eta}) )
\end{align*}
where (\emph{a}) follows from substitution of \eqref{eq:p-def}.

\subsection{Proof of \eqref{eq:lim_E_Ytilde}}

\vspace{\appEnegspace}

\begin{align*}
  \lim_{n\to\infty} \iE{\Ytilde_k}
  &\eqlabel{a} \lim_{n\to\infty} \frac{\lambda \eta}{n p} \\
  &\eqlabel{b} \lim_{n\to\infty} \frac{\lambda \eta}{\lambda(1 - e^{-\eta})} 
  = \frac{\eta}{1-e^{-\eta}} ,
\end{align*}
where (\emph{a}) follows from substitution of \eqref{eq:ytilde-mean}; and
(\emph{b}) from substitution of \eqref{eq:lim_np}.

\subsection{Proof of \eqref{eq:lim_var_Ytilde}}

\vspace{\appEnegspace}

\begin{align*}
  \lim_{n\to\infty} \ivar{\Ytilde_k}
  &\eqlabel{a} \lim_{n\to\infty}
         \frac{1}{p}   \!\left(\frac{\lambda}{n}\eta + \frac{\lambda}{n}\eta^2 + \left(\frac{\lambda}{n}\eta\right)^{\!2}\right)
       - \frac{1}{p^2} \!\left(\frac{\lambda}{n}\eta\right)^{\!2} \\
  &= \lim_{n\to\infty}
           \frac{\lambda\eta}{n p} + \frac{\lambda\eta^2}{n p} + \frac{\lambda^2\eta^2}{n^2 p}
           - \frac{\lambda^2\eta^2}{n^2p^2} \\
  &\eqlabel{b} \lim_{n\to\infty}
           \frac{\lambda\eta + \lambda \eta^2}{\lambda(1-e^{-\eta})} 
           - \frac{\lambda^2\eta^2}{\lambda^2 (1-e^{-\eta})^2} \\
  &= \frac{\eta - (\eta+\eta^2)e^{-\eta}}{(1-e^{-\eta})^2} ,
\end{align*}
where (\emph{a}) follows from substitution of \eqref{eq:ytilde-var}; and
(\emph{b}) from substitution of \eqref{eq:lim_np} and noting that the third term vanishes.

\bibliography{TCI_Main}

\end{document}